\documentclass[twocolumn,tighten,iop]{aastex63}
\usepackage{amsmath}
\usepackage{iondefs, farhan-defs}
\usepackage[super]{nth}

\hypersetup{colorlinks=true, citecolor=blue, filecolor=magenta,urlcolor=blue, linkcolor=blue}

\begin{document}


\title{\sc Evolution of {\CIV} Absorbers. I.~The Cosmic Incidence}


\author[0000-0002-0072-0281]{Farhanul Hasan}
\affiliation{Department of Astronomy, New Mexico State University, Las Cruces, NM 88003, USA}

\author[0000-0002-9125-8159]{Christopher W. Churchill}
\affiliation{Department of Astronomy, New Mexico State University, Las Cruces, NM 88003, USA}

\author[0000-0002-6434-4684]{Bryson Stemock}
\affiliation{Department of Astronomy, New Mexico State University, Las Cruces, NM 88003, USA}

\author[0000-0002-0146-2368]{Nigel L. Mathes}
\affiliation{Department of Astronomy, New Mexico State University, Las Cruces, NM 88003, USA}

\author[0000-0003-2377-8352]{Nikole M. Nielsen}
\affiliation{Centre for Astrophysics and Supercomputing, Swinburne University of Technology, Hawthorn, Victoria 3122, Australia}
\affiliation{ARC Centre of Excellence for All Sky Astrophysics in 3 Dimensions (ASTRO 3D)}

\author[0000-0002-0496-1656]{Kristian Finlator}
\affiliation{Department of Astronomy, New Mexico State University, Las Cruces, NM 88003, USA}

\author[0000-0002-9009-6768]{Caitlin Doughty}
\affiliation{Department of Astronomy, New Mexico State University, Las Cruces, NM 88003, USA}

\author[0000-0002-2908-9702]{Mark Croom}
\affiliation{Department of Astronomy, New Mexico State University, Las Cruces, NM 88003, USA}

\author[0000-0003-1362-9302]{Glenn G. Kacprzak}
\affiliation{Centre for Astrophysics and Supercomputing, Swinburne University of Technology, Hawthorn, Victoria 3122, Australia}
\affiliation{ARC Centre of Excellence for All Sky Astrophysics in 3 Dimensions (ASTRO 3D)}

\author[0000-0002-7040-5489]{Michael T. Murphy}
\affiliation{Centre for Astrophysics and Supercomputing, Swinburne University of Technology, Hawthorn, Victoria 3122, Australia}

\correspondingauthor{Farhanul Hasan} 
\email{farhasan@nmsu.edu}


\shorttitle{Evolution of {\CIV} Absorbers I}\shortauthors{Hasan et al.}


\begin{abstract}

We present a large high-resolution study of the distribution and evolution of {\CIV} absorbers, including the weakest population with equivalent widths ${\EWr} < 0.3$ {\AA}. By searching 369 high-resolution, high signal-to-noise spectra of quasars at $1.1 \leq {\zem} \leq 5.3$ from Keck/HIRES and VLT/UVES, we find $1268$ {\CIV} absorbers with {\wvweak} (our $\sim\!50\%$ completeness limit) at redshifts {\zrangeall}. A Schechter function describes the observed equivalent width distribution with a transition from power-law to exponential decline at ${\EWr} \gtrsim 0.5$ {\AA}. The power-law slope $\alpha$ rises by $\sim7\%$ and transition equivalent width $W_{\star}$ falls by $\sim\!20\%$ from $\langle z \rangle=1.7$ to $\langle z \rangle=3.6$. 
We find that the co-moving redshift path density, {\dndx}, of {\wvweak} absorbers rises by $\sim1.8$ times from $z\simeq 4.0$ to $z\simeq 1.3$, while the {\wstrong} {\dndx} rises by a factor of $\sim8.5$. We quantify the observed evolution by a model in which {\dndx} decreases linearly with increasing redshift. The model suggests that populations with larger {\EWr} thresholds evolve faster with redshift and appear later in the universe. The cosmological {\sc Technicolor Dawn} simulations at $z=3-5$ over-produce the observed abundance of absorbers with ${\EWr} \leq 0.3$ {\AA}, while yielding better agreement at higher {\EWr}. Our empirical linear model successfully describes {\CIV} evolution in the simulations and the observed evolution of {\wstrong} {\CIV} for the past $\sim\!\!12$ Gyr. Combining our measurements with the literature gives us a picture of {\CIV}-absorbing structures becoming more numerous and/or larger in physical size over the last $\approx\!\!13$ Gyr of cosmic time ($z\sim6$ to $z\sim0$).

\end{abstract}


 \received{July 22, 2020}
 \revised{September 28,2020}
 \accepted{October 2, 2020}
 \submitjournal{\apj}

\section{Introduction}

The processing of baryons within and between galaxies is fundamental to how galaxies form and evolve, and is an important phenomenon in shaping the universe as we observe it today. Obtaining a detailed census of the cosmic distribution and evolution of metal-enriched gas is a key step in understanding this ``baryon cycle''. 

According to our current understanding, metals form in stars within galaxies and are expelled out to the circumgalactic medium (CGM) and the intergalactic medium (IGM) via a variety of feedback processes \cite[e.g.,][]{Lilly13,DM14,SD15,KF17}. Over billions of years, some of these metals are recycled back into the galaxies whence they originated, possibly repeating the cycle several times, whereas others may escape the galaxy and enter the IGM \citep[e.g.,][]{OD08,Ford14,Muratov15,Muratov17}. Some fraction of these metals in the IGM may eventually accrete into the CGM or interstellar medium of other galaxies \citep[e.g.,][]{Oppenheimer10,Brook14,AA17}. Metals in the gas phase are therefore an invaluable and crucial tracer of the baryon cycle \citep[e.g.,][]{Steidel93, cwc99, glenn12, nikki15, Christensen16, Tumlinson17,  lehner19}. 

Fortunately, CGM and IGM gas-phase metals are directly observable using quasar absorption-line spectroscopy, as long as the gas is intervening to the line of sight of a background quasar. The {\CIVdblt} resonant fine-structure doublet is one of the most commonly observed metal absorption features in quasar spectra, and has been studied extensively over the redshift range $0\lesssim z \lesssim 6.5$ \citep[e.g.,][]{Sargent88, Steidel90, PB94, Rauch96, Chen01, Schaye03, Adelberger05, Songaila05, Scanna06, Becker09, Cooksey10, Cooksey13, Dodo10, Dodo13, Bordoloi14, Shull14, Burchett15, BS15, Codoreanu18, Cooper19}. 

The measured global cosmic evolution of {\CIV} absorbers provides insights into the universal carbon content of the universe \citep[e.g.,][]{DS08, Simcoe11, Rafelski12} and the ionizing ultraviolet background \citep[UVB, e.g.,][]{FG09, HM12, Becker13, Doughty18} as traced by moderately low density warm temperature gas structures, i.e., $\log n_{\hbox{\tiny H}} \sim -3$ [cm$^{-3}$] and $\log T \sim 4.8$~[K] \citep[e.g.,][]{Steidel90, Bergeron05, cwc15}. Two of the most important quantities that quantify the distribution and evolution of {\CIV} absorbing gas are the co-moving path number density, $dN/dX$, which is the number of absorbers per unit ``absorption path'' \citep{BP69}, and the equivalent width distribution (EWD), which provides the number of absorbers per unit equivalent width per unit absorption path ($d^2\!N/dWdX$).

To date, the largest survey of {\CIV} absorbers was conducted by \citet[][hereafter \citetalias{Cooksey13}]{Cooksey13}, who located ${\sim\!15,000}$ absorbers at $1.5\!\lesssim\! z \!\lesssim\!4.5$ in ${\sim\!26,000}$ quasar spectra from the Sloan Digital Sky Survey \citep[SDSS,][]{York00}, comprising a co-moving path length $\Delta X \!\simeq\!38,600$. They placed robust constraints on the evolution of strong absorbers, i.e., those with rest-frame {\CIV}~{\strongdblt} equivalent widths ${\wrciv}\!\geq\!0.6$~{\AA} (hereafter, $W_r$). Their sensitivity to detecting smaller $W_r$ absorbers decreased quickly for $W_r < 0.6$~{\AA}, reaching $\sim\!20\%$ completeness at ${W_r = 0.3}$~{\AA}.  

For {\wstrong} absorbers, \citetalias{Cooksey13} found that {\dndx} increased by a factor of  $\approx\!2.5$ from $z = 4.5$ to $z = 1$. Their highly constrained measurement of the increase in {\dndx} with cosmic time is consistent with the trends of less constraining findings from smaller surveys \citep[e.g.,][]{Sargent88, Misawa02, Peroux04, Dodo16}, some of which probed smaller $W_r$ thresholds.  Incorporating absorbers with $W_r \geq 0.3$~{\AA}, \citetalias{Cooksey13} found that the EWD is well-described by an exponential function, a result consistent with the earliest surveys \citep{Sargent88, Steidel90}.

Unfortunately, the available information characterizing ``weak'' {\CIV} absorbers with $W_r < 0.3$~{\AA} across various redshift ranges is limited \citep[]{Schaye03, Schaye07, Scanna06, Cooksey10, Dodo10, BS15, Burchett15}. 
As such, the distribution, statistical properties, and evolution of the $W_r < 0.3$~{\AA} regime of {\CIV} is yet to be fully characterized. Charting this regime is important, as it probes moderately low density warm gas structures optically thin to hydrogen and helium ionizing photons from the ionizing spectrum, which means sensitivity to the evolving shape of the UVB can be directly probed with cosmic time. Furthermore, {\dndx} of weak {\CIV} absorbers can, in a statistical sense, inform us of the distribution of carbon and other metals in optically thin gas structures relative to galaxies as a function of redshift.

Hydrodynamic cosmological simulations have often used {\CIV} as a target metal-line absorber to test theoretical predictions against observations, hence improving our understanding of the baryon cycle in the broader context of galaxy evolution \citep[e.g.,][]{OD08,ODF09,CC11,Rahmati16,Bird16,KF15,KF20}. One of the reasons for this is the high cosmic abundance and oscillator strength of the C$^{+3}$ ion. Moreover, C$^{+3}$ serves as a higher-energy complement to the {\Lya} forest as a probe of the metagalactic UVB, due to the ground-state ionization potential of C$^{+2}$ and C$^{+3}$ being 3.5~Ryd and 4.7 Ryd, respectively. {\CIV} is also an independent probe in the ionization energy range of He$^{0}$ (1.8~Ryd) and He$^{+}$ (4.0~Ryd). Providing as stringent observational constraints as possible on the distribution, incidence, and evolution of the broadest range of {\CIV} absorbing gas structures will be key to driving such studies and progressing our understanding of the distribution of metals and the evolution of the ionizing UVB.

To improve the global statistics of {\CIV} absorbers and provide robust statistics of weak absorbers, we searched for {\CIVdblt} doublet absorption in several hundred high-resolution quasar spectra from both the Keck/HIRES \citep{hires94} and VLT/UVES \citep{uves00} instruments. We employed an automated line detection approach similar to that of \citet{Zhu13} and intensive visual inspection to identify and verify each absorption doublet. The high sensitivity of our survey allows us to probe the distribution and evolution of {\CIV} absorbers an order of magnitude weaker than those studied by \citetalias{Cooksey13} (whose spectral resolution was $\gtrsim\!20$ times lower). In summary, we analyzed over 1300 {\CIV} absorbers with $W_r \geq 0.05$~{\AA} spanning redshifts {\zrangeall} identified in the HIRES and UVES spectra of 369 quasars distributed over redshifts $1.1 \leq {\zem} \leq 5.3$.


In this paper, we present our survey methods and results, and our inferences about the nature of {\CIV} absorbers. The survey design, methods for identifying {\CIV} absorption doublets, calculation of the redshift-dependent sensitivity and detection completeness limits, and measurements of the absorption properties are presented in Section~\ref{data}. In Section~\ref{results}, we describe our analysis methods and present the observed EWD and {\dndx} of {\CIV} absorbers. We also introduce an empirical model to quantify the observed evolution. In Section~\ref{disc}, we discuss the observed evolution and our model extrapolations, draw from the literature to augment the redshift range of our survey, and compare our observations with a mock {\CIV} survey generated from the {\sc Technicolor Dawn} hydrodynamic cosmological simulations \citep{KF18}. We provide our concluding remarks in Section~\ref{conc}. When necessary, and for consistent comparison with \citetalias{Cooksey13}, we adopt the WMAP5 cosmology, with $H_{0}=71.9$ {\kmsmpc}, $\Omega_{\hbox{\tiny M}} = 0.258$, and $\Omega_{\Lambda} = 0.742$ \citep{WMAP09}. 

\section{Data and analysis} 
\label{data}

\subsection{Quasar spectra} 
\label{sec:spec}

We have analyzed 369 high resolution, high signal-to-noise quasar spectra obtained with the HIRES spectrograph \citep{hires94} and the UVES spectrograph \citep{uves00} from the Keck and Very Large Telescope (VLT) observatories, respectively. The wavelength coverage of the spectra spans approximately $3000-10000$~{\AA}, though there is variable coverage in this range from spectrum to spectrum. 
Since these spectra were acquired using various decker settings and slit widths, the spectral resolutions ranged between $R = \lambda/\Delta\lambda = 36,000$--$103,000$. Only a minority of the spectra were not obtained using the most commonly adopted resolving power of both HIRES and UVES, which is $R \approx 45,000$, or $\sim 6.6$ {\kms}, with 3 pixels per FWHM resolution element.
The FWHM velocity resolution is approximately constant as a function of observed wavelength. The signal-to-noise (S/N) ratios in the regions of {\CIV} search space range between 2 and 150 per pixel, with a mean S/N of $\sim20$. The \nth{25} and \nth{75} percentile S/N are 10 and 25, respectively. 
We do not include Broad Absorption Line (BAL) quasars in our sample.

Roughly half of the Keck/HIRES spectra were obtained from various observing programs prior to the creation of the Keck Observatory Archive,\footnote{\url{https://www2.keck.hawaii.edu/koa/public/koa.php}} some having been donated in science-ready form by Charles Steidel, J. Xavier Prochaska, Christopher Churchill, Michael Rauch and the late Wallace Sargent. The remaining Keck/HIRES spectra were obtained from Data Release 1 of the KODIAQ project \citep{OMeara15}. The VLT/UVES spectra were acquired through the efforts of the UVES SQUAD prior to their Data Release 1 \citep{Murphy19}. 

The journal of quasar spectra used in this study is listed in Table~\ref{tab:journal}. For each quasar, the columns list (1) the quasar name, taken from \citet{vcv01}, (2) the emission redshift, (3) the instrument with which the spectrum was obtained, (4) the minimum wavelength of the {\CIV} search space in the spectrum, set by {\Lyawave} emission of the quasar (see Section~\ref{sec:detect}), and (5) the maximum wavelength of the {\CIV} search space, set by 5000 {\kms} blueward of the {\CIV} $\lambda1548$ emission of the quasar (Section~\ref{sec:detect}).

\begin{deluxetable}{lclcc}[t]
\centering
\tabletypesize{\small}
\tablewidth{0pt}
\tablecaption{Journal of Observations\label{tab:journal}}
\tablehead{
\colhead{Quasar} &
\colhead{$z_{em}$} &  \colhead{Instrument(s)} & 
\colhead{$\lambda_{min}$\tablenotemark{\scriptsize a}}  & \colhead{$\lambda_{max}$\tablenotemark{\scriptsize b}} \\[-5pt]
\colhead{} &
\colhead{} &  
\colhead{} & 
\colhead{({\AA})} & 
\colhead{({\AA})}
} 
\startdata
J002830$-$281704 & 2.400 & HIRES & 4133 & 5176 \\ [-3pt]
J010104$-$285801 & 3.070 & UVES & 4947 & 6196 \\ [-3pt]
J064632$+$445116 & 3.408 & UVES & 5358 & 6710 \\ [-3pt]
J102325$+$514251 & 3.447 & HIRES & 5406 & 6770 \\ [-3pt]
J105744$+$062914 & 3.142 & UVES & 5035 & 6305 \\ [-3pt]
J110045$+$112239 & 4.707 & HIRES & 6937 & 8688 \\ [-3pt]
J120207$+$323538 & 5.292 & HIRES & 7648 & 9578 \\ [-3pt]
J155556$+$480015 & 3.299 & HIRES & 5226 & 6544 \\ [-3pt]
J201717$-$401924 & 4.131 & UVES & 6237 & 7811 \\ [-3pt]
J235714$-$273659 & 1.732 & UVES & 3321 & 4159 \\
\enddata 
\vspace{+5pt}
$^{\scriptsize a}$ Minimum wavelength of {\CIV} search space, defined by {\Lyawave} emission of the quasar (see Section~\ref{sec:detect}) $^{\scriptsize b}$ Maximum wavelength of {\CIV} search space, 5000 {\kms} blueward of the {\CIV} $\lambda1548$ emission of the quasar
\tablecomments{Table~\ref{tab:journal} is published in its entirety in machine- readable format. A portion containing a representative selection of quasars is shown here for guidance regarding its form and content.}
\vspace{-25pt}
\end{deluxetable}

\begin{figure*}[htb]
\centering
\includegraphics[width=0.96\textwidth]{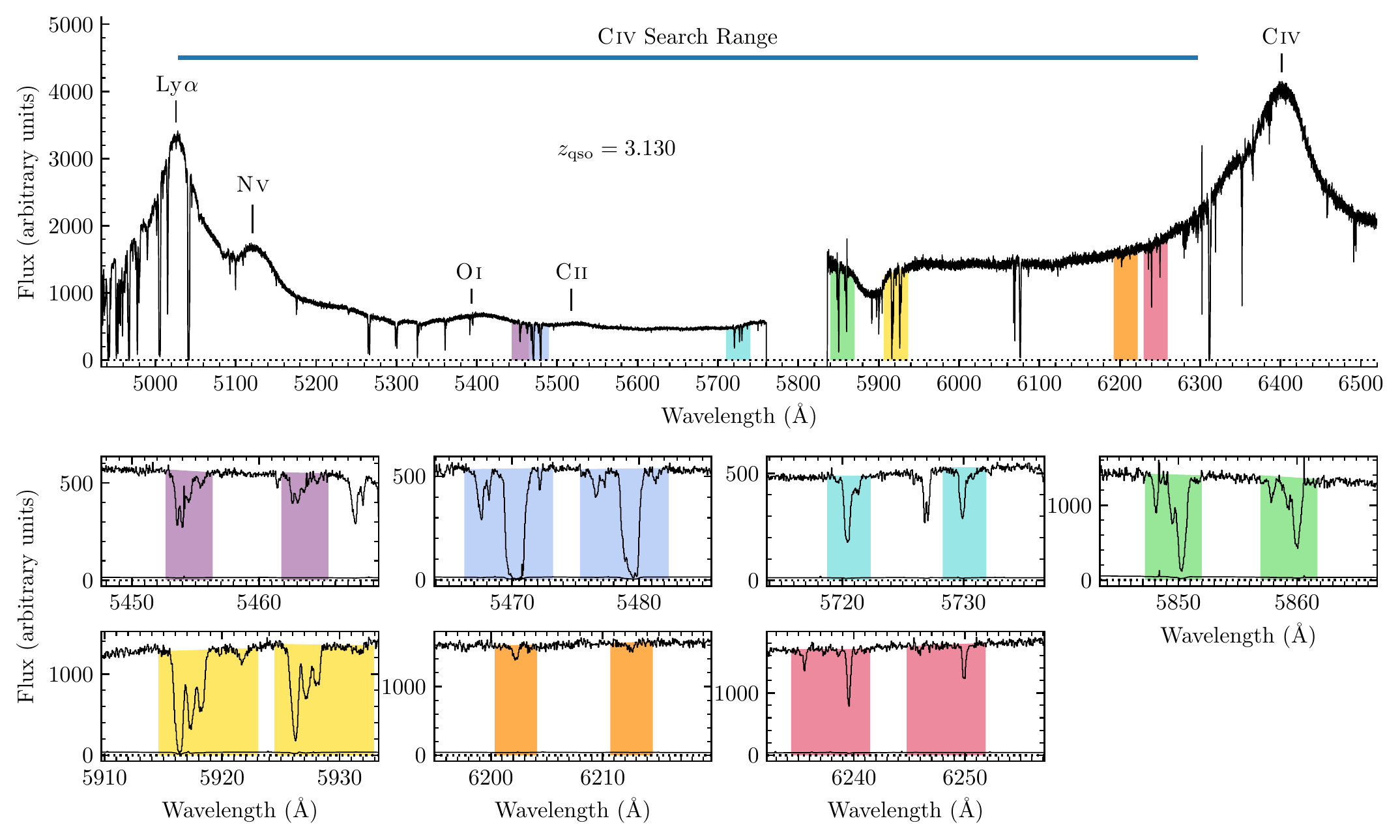}
\caption{An example spectrum of the quasar J103909$-$231326 at {\zem}$=3.130$ observed with VLT/UVES and seven of its {\CIVdblt} doublets presented in redshift order, $z=2.5228$ (purple), $z=2.5333$ (blue), $z=2.6949$ (teal), $z=2.7787$ (green), $z=2.8216$ (yellow), $z=3.0061$ (orange), and $z=3.0302$ (red). The search range for {\CIV} doublets is indicated, as are some emission lines from the quasar. Note the coverage gap at $\sim5760$--$5840$ {\AA}.}
\label{fig:spectrum}
\vglue +5pt
\end{figure*}

\vspace{+5pt}

\subsection{Reduction of Spectra}
\label{subsec:reduction}

The KODIAQ spectra were reduced according to the prescriptions of \citet{OMeara15}. The spectra of Churchill were reduced using the standard Image Reduction and Analysis Facility (IRAF\footnote{IRAF is distributed by the National Optical Astronomy Observatories, which are operated by the Association of Universities for Research in Astronomy, Inc., under cooperative agreement with the National Science Foundation.}), as outlined in \cite{cwcthesis}. Those of Sargent, Rauch, Prochaska, and Steidel were reduced using the Mauna Kea Echelle Extraction (MAKEE) data reduction package, developed by Tom Barlow \citep{barlow05}, which is optimized for the spectral extraction of single, unresolved point sources. All HIRES spectra were wavelength calibrated using ThAr lamps to the vacuum heliocentric standard at rest and were continuum fit by their respective donors.

The UVES spectra were reduced using the UVES pipeline in the MIDAS environment \citep{uves00}, which is provided by the European Southern Observatory (ESO).  The wavelength solution is determined using a standard ThAr lamp exposure corrected to vacuum heliocentric velocities. The individual exposures were then combined into one-dimensional spectra using the UVES POst-Pipeline Echelle Reduction ({\sc uves\_popler}) software \citep{MurphyPOPLER, Murphy19}.

Using the interactive mode of the {\sc Popler} software, we examined and refined the continuum fits for all quasar spectra with particular attention to the spectral regions redward of the quasar {\Lyawave} emission line following the prescriptions of \citet{King12} and \citet{Murphy16}. An example UVES spectrum of the {\zem}$=3.130$ quasar J103909$-$231326 is shown in Figure~\ref{fig:spectrum}. Strong emission lines are labelled, and the spectral region where we searched for {\CIVdblt} doublet lines is shown (see Section~\ref{sec:detect}). {\CIV} absorbers detected at different redshifts are highlighted.

\subsection{Identifying and Verifying Absorption Doublets} 
\label{sec:detect}

We surveyed the spectra for the {\CIVdblt} resonance doublet. Such high quality spectra allow for an in-depth investigation into the cosmic frequency of the very weakest {\CIV} absorbers, a regime that heretofore has not been explored in great detail.

To detect intervening {\CIVdblt} absorption in our quasar spectra, we confined the search range to regions of the spectrum redward of the {\Lya} emission, as {\Lya} forest contamination would render automated detection of weak metal lines nearly impossible. 
To avoid selecting {\CIV} absorbers associated with the quasar environment, we also restrict our search range to velocities blueshifted by ${\dvqso} > 5000~{\kms}$ from the quasar emission redshift. This criterion is consistent with that of \citetalias{Cooksey13}, allowing a direct comparison. We find that choosing ${\dvqso} > 3000~{\kms}$ instead yields a final science sample (see Section~\ref{sec:CofW}) that is 4\% larger, changing our measured properties of {\CIV} absorbers by no more than $\approx\!1\%$.  This level of sensitivity to the choice of the velocity cutoff is consistent with \citetalias{Cooksey13}, who found that their statistics were sensitive at no more than the $2\%$ level.

Moreover, strong telluric absorption bands at $6868$--$6932$~{\AA} (B-band) and $7594$--$7700$~{\AA} (A-band) are excluded as these molecular lines can have separations and ratios leading to false positives and false negatives when searching for {\CIV} doublets. 
Though additional telluric features are present at red wavelengths in the spectra, we searched these regions.  These weaker telluric features could also result in false positives and false negatives. With regards to false positives, we visually inspected all candidate absorbers, a process we describe in detail below. With regards to false negatives, we can treat these using completeness corrections, which we obtained using Monte Carlo methods to determine the detection sensitivity threshold (described in Section~\ref{sec:gwz}) as a function of wavelength in each individual spectrum.

Our automated search routine is based on the methods of \citet{Zhu13}. We performed a matched filter search for {\CIVdblt} doublet candidates detected by signals exceeding $5\sigma$ in local noise of the resulting output signal spectrum. We used a two-component top hat filter separated by the wavelength difference of the {\CIV} doublet, where each component has a width of seven pixels, which is 2.33 times the FWHM of an unresolved absorption feature. We convolved the filter with the normalized spectrum to generate the output signal spectrum in redshift space.

To measure the uncertainty in the output signal spectrum at a given pixel, we examined a $\Delta z \simeq 0.01$ region corresponding to roughly 3000 pixels centered on the pixel. We used iterative sigma-clipping ($1\sigma$) to remove outlier signals, leaving only the continuum of the output signal spectrum. We calculated the standard deviation of this continuum and adopted it as the noise in the output spectrum at the centered pixel. Further details are provided in \citet{nigelthesis}.

We also measured the equivalent width detection limit as a function of redshift for all quasar spectra. For each spectrum, we inserted Gaussian {\CIV} absorption doublets in intervals of seven pixels. We assumed a Gaussian width, $\sigma_l$, equal to the instrumental line spread function to represent unresolved absorption lines. We then solved for the amplitude, $A_l$, of the Gaussian required to detect the unresolved line at a $5\sigma$ significance level with our matched filtering technique.  Finally, we integrated to find the equivalent width, $W = \sqrt{2\pi}A_l \sigma_l$, which we adopted as the minimum detectable {\it observed\/} equivalent width at that redshift. Converting to the rest-frame equivalent width at each redshift yielded the detection threshold spectra.  These spectra allow us to accurately characterize the completeness of our sample and are instrumental in quantifying the redshift path covered by the survey.

Once candidate {\CIV} features are flagged, we define an ``absorber'' to include all absorption in {\CIVdblt} lines detected within $\pm 500$~{\kms} of one another. 
As the rest-frame velocity separation of the {\CIV} doublet is $\Delta v =498$~{\kms}, a $\pm 500$~{\kms} window naturally captures all velocity components within the velocity window in which the {\CIV} $\lambda 1548$ and $\lambda 1550$ become self-blended. Further, a window of $\pm 500$~{\kms} captures all velocity components within the observed velocity dispersion of CGM gas in individual halos and/or small bound groups \citep[e.g.,][]{ PB90, PB94, cwc01, cos-halos11, nikki15, nikki16, nikki17, nikki18, pointon17, glenn19, mason19, cwc20}.  Smaller velocity windows \citep[e.g. ][]{Dodo13}, will likely arbitrarily break single absorbing structures into multiple absorbers. Our $\pm 500$~{\kms} window is not considered to be too large because the line-of-sight velocity clustering function of {\CIV} absorbers has a peak within 500~{\kms} and is flat beyond that velocity \citep{Sargent88}, so there is little chance of arbitrarily grouping multiple {\CIV} systems into a single absorber. We note that there were no systems in our sample for which the 500~{\kms} window resulted in a borderline or ambiguous case of assigning velocity components with a single absorber.

Each candidate absorber was examined visually using our interactive graphical software {\sc Sysanal} \citep[see][]{cwcthesis,cwc99,cwc01}. In {\sc Sysanal}, the absorption profiles of the doublets are aligned in velocity space. Following the methods of \citet{Schneider93}, the software  employs the ``equivalent width spectrum'' to objectively locate the absorption features. We used a $5\sigma$ significance level for the {\strongdblt} transition and a $3\sigma$ significance level for the {\weakdblt} transition, where the latter is examined only in the velocity range over which {\strongdblt} absorption is detected. Adopting the method described in \citet{cwc99} and \citet{cwc01}, the velocity extremes of each absorption feature are determined when the equivalent width spectrum returns to the $1\sigma$ level. Following visual inspection and possible editing of the defined absorption regions, the software calculates the optical depth-weighted median absorption redshift, {\zabs}, setting the final velocity zero point for the absorption \citep[see][]{cwc01}. Also computed are the equivalent widths of the doublet members, apparent optical depth (AOD) column density, and doublet ratios. 
Uncertainties due to the placement of the quasar continuum are propagated into the equivalent width uncertainty measurements in {\sc Sysanal} \cite[see][]{cwcthesis,cwc99,cwc01}.

With the absorber redshift determined, we examined the full quasar spectrum using our software {\sc Search} \citep[see][]{cwc99} to examine whether associated absorption accompanied the {\CIV} absorption. This helped us identify associated transitions such as {\Lyawave}, {\SiIVdblt}, {\MgIIdblt}, or other quasar absorption lines commonly found in metal-bearing gas clouds.

For an absorber to be included in our final science sample, we employed a three-person iterative inspection system. Each of the three individuals (F.H., B.S., and C.W.C.) was assigned a subset of candidate absorbers for which they were the primary inspector. The primary inspector would assign a rating of 0 to 10 for each of the candidate {\CIV} absorbers, where 0 indicated zero confidence and 10 indicated total confidence in the reality of the {\CIV} absorber.
In particular, it was checked whether the doublet ratios, ${\wrciv}/W_r (1551)$,  were between 1 and 2 (within uncertainties) as expected from the oscillator strengths of the doublet members. Alignment of the absorption profile shapes in velocity space was another critical aspect of this check. 
The primary inspector also provided a second rating of 0 to 10 in their confidence in the velocity limits obtained with {\sc Sysanal} as described above. These limits are critical in defining which pixels are included in the equivalent width calculations.

\begin{figure*} [ht]
\centering
\gridline{\fig{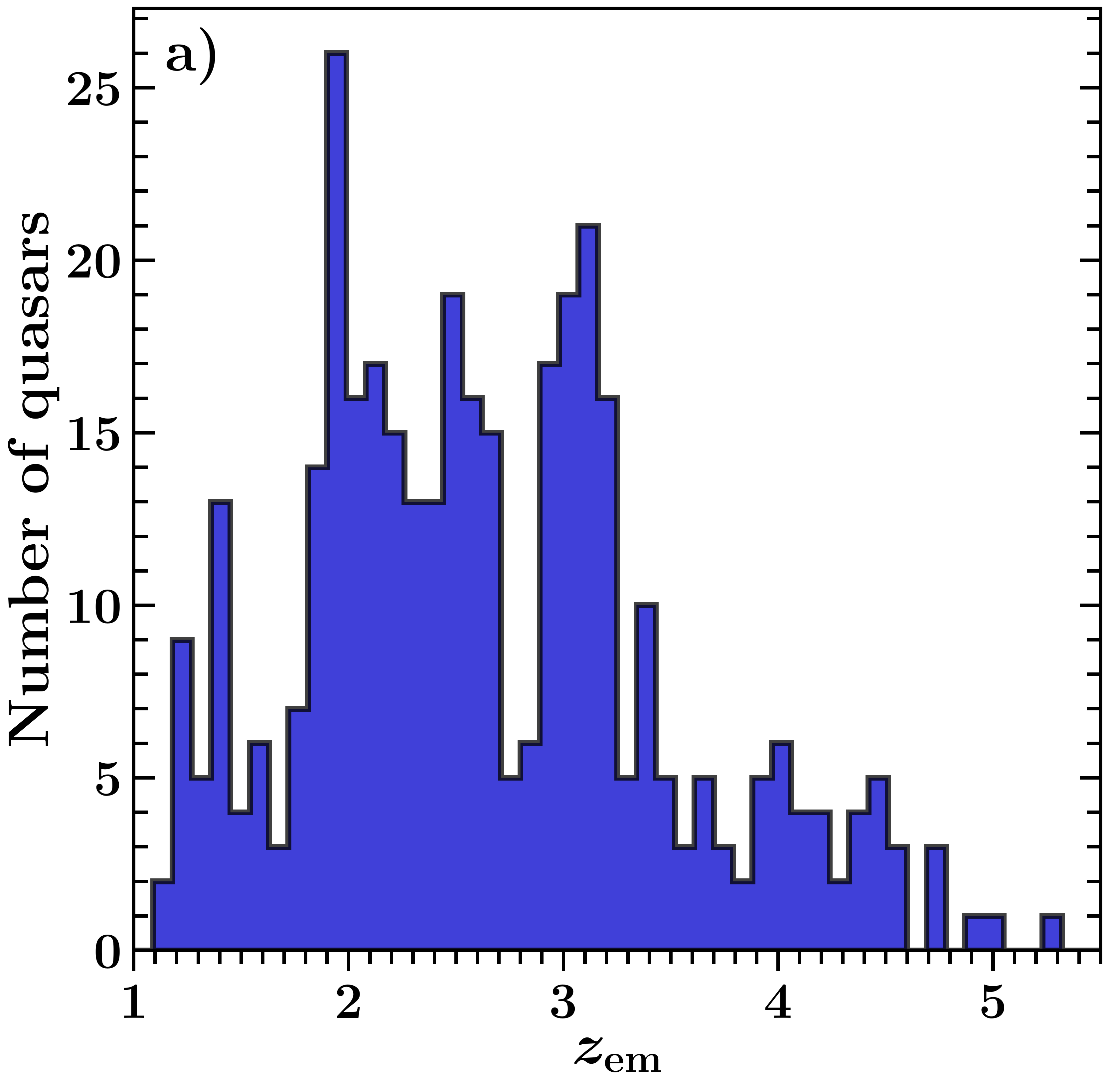}{0.32\textwidth}{} 
        \fig{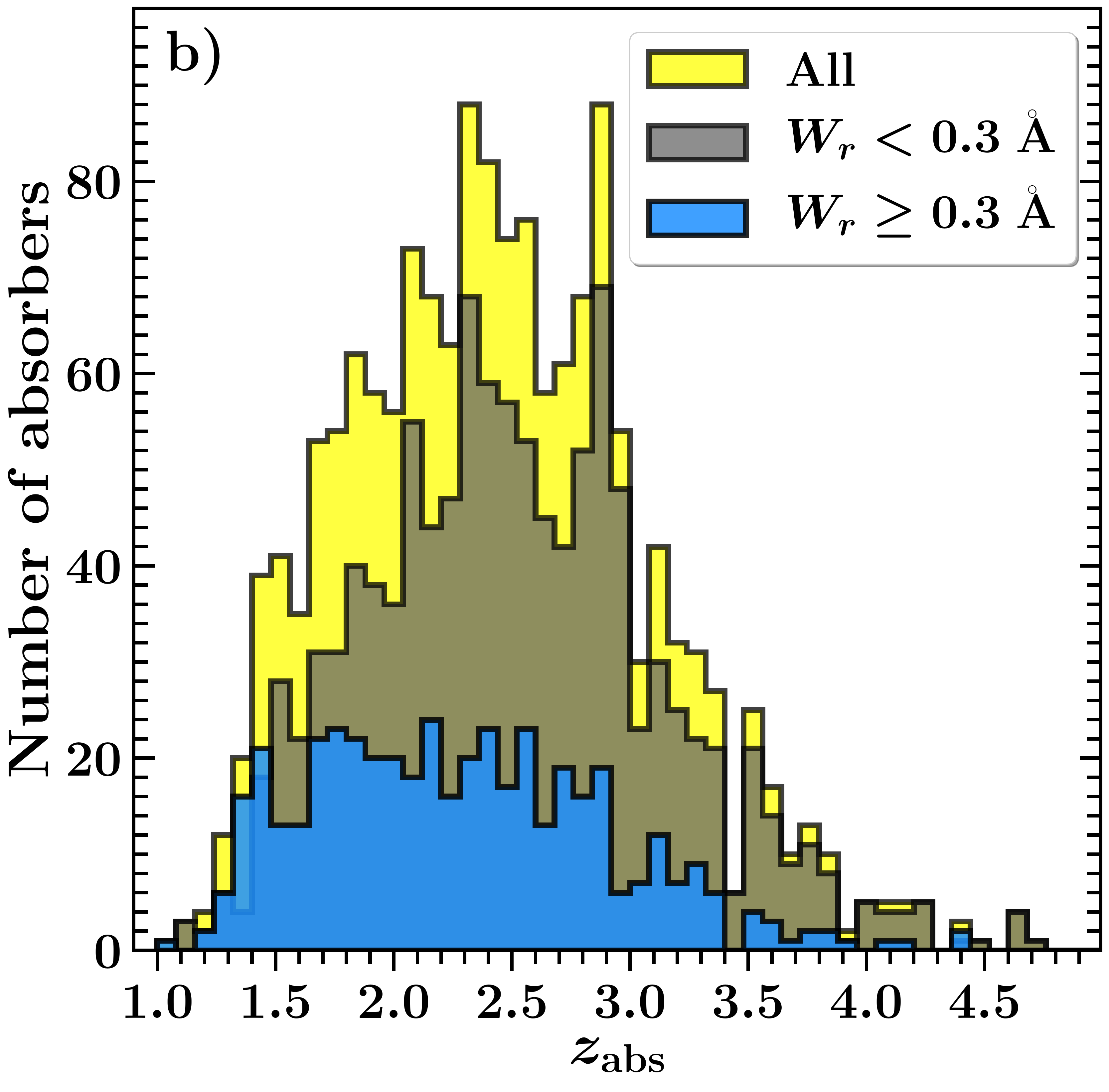}{0.32\textwidth}{}
        \fig{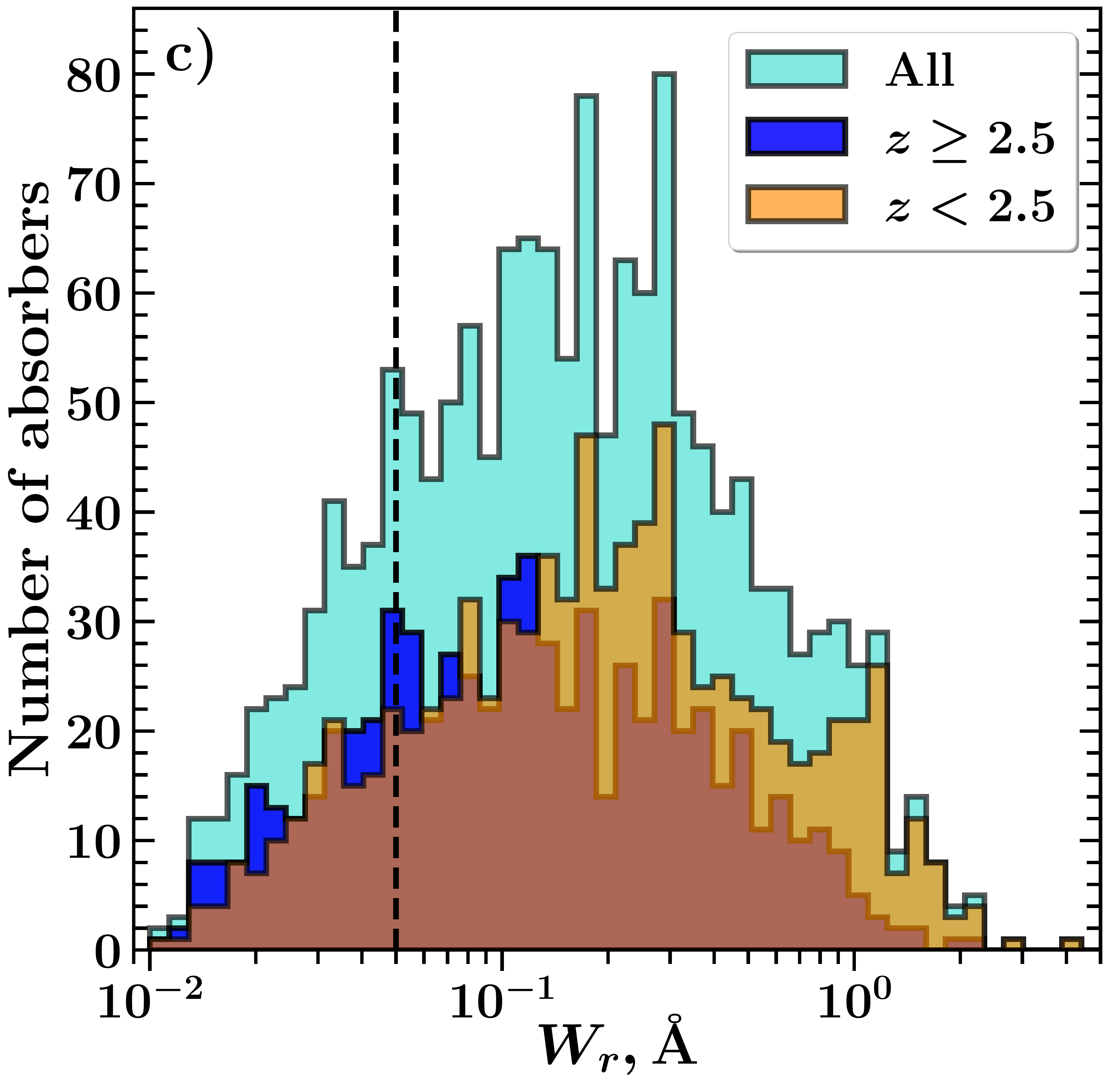}{0.32\textwidth}{}} \vspace{-25pt}
\caption{(a) The distribution of quasar emission redshifts, {\zem}. (b) The distribution of {\CIV} absorber redshifts, {\zabs}. Weak absorbers, i.e., those with ${\EWr} < 0.3$~{\AA} (shown in dark grey), are unique to this survey. (c) The distribution of rest-frame equivalent widths, {\EWr}, for all redshifts ({\zrangeall}), $z < 2.5$, and $z \geq 2.5$. The vertical dashed line is $W_r = 0.05$~{\AA}, where the full survey is $50\%$ complete (see Figure~\ref{fig:comp}).}
\label{fig:sample}
\end{figure*}

A rating of (10, 10) ensured inclusion into the science sample without further inspection, and a rating of (0, 0) ensured exclusion without further inspection. Any other ratings would require an independent assessment from a secondary inspector. The corroboration of associated transitions was taken as a strong indicator for a true {\CIV} absorber. If consensus could not be achieved as to the reality of the absorber, then the third inspector would arbitrate while all three individuals inspected and discussed the absorber. 
While conducting visual inspection of each absorption system, each of the multiple individuals who inspected a system made detailed notes on the possible blends with other lines. 
Roughly $3\%$ of our sample contained potential blends with other lines. The overwhelming majority of these were minor blends, so the measured  equivalent widths would not change to any significant degree. Therefore, the resulting statistics of our sample and the scientific conclusions of this paper would not be affected appreciably by these blends.

Through this process, we located a total of 1565 {\CIV} absorbers. For our final science sample, we use 1268 absorbers with equivalent widths above our $50\%$ completeness level of ${\EWr}=0.05$~{\AA} (see Section~\ref{sec:CofW}).
In Table~\ref{tab:abs}, we list the 1268 absorbers included in our final science sample. The columns are (1) the quasar name, (2) the absorption redshift, (3) the rest-frame equivalent width of the {\CIV} {\strongdblt} transition, {\EWr}, and (4) the AOD column density, $\log N$. Saturated lines are measured to have lower limits on column densities.  We do not utilize the measured column densities in our analysis.

\begin{deluxetable}{lccc}[bht]
\centering
\tabletypesize{\small}
\tablewidth{0pt}
\tablecaption{{\CIV} Absorbers \label{tab:abs}}
\tablehead{
\colhead{Quasar} &
\colhead{$z_{abs}$\tablenotemark{\scriptsize a} } &  
\colhead{$W_r$ ({\AA})\tablenotemark{\scriptsize b}} &  
\colhead{$\log N$\tablenotemark{\scriptsize c}} }
\startdata
J000149$-$015940 & 2.0107 & $0.082 \pm 0.002$ & $13.44^{+0.09}_{-0.10}$ \\
J003501$-$091817 & 2.2211 & $0.261 \pm 0.009$ & $13.92^{+0.03}_{-0.04}$ \\
J011150$+$140141 & 2.1365 & $0.504 \pm 0.012$ & $>14.37$\\
J074927$+$415242 & 2.7650 & $0.153 \pm 0.009$ & $13.68^{+0.02}_{-0.03}$ \\
J115411$+$063426 & 2.5634 & $0.058 \pm 0.003$ & $13.24^{+0.06}_{-0.07}$ \\
J123055$-$113909 & 3.2060 & $0.088 \pm 0.007$ & $13.41^{+0.04}_{-0.06}$ \\
J164656$+$551445 & 3.2887 & $0.167 \pm 0.007$ & $13.68^{+0.02}_{-0.03}$ \\
J214222$-$441929 & 2.8526 & $0.498 \pm 0.013$ & $14.30^{+0.02}_{-0.03}$ \\
J225719$-$100104 & 1.6613 & $0.259 \pm 0.007$ & $14.07^{+0.03}_{-0.04}$ \\
J235702$-$004824 & 2.7448 & $0.054 \pm 0.006$ & $13.16^{+0.04}_{-0.07}$ \\
\enddata 
\vspace{+5pt}
$^{\scriptsize a}$ Absorber redshift     $^{\scriptsize b}$ Equivalent width of the {\CIV} {\strongdblt} transition    $^{\scriptsize c}$ AOD column density. Lower limits are denoted with $>$.
\tablecomments{Table~\ref{tab:abs} is published in its entirety in machine- readable format. A portion containing a representative selection of absorbers is shown here for guidance regarding its form and content.}
\vspace{-25pt}
\end{deluxetable}

\begin{figure*}[thp]
\centering
\includegraphics[width=0.95\textwidth]{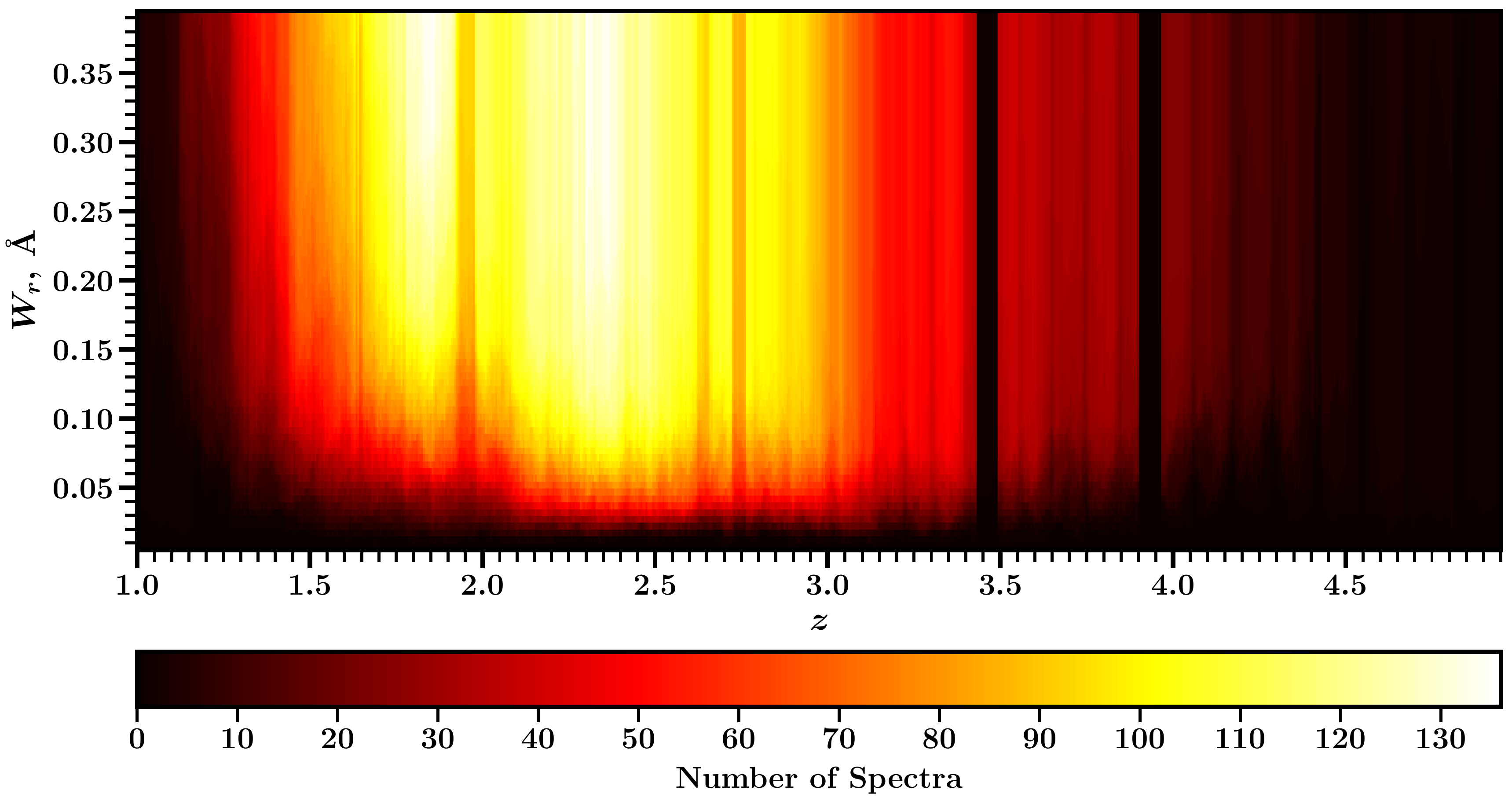} 
\caption{ The sensitivity function, $g(W_r,z)$. The heat map details the number of quasar spectra in which we could have detected {\CIV} absorption with a rest-frame equivalent width of $W_r$ or greater at redshift $z$ at the $3\sigma$ significance level. Note the dark vertical bands at $z\sim3.5$ and $z\sim3.9$ are due to the excluded atmospheric B and A bands, respectively.}
\label{fig:gwz}
\end{figure*}

\subsection{Sample Characteristics} \label{sec:sample}

In Figure~\ref{fig:sample}(a), we show the distribution of quasar emission redshifts, {\zem}. In Figure~\ref{fig:sample}(b), we present the distribution of absorber redshifts, {\zabs}, for our sample of {\CIV} absorbers. As this survey is the first large survey sensitive to ``weak" absorbers, i.e., those with $W_r < 0.3$~{\AA}, we indicate the distribution of redshifts for those absorbers below and above ${\EWr} = 0.3$~{\AA}. 
The weak absorbers dominate our distribution. In Figure~\ref{fig:sample}(c), we present the distribution of {\CIV}~{\strongdblt} rest-frame equivalent widths, {\EWr}, of the detected absorbers for all redshifts ({\zrangeall}), $z < 2.5$, and $z \geq 2.5$.

\subsection{Redshift and Co-moving Path of the Survey} 
\label{sec:gwz}

From the equivalent width detection limit mentioned in Section~\ref{sec:detect}, we calculate the detection threshold of our survey. Namely, we construct the redshift path sensitivity function, {\gwz} \citep[e.g.,][]{Lanzetta87}, which is the number of quasars in which an absorber with rest-frame equivalent width {\EWr} or greater could be detected at redshift $z$ in our survey,
\begin{equation}
    g(W_r, z) = \sum_{q} H(z\!-\!z_{q}^{-}) 
    H (z_{q}^{+}\!-\!z) 
    H (W_r\!-\!3\sigma_q(z) ) \, ,
\label{eq:gofWz}
\end{equation}
where $H(x)$ is the Heaviside step function, $z_{q}^{-}$, and $z_{q}^{+}$ are the minimum and maximum redshifts of the $q^{\mathrm{th}}$ quasar spectrum, respectively, and $\sigma_q(z)$ is the uncertainty in rest-frame equivalent width at redshift $z$ in the $q^{\rm th}$ quasar spectrum. The sum extends over all quasars in the sample. We generate the {\gwz} function for ${\wrlim} \geq 0.005$~{\AA} over the range $z\!=\!1$ to 5. The ``heat map'' shown in Figure~\ref{fig:gwz} represents the number of quasar spectra in which a {\CIVdblt} feature with a given equivalent width or greater could be detected at the 3$\sigma$ level at redshift $z$. The dark vertical bands around $z\sim3.5$ and $z\sim3.9$ represent excluded telluric bands at 6868--6932~{\AA} and 7594--7700~{\AA}, respectively. For a fixed $W_r$, the integral of $g(W_r,z)$ over a given redshift range 
\begin{equation}
    \Delta Z (W_r) = \int_{z_1}^{z_2} \! \!\! g(W_r,z) \, dz \, ,
\label{eq:DeltaZ}
\end{equation}
yields the redshift path over which absorbers with equivalent width $W_r$ or greater could be detected in our survey at a $3\sigma$ significance level between $z_1$ and $z_2$. Similarly, the ``absorption distance'', which we will call the co-moving path, is defined as
\begin{equation}
   \Delta X (W_r) = \int_{z_1}^{z_2} g(W_r,z) \frac{dX}{dz} \, dz \,
  \label{eq:DeltaX}
\end{equation}
where 
\begin{equation}
  \frac{dX}{dz} = \frac{(1+z)^2}
  {\sqrt{\Omega_{\hbox{\tiny M}} (1+z)^3 + \Omega_{\Lambda}}} \, 
  \label{eq:dxdz}
\end{equation}
accounts for both the radial and transverse components of the Hubble expansion. In the modern era of precision cosmology, the cosmological parameters $\Omega_{\hbox{\tiny M}}$ and $\Omega_{\Lambda}$ are known to a few percent accuracy \citep[e.g.,][]{WMAP09}, so that the function $dX/dz$ is tightly constrained.  As such, all analysis in this work will be conducted in terms of the co-moving path, as this allows any cosmic evolution in the absorber population to be directly measured (see Section~\ref{sec:dndx} for details).

\subsection{Survey Completeness and Science Sample} 
\label{sec:CofW}

\begin{figure}[hbtp]
\centering
\vglue 5pt 
\includegraphics[width=0.45\textwidth]{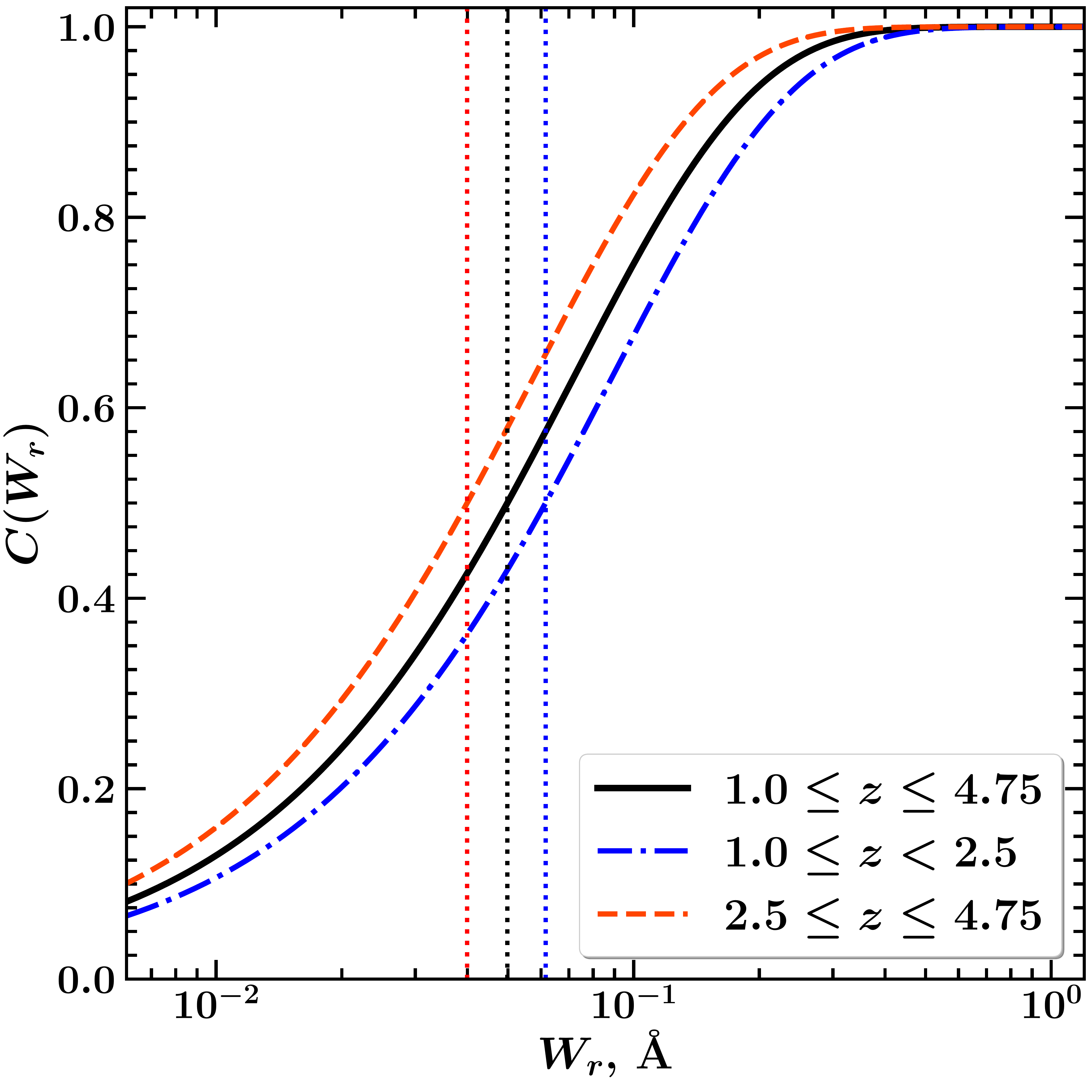}
\caption{The $3\sigma$ detection threshold completeness, $C(W_r)$, as a function of rest-frame equivalent width. The black curve is $C(W_r)$ for the full redshift coverage of the survey. The dot-dashed blue curve shows $C(W_r)$ for redshifts {\zrangel}, while the dashed red curve shows $C(W_r)$ for {\zrangeh}. The blue, red, and black vertical dotted lines represent the $50\%$ completeness limits for {\zrangel}, {\zrangeh}, and {\zrangeall}, respectively.}
\label{fig:comp}
\vglue 5pt 
\end{figure}

The survey detection sensitivity function is defined as the co-moving path over which an absorber with equivalent width {\EWr} or greater could be detected in our survey normalized to the total co-moving path of the survey. In a given redshift range $z_1$ to $z_2$, this function is
\begin{equation}
   C(W_r)\bigg| _{z_1}^{z_2}  = \frac{\Delta X (W_r)}{\Delta X_{\rm max}} \, ,
  \label{eq:Cw}
\end{equation}
where $\Delta X(W_r)$ is computed from Eq.~\ref{eq:DeltaX}, and $\Delta X_{\rm max}$, the total co-moving path of our survey, is computed from Eq.~\ref{eq:DeltaX} with $g(W_r,z)$ calculated such that the Heaviside function $H(W_r\!-\!3\sigma_q(z))$ in Eq.~\ref{eq:gofWz} is fixed at unity.

In Figure~\ref{fig:comp}, we show selected survey detection sensitivity functions as a function of {\EWr}. To a high approximation, this function is the $3\sigma$ detection completeness fraction for our survey. We divide our sample into the two redshift ranges, {\zrangel} and {\zrangeh}, because they have equal co-moving path coverage. Over our total redshift range, {\zrangeall}, the survey is $\sim50$\% complete to ${\EWr} = 0.05$~{\AA}. The {\zrangel} redshift interval reaches $\sim50$\% completeness at ${\EWr} = 0.04$ {\AA}, while the {\zrangeh} interval is $\sim50$\% complete at ${\EWr} = 0.06$ {\AA}. The survey is effectively $100\%$ complete over all redshifts at ${\EWr} = 0.3$~{\AA}.

All subsequent analysis is restricted to {\CIV} absorbers with {\wvweak}, which is the $50\%$ completeness limit of the full redshift range of the survey. Applying this selection criterion, of the total 1565 {\CIV} absorbers we located, we include 1268 {\CIV} absorbers with {\wvweak} in our final science sample. 
We compute the equivalent width distribution and the co-moving path density of {\CIV} using this final sample.

\section{Results} 
\label{results}

\subsection{Equivalent Width Distribution} 
\label{sec:ewd}

\begin{figure*}[ht]
\centering
\gridline{\fig{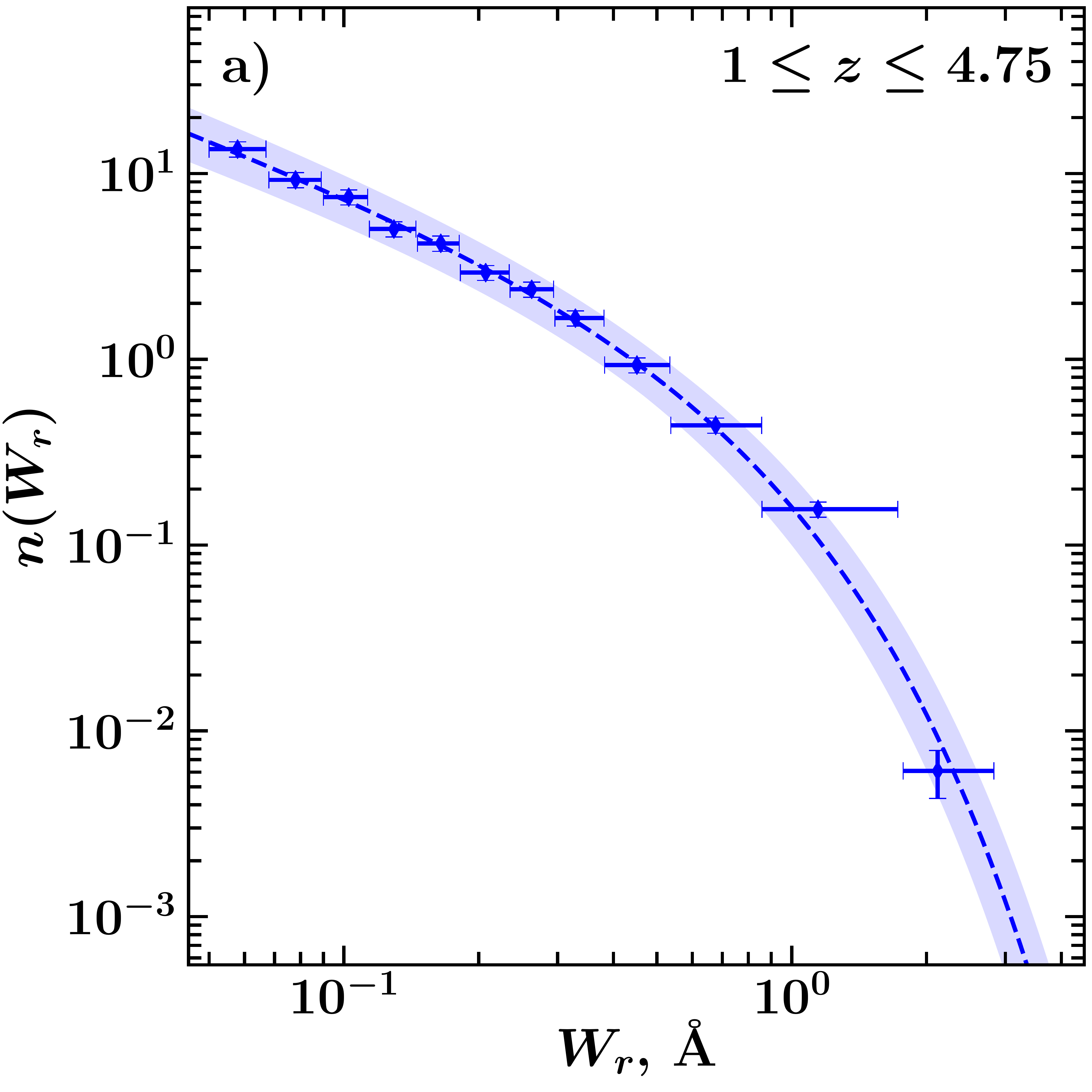}{0.32\textwidth}{}
        \fig{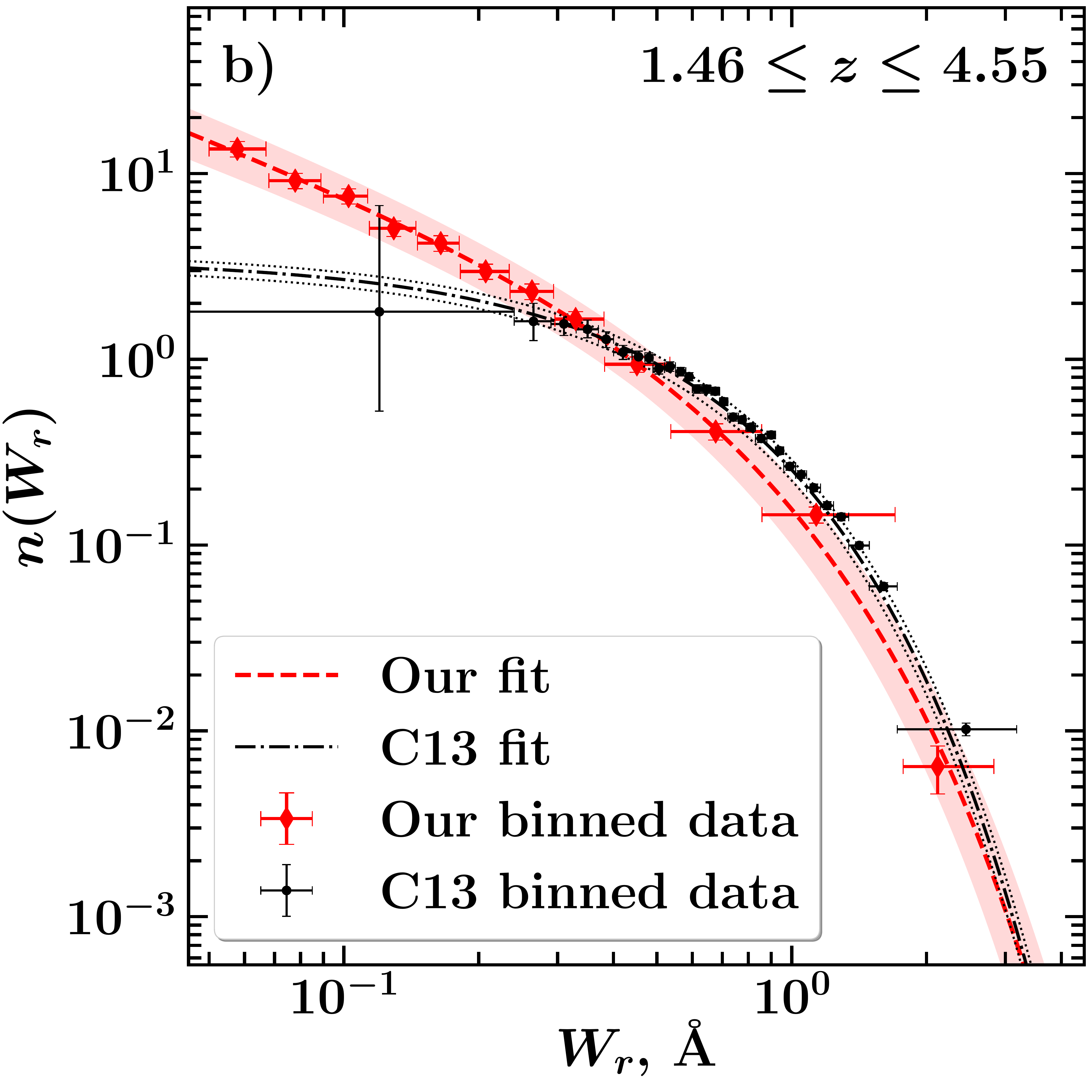}{0.32\textwidth}{}
        \fig{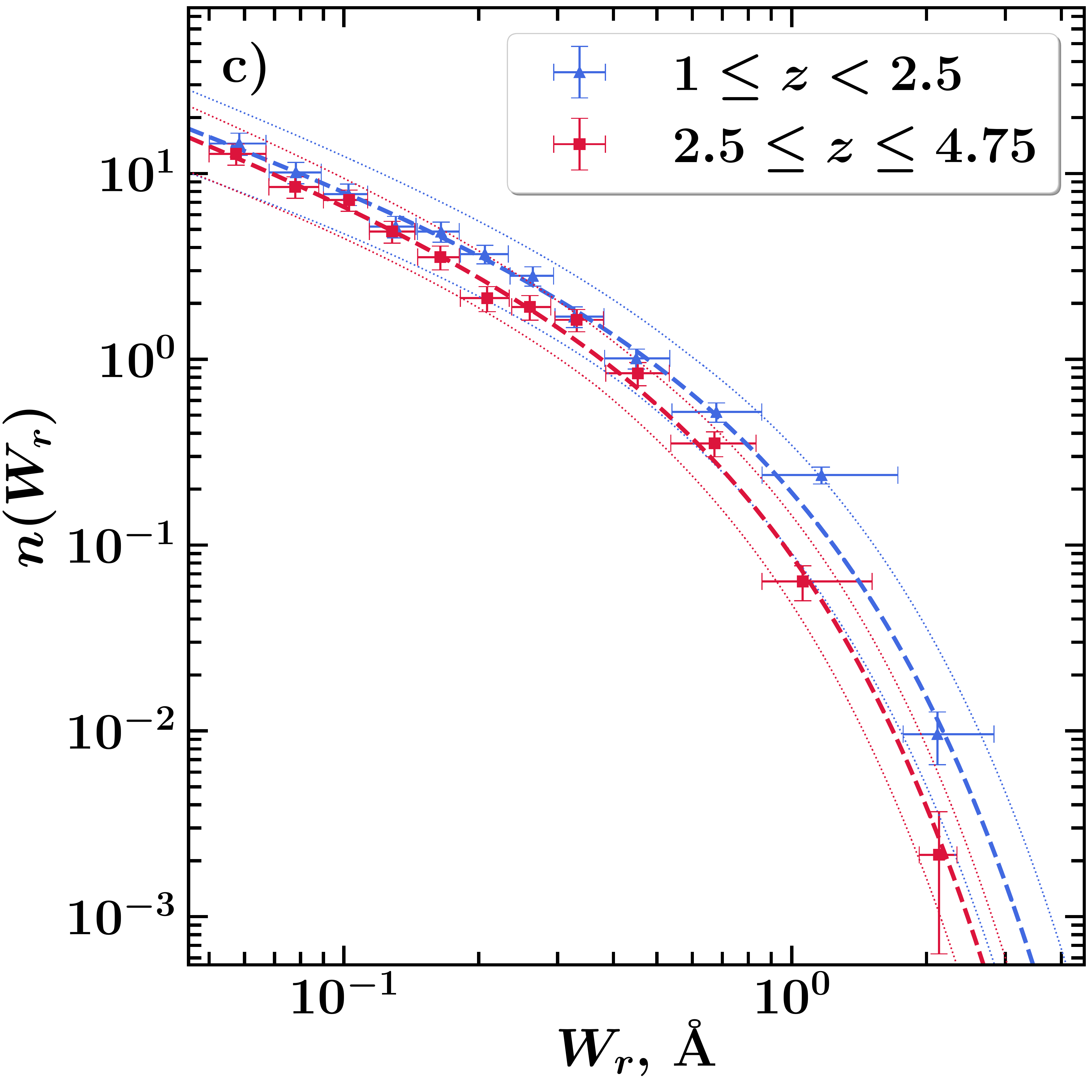}{0.32\textwidth}{}
        }  \vspace{-25pt}
\caption{(a) The {\CIV}~{\strongdblt} equivalent width distribution, {\fw}, for the full redshift range of our survey, {\zrangeall}. The blue points show our observed \fw{} in \EWr{} bins, the dashed blue curve shows the Schechter function fit to these binned data, and the shading represents the $\pm 1\sigma$ confidence interval for our fit. (b) Comparison of our {\fw} with that of \citetalias{Cooksey13} for \citetalias{Cooksey13}'s redshift range: {\zrangec}. Red represents our data and fit; black points and error bars represent \citetalias{Cooksey13}'s {\fw} and associated errors in each {\EWr} bin, while the black curves show \citetalias{Cooksey13}'s exponential fit to their data. (c) {\fw} for two different redshift ranges, {\zrangel} (blue) and {\zrangeh} (red), with the best-fit Schechter functions shown by dashed curves and the $\pm1 \sigma$ uncertainties shown by dotted curves. All of our EWD fit parameters are presented in Table~\ref{tab:schparams}.}
\label{fig:ewd}
\vglue +5pt
\end{figure*}

We first characterized $\fw$, the underlying (or true) equivalent width distribution (hereafter, EWD) of {\CIV} absorbers. This is defined as the number of absorbers per unit equivalent width per unit co-moving path,
\begin{equation}
\fw = \frac{d^2\!N}{dW_r dX} \, ,
\end{equation}
such that 
\begin{equation} \label{eq:ewdnorm}
\int _{W_{\rm min}}^{\infty} \!\!\!\!\!  \fw \, dW_r = \frac{dN}{dX} \, ,
\end{equation}
where {\dndx} is the co-moving path density of absorbers with $W_r \geq W_{\rm min}$ (see Section~\ref{sec:dndx}). We are especially interested in the distribution when we account for the weak absorbers, i.e., those with ${\EWr}<0.3$~{\AA}, as these have not been studied in previous investigations. We examined the {\CIV} absorbers in equivalent width bins defined such that an equal number of absorbers is contained in each bin. The value of {\fw} in a bin centered on {\EWr} between $W_l$ and $W_h$ is
\begin{equation}
    n(W_l\!\leq\!W_r\!<\!W_h)  = \frac{N({\EWr})}{\Delta W_r ~\Delta X (W_r)} \, ,
\label{eq:obsnofW}
\end{equation}
where $N(\EWr)$ is the number of absorbers in the bin, $\Delta W_r = W_h\! -\! W_l$ is the bin width, and the quantity $\Delta X (\EWr) = [\Delta X (W_l)\!+\!\Delta X (W_h)]/2 $ is the averaged total co-moving path length available for the bin as computed using Eq.~\ref{eq:DeltaX}. 

We present the observed EWD, for different ranges, in Figures~\ref{fig:ewd}(a)--(c). The fitted data points are ($\overline{\EWr}$, {\fw}), where $\overline{\EWr}$ is the mean equivalent width of absorbers in the bin and {\fw} is given by Eq.~\ref{eq:obsnofW}. The uncertainties in $\overline{\EWr}$ are the standard deviations of the equivalent widths of the absorbers in the bin. We adopt the standard deviations as a true reflection of the distribution of values inside the bin, rather than the bin width $\Delta {\EWr}$, which would over-estimate the distribution of points in the bin. The uncertainties in {\fw} are obtained assuming Poisson fluctuations; there are roughly 100 absorbers per binned data point.

To explore any difference in the EWD with redshift, we fit the EWD for {\zrangeall} (the full survey range; Figure~\ref{fig:ewd}(a)), and for {\zrangel} and {\zrangeh} (Figure~\ref{fig:ewd}(c)), which each have equal total co-moving path lengths, $\Delta X$, for $W_r \geq 0.05$~{\AA}. We also fit the EWD over {\zrangec} in order to directly compare our EWD with that of \citetalias{Cooksey13} over their redshift range (Figure~\ref{fig:ewd}(b)).

The shape of our observed completeness-corrected EWD indicates a slow power-law decline with {\EWr} at the weak end and a rapid exponential decline at the strong end of the frequency distribution. This is not unexpected, as sensitive high-resolution surveys that probe down to the linear part of the curve of growth found power-law distributions for the column density \citep[e.g.,][]{Songaila01,Songaila05,Dodo10,BS15}, whereas larger surveys sensitive to {\wweak} reported an exponential distribution for equivalent widths \citep[e.g.,][\citetalias{Cooksey13}]{Sargent88,Cooksey10}. Thus, following the parameterization of \citet{glenncwc11} of the distribution of strong and weak {\MgII} absorbers, we fit the observed {\CIV} EWD by a Schechter \citeyearpar{Sch76} function
\begin{align}
    n(W_r) \, dW_r = n_{\star} 
    \left( \frac{W_r}{W_{\star}} \right)^{\alpha} 
    \exp \left( -\frac{W_r}{W_{\star}} \right) \, dW_r \, ,
\label{eq:EWD}
\end{align}
where $\alpha$ is the weak-end power-law slope, $W_{\star}$ is the characteristic equivalent width where the function transitions from a power-law to an exponential function, and $n_{\star}$ is the normalization.

We obtained best-fit values for $n_{\star}$, $W_{\star}$, and $\alpha$, using the least-squares orthogonal distance regression (ODR) method in the {\sc SciPy} package,\footnote{\url{https://docs.scipy.org/doc/scipy/reference/odr.html}} which accounts for uncertainties in both the dependent and independent variables. We use the ODR method for all subsequent functional fits in this paper.

Before adopting our final best-fit parameters, we perform a series of experiments to explore sensitivity to the number of bins used in the least-squares fit. We fit the data using 10, 11, 12, 13, and 14 bins, while always enforcing an equal number of absorbers in each bin. We also explored the use of $1\sigma$ and $3\sigma$ uncertainties in the binned $\overline{W_r}$, {\fw} ordered pairs. The fitted parameters $n_{\star}$, $W_{\star}$, and $\alpha$ are found to be highly robust against changes in the number of bins and the use of $1\sigma$ or $3\sigma$ uncertainties, meaning that variations in the values of the best-fit parameters are within the $\pm1\sigma$ uncertainties for all experiments. The 12 bin fit yielded the smallest fitted-parameter uncertainties, whereas the 10 and 11 bin fits yielded the largest. The adopted best-fit Schechter function parameters were obtained by computing the $1\sigma$ variance-weighted means of the 10, 11, 12, 13, and 14 bin experiments.

\begin{deluxetable}{llccc}[hbtp]
\vglue 5pt
\centering
\tablewidth{0pt}
\tabletypesize{\small}
\tablecaption{EWD Fit Parameters (see Figure~\ref{fig:ewd})}
\label{tab:schparams}
\tablehead{
\colhead{$z_1$} & 
\colhead{$z_2$} &
\colhead{$\alpha$} &
\colhead{$W_{\star}$ ({\AA})} &
\colhead{$n_{\star}$ ({\AA}$^{-1}$)} 
}
\startdata
$1.00$ & $4.75$ & $-0.89 \pm 0.04$ & $0.51 \pm 0.05$ & $2.05 \pm 0.32$ \\ [-3pt]
$1.46${\tablenotemark{\scriptsize a}} & $4.55${\tablenotemark{\scriptsize a}} & $-0.89 \pm 0.04$ & $0.50 \pm 0.04$ & $2.12 \pm 0.31$ \\ [-3pt]
$1.00$ & $2.50$ & $-0.85 \pm 0.06$ & $0.51 \pm 0.07$ & $2.38 \pm 0.56$ \\ [-3pt]
$2.5$ & $4.75$ & $-0.91 \pm 0.06$ & $0.40 \pm 0.04$ & $2.39 \pm 0.43$ \\
\enddata
\tablenotetext{\scriptsize a}{The redshift range of \citetalias{Cooksey13}, for comparison.}
\vspace{-8pt}
\end{deluxetable}

In Figure~\ref{fig:ewd}(a), we present the best-fit Schechter function for the full redshift range of our sample, {\zrangeall}. The shaded region represents the $\pm1\sigma$ confidence interval given the uncertainties in the three fitted parameters. These best-fit parameters are listed in Table~\ref{tab:schparams}.  We obtained Schechter parameters $\alpha \simeq -0.9$, $W_{\star} \simeq 0.5$ {\AA}, and $n_{\star} \simeq 2.1$. Note that $n_{\star}$ is the most uncertain fitted parameter, while $W_{\star}$ and $\alpha$ are more highly constrained.

In order to compare our EWD to that of \citetalias{Cooksey13} over the redshift range of their sample, {\zrangec}, we show both distributions in Figure~\ref{fig:ewd}(b).  The \citetalias{Cooksey13} distribution is fit well by an exponential function, but their sample is only $\sim\!\!20\%$ complete at ${\EWr}=0.3$~{\AA}.  Our sample is $50\%$ complete at ${\EWr}=0.05$~{\AA} and is well-fit to a Schechter function. Note that the power-law portion of the Schechter function is clearly established for ${\EWr} \leq 0.3$~{\AA}. For {\zrangec}, we obtain virtually identical Schechter parameters ($\alpha \simeq -0.9$, $W_{\star} \simeq 0.5$ {\AA}, and $n_{\star} \simeq 2.1$) as for our full redshift range.

It is important that we establish that our survey results are consistent with those of \citetalias{Cooksey13} in the regions of redshift and equivalent width overlap. This provides confidence that our methods reproduce the currently most definitive works on {\CIV} absorber statistics.  We performed Kolomogorv-Smirnov (K-S) tests comparing our observed sample limited to {\zrangec} and {\wweak} to the exponential probability density function fitted to the data of \citetalias{Cooksey13}.  We adopt a Monte Carlo approach in order to account for the three orders of magnitude disparity in the sizes of the two samples.  We created 1 million Monte Carlo realizations of the \citetalias{Cooksey13} data having the same number of elements as our sample limited to {\zrangec} and {\wweak}. 
For each realization, we performed a K-S test between our observed sample and the Monte Carlo sample. Formally, one can think of our sample as a single realization that is compared to 1 million different realizations of the \citetalias{Cooksey13} sample.  We adopted $P(KS) \leq 0.0027$ for rejecting the null hypothesis that our sample was drawn from the same distribution as that of \citetalias{Cooksey13}. This corresponds to a $3\sigma$ confidence level.

\begin{figure}[htbp]
\centering
\vglue +10pt
\includegraphics[width=0.42\textwidth]{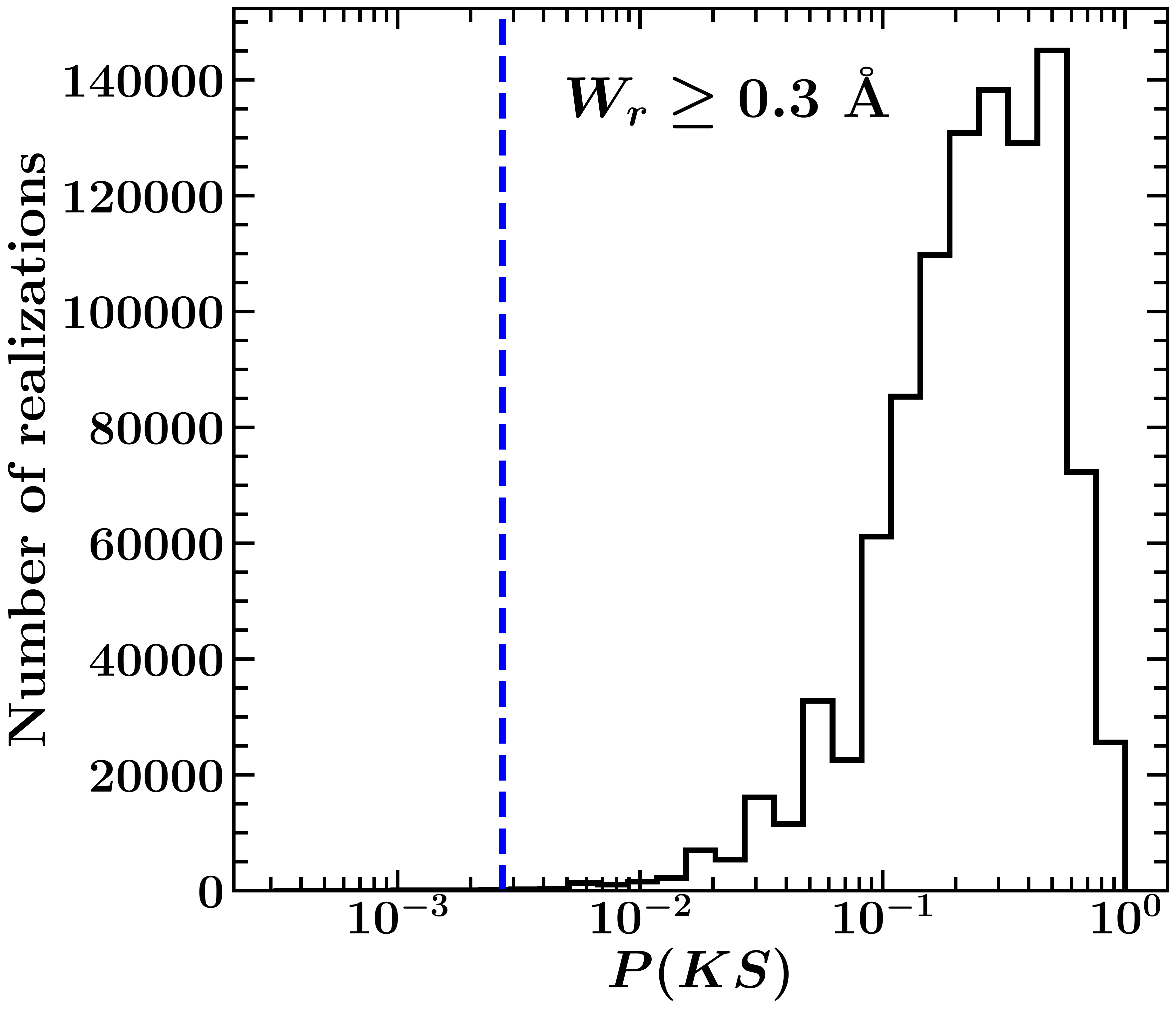}
\caption{Distribution of $P(KS)$ values from 1 million realizations of K-S tests comparing \citetalias{Cooksey13}'s exponential EWD to our observed {\EWr} values at {\zrangec}, for {\wweak}. The vertical blue dashed line shows $P(KS)=0.0027$. See text for details.}
\label{fig:ks_exp}
\end{figure}

In Figure~\ref{fig:ks_exp}, we present the distribution of $P(KS)$ values returned by 1 million Monte Carlo realizations. The dashed blue line represents $P(KS)=0.0027$. We find that $P(KS)\leq 0.0027$ in $0.05\%$ of the realizations (or 500 out of 1 million times). We interpret this to indicate that we our {\wweak} sample is not inconsistent with that of \citetalias{Cooksey13} at the $99.95\%$ confidence level. Interestingly, when we conduct the Monte Carlo experiment above for absorbers with {\wvweak} at {\zrangec}, we find that $P(KS)<0.0027$ in $100\%$ of the realizations.  Thus, the fitted exponential distribution function describing the \citetalias{Cooksey13} data is ruled out to a confidence level greater than six nines for {\wvweak}.

In Figure~\ref{fig:ewd}(c), we present the EWDs for two redshift ranges, $1 \leq z < 2.5$ and $2.5 \leq z \leq 4.75$, chosen such that both redshift ranges have equal survey co-moving path lengths. 
Our data indicate a slightly steeper weak-end power-law slope at higher redshift, $\alpha \simeq -0.9$, than at lower redshift, $\alpha \simeq -0.85$. At higher redshift, we also find that the characteristic transition from a power-law to exponential distribution, $W_{\star}$, falls by $\sim20\%$. The normalization, $n_{\star}$, is seen to be virtually unchanged from higher to lower redshift, though the uncertainty on this parameter is about $20\%$.
Taken together, these best fit parameters indicate redshift evolution in the relative frequencies of weaker and stronger absorbers; at higher redshifts, the frequency of weaker absorbers relative to stronger absorbers is higher than at lower redshifts.

\subsection{Co-moving Path Density, {\dndx}} 
\label{sec:dndx}

For a given population of absorbers, the observed co-moving path density is proportional to the product of the absorber cosmic number density, $n(z)$, and the absorbing structure physical cross-section, $\sigma(z)$,
\begin{equation} 
\frac{dN(z)}{dX} \equiv N_{\hbox{\tiny X}}(z) = \frac{c}{H_0} n(z) \sigma(z) \, ,
\label{eq:dndx2}
\end{equation}
where we hereafter adopt ${\Nx}(z)$ to designate the {\it measured value\/} of co-moving path density. Examination of $dN(z)/dX$ provides direct insights into the redshift evolution of the product $n(z)\sigma(z)$; a non-evolving population of absorbers will have a constant co-moving path density as a function of redshift.

Historically, it was the redshift path density, {\dndz}, that has been employed to measure the cosmic evolution of a population of absorber, where {\dndz} is defined as the observed number of absorbers with equivalent width $W_r$ or greater per unit of redshift \citep[e.g.][]{Sargent88,Steidel90,Misawa02,Peroux04}. The co-moving path density is related to {\dndz} through the relation
\begin{equation} \label{eq:dndz}
    \frac{dN(z)}{dz} = \frac{dN(z)}{dX} \frac{dX}{dz}  \, ,
\end{equation}
where the function $dX/dz$ is defined in Eq.~\ref{eq:dxdz}. Unlike {\dndx}, {\dndz} is not a constant for a non-evolving population of absorbers. Thus, we use {\dndx} to characterize the evolution of {\CIV} absorbers rather than {\dndz}.

We examined the co-moving path density as a function of both redshift and equivalent width. In a given redshift range spanning $z_1$ to $z_2$, the co-moving path density and its variance are obtained by summing over all $N_{\mathrm{abs}}$ absorbers in the redshift range whose rest-frame equivalent widths are greater than or equal to {\wrlim}. Thus, we measure {\Nx} as
\begin{equation}
\begin{aligned}
    {\Nx}(W_i\! \geq \!W_{r,\mathrm{lim}}) \bigg| _{z_1}^{z_2}  
    &= \sum_{i=1}^{N_{\mathrm{abs}}} \frac{1}{\Delta X (W_i)} \, , \\[8pt] 
    \sigma_{{\Nx}}^{2} (W_i\! \geq \!W_{r,\mathrm{lim}}) \bigg| _{z_1}^{z_2}  
    &= \sum_{i=1}^{N_{\mathrm{abs}}} \left[ \frac{1}{\Delta X (W_i)} \right]^2 \, .
\end{aligned}
\label{eq:dNdX}
\end{equation}
where $\Delta X(W_i)$, as computed using Eq.~\ref{eq:DeltaX}, is the total co-moving path over which absorber $i$ with rest-frame equivalent width $W_i$ could be detected at the $3\sigma$ significance level accounting for all quasars in the survey.

\begin{figure}[htbp]
\centering
\vglue 10pt 
\includegraphics[width=0.45\textwidth]{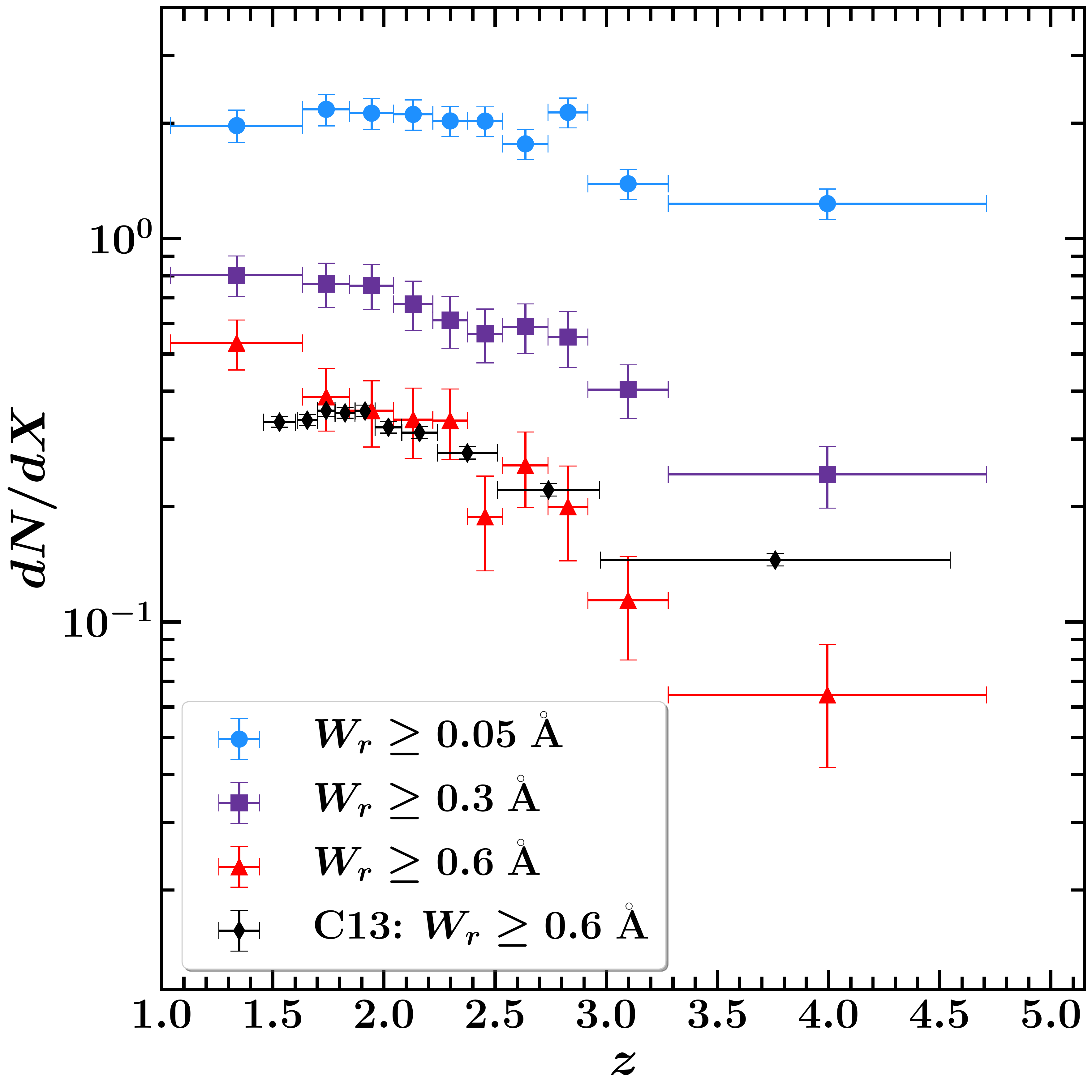}
\caption{The measured {\CIV} absorber co-moving path density ${\Nx}(z) = {\dndx}$ as a function of redshift for the rest-frame equivalent width thresholds of 0.05~{\AA} (light blue circles), 0.3~{\AA} (purple squares), and 0.6~{\AA} (red triangles), and associated errors. For comparison, we show ${\Nx}(z)$ for $W_r \geq 0.6$ {\AA} from \citetalias{Cooksey13} (black diamond points). Our measured values are presented in Table~\ref{tab:dndx}.}
\label{fig:dndx}
\end{figure}

In applying Eq.~\ref{eq:dNdX}, we resolve redshift into bins $(z_1,z_2]$ such that an equal number of absorbers reside in each bin for the minimum equivalent width threshold ${\wrlim}\! =\! 0.05$~{\AA}. We use 10 redshift bins, each of which has $\sim\!130$ absorbers. These redshift bins are also adopted for the ${\wrlim}\! =\! 0.3$~{\AA} and ${\wrlim}\! =\! 0.6$~{\AA} populations. In Figure~\ref{fig:dndx}, we present the observed ${\Nx}(z)$ over the redshift range {\zrangeall} for three minimum equivalent width thresholds, ${\wrlim} = 0.05$~{\AA}, $0.3$~{\AA}, and $0.6$~{\AA}. We list our observed ${\Nx}(z)$ values in Table~\ref{tab:dndx}. For comparison, we also present ${\Nx}(z)$ values from \citetalias{Cooksey13} for ${\wrlim}=0.6$~{\AA} for the redshift range {\zrangec}; our values are consistent with those of \citetalias{Cooksey13}, who were $\sim\!50\%$ complete at ${\EWr} \simeq 0.6$~{\AA}.

\begin{deluxetable}{ccccc}[bhtp]
\vglue 5pt
\centering
\tablewidth{0pt}
\tablecaption{Measured Co-moving Path Density (see Figure~\ref{fig:dndx})}
\label{tab:dndx}
\tablehead{
\colhead{$\langle z \rangle$} & 
\colhead{$\Delta z$\tablenotemark{\scriptsize a} } &
\colhead{${\Nx}(z)$} &
\colhead{${\Nx}(z)$} &
\colhead{${\Nx}(z)$} \\[-5pt]
\colhead{} &
\colhead{} &
\colhead{({\footnotesize {\wvweak}})} & 
\colhead{({\footnotesize {\wweak}})} & 
\colhead{({\footnotesize {\wstrong}})} 
}
\startdata
$1.338$ & $0.594$ & $1.97 \pm 0.19$ & $0.80 \pm 0.10$ & $0.53 \pm 0.08$ \\ [-3pt]
$1.740$ & $0.211$ & $2.17 \pm 0.21$ & $0.76 \pm 0.10$ & $0.39 \pm 0.07$ \\ [-3pt]
$1.944$ & $0.197$ & $2.12 \pm 0.20$ & $0.75 \pm 0.10$ & $0.36 \pm 0.07$ \\ [-3pt]
$2.131$ & $0.177$ & $2.11 \pm 0.19$ & $0.67 \pm 0.10$ & $0.34 \pm 0.07$ \\ [-3pt]
$2.298$ & $0.156$ & $2.03 \pm 0.18$ & $0.61 \pm 0.09$ & $0.34 \pm 0.07$ \\ [-3pt]
$2.455$ & $0.158$ & $2.02 \pm 0.18$ & $0.56 \pm 0.09$ & $0.19 \pm 0.05$ \\ [-3pt]
$2.636$ & $0.204$ & $1.77 \pm 0.16$ & $0.59 \pm 0.09$ & $0.26 \pm 0.06$ \\ [-3pt]
$2.828$ & $0.179$ & $2.13 \pm 0.19$ & $0.55 \pm 0.09$ & $0.20 \pm 0.06$ \\ [-3pt]
$3.098$ & $0.361$ & $1.39 \pm 0.12$ & $0.40 \pm 0.06$ & $0.11 \pm 0.03$ \\ [-3pt]
$3.995$ & $1.432$ & $1.23 \pm 0.11$ & $0.24 \pm 0.04$ & $0.06 \pm 0.02$ \\
\enddata
\tablenotetext{\scriptsize a}{The redshift bins, $\Delta z\!=\!z_2\!-\!z_1$ with $\langle z \rangle \!=\! (z_1 \!+\! z_2)/2$, are defined such that the ${\wrlim} \!=\! 0.05$~{\AA} sample has equal numbers of absorbers in each bin.}
\vspace{-25pt}
\end{deluxetable}

Examination of Figure~\ref{fig:dndx} clearly shows two types of evolution. First, for all three equivalent width thresholds, the co-moving path density decreases with increasing redshift. Second, this evolution is more pronounced as the equivalent width threshold is increased. The ratio of ${\Nx}(z)$ of the ${\wrlim} = 0.05$~{\AA} sample to the ${\wrlim} = 0.6$~{\AA} sample changes from $\sim\!3.5$ at $\langle z \rangle = 1.3$ to $\sim\!18$ at $\langle z \rangle = 4$.

From Eq.~\ref{eq:dndx2}, we infer that the product of cosmic number density and absorber cross-section, $n(z)\sigma(z)$, increases for all of the populations of absorbers we study at {\zrangeall}, with the increase being more rapid for absorbers with larger {\EWr}.
For the absorbers defined by {\wvweak}, a shallow rise in ${\Nx}(z)$ of a factor of $\simeq\! 1.8$ is observed. The co-moving path density for {\wweak} absorbers rises by a factor of $\simeq\! 3.3$ over this period. For absorbers with {\wstrong}, the co-moving path density increases by a factor of $\simeq\! 8.5$ over this period, which is $\simeq\! 2.5$ times the rise for {\wweak} absorbers and $\simeq\! 4.5$ times the rise for {\wvweak} absorbers. These results reinforce our finding that the EWD is evolving such that at higher redshifts there are relatively more weak absorbers to strong absorbers than there are at lower redshifts (also see Figure~\ref{fig:ewd}).

\begin{figure}[htbp]
\vglue 5pt
\centering
\includegraphics[width=0.42\textwidth]{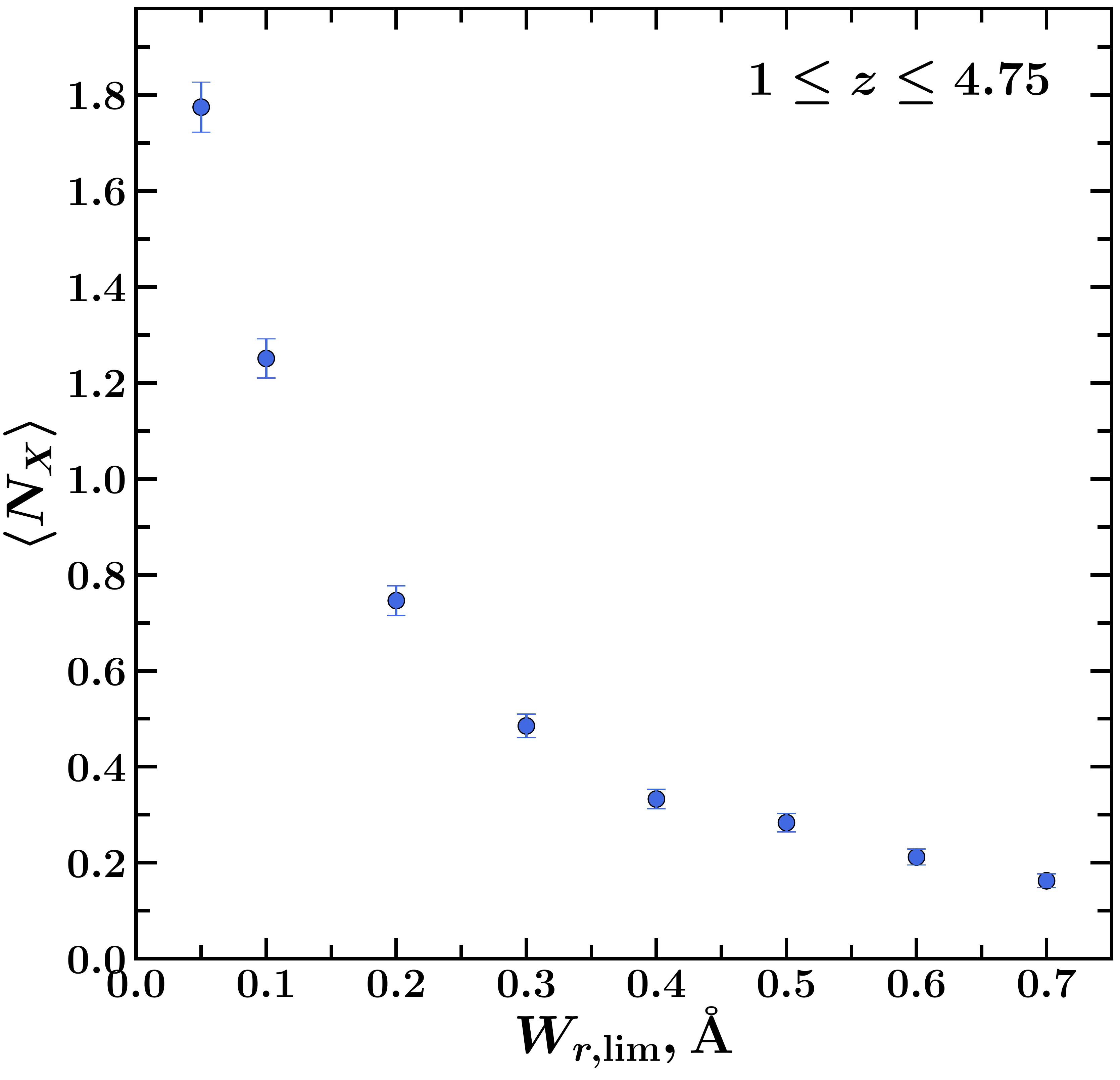}
\caption{The observed cosmic mean co-moving path density of {\CIV} absorbers, $\langle N_{\hbox{\tiny X}}\rangle = (c/H_0) \langle n\sigma \rangle$, over the redshift range {\zrangeall} as a function {\wrlim}.}
\label{fig:mean}
\end{figure}

\begin{figure*}[htbp]
\centering 
\includegraphics[width=0.97\textwidth]{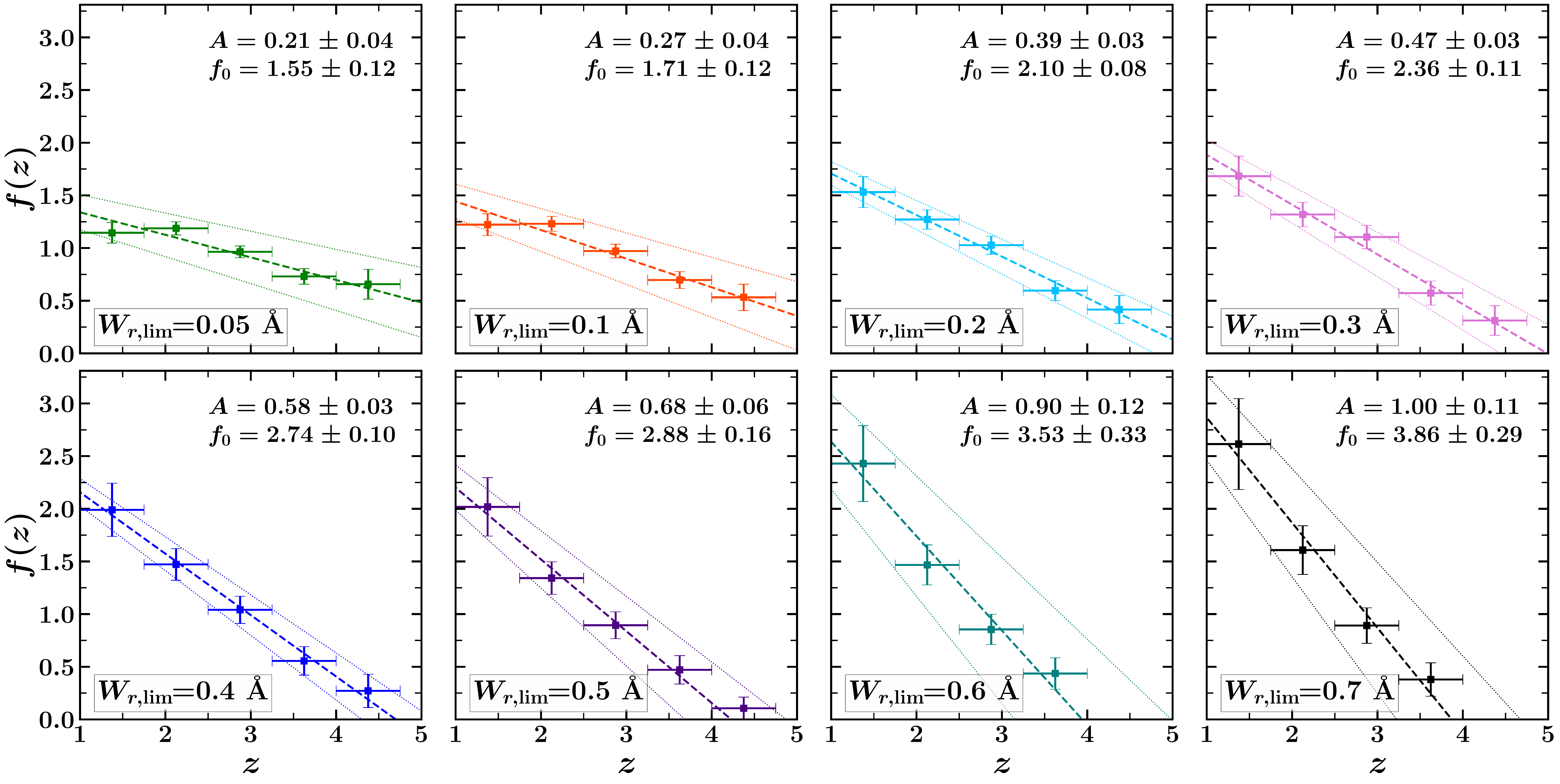}
\caption{The observed ratio $f(z) = N_{\hbox{\tiny X}}(z)/ \langle N_{\hbox{\tiny X}} \rangle$ is plotted as a function of redshift for subsamples having differing minimum rest-frame equivalent widths. The dashed lines through the data are the least-squares fits to Eq.~\ref{eq:fz}, with the dotted lines representing the $\pm1\sigma$ confidence intervals. The individual panels are for absorber populations defined by
(a) ${\wrlim} = 0.05$~{\AA}, 
(b) ${\wrlim} = 0.1$~{\AA},
(c) ${\wrlim} = 0.2$~{\AA},
(d) ${\wrlim} = 0.3$~{\AA},
(e) ${\wrlim} = 0.4$~{\AA},
(f) ${\wrlim} = 0.5$~{\AA},
(g) ${\wrlim} = 0.6$~{\AA},
(h) ${\wrlim} = 0.7$~{\AA}.
}
\label{fig:fz}
\vglue 5pt
\end{figure*}

To further characterize the incidence of {\CIV} absorbers, we define the measured cosmic mean co-moving path density 
\begin{equation} 
\langle N_{\hbox{\tiny X}} \rangle  = \frac{c}{H_0} \langle n \sigma \rangle  \, ,
\label{eq:dndxmean}
\end{equation}
as the arithmetic mean of the measured ${\Nx}(z)$ of an absorber population, weighted by the variance within each redshift bin, in the redshift range {\zrangeall}. In Figure~\ref{fig:mean}, we show {\meanNx} over the redshift range {\zrangeall} as a function {\wrlim}. 
Over this redshift range, we see that the cosmic mean co-moving path density decreases as ${\wrlim}$ increases; the mean incidence of the weakest absorbers with {\wvweak} is almost an order of magnitude higher than the strongest absorbers we studied, with ${\EWr} \geq 0.7$ {\AA}.

\subsection{Quantifying Evolution}
\label{sec:quantevolve}

We aim to further quantify the evolutionary characteristics in ${\Nx}(z)$, and therefore the product $n(z)\sigma(z)$, by examining redshift evolution 
as a function of minimum equivalent width threshold, {\wrlim}.  We examine the ratio
\begin{equation} 
\frac{N_{\hbox{\tiny X}}(z)}{\langle N_{\hbox{\tiny X}} \rangle } =
\frac{n(z) \sigma(z)}{\langle n \sigma \rangle} = f(z) \, ,
\label{eq:nsig}
\end{equation}
which we can parameterize with an arbitrary function of redshift, $f(z)$.
We considered five equal redshift bins of $\Delta z = 0.75$ in the range {\zrangeall}, for absorber populations with {\wrlim} = 0.05~{\AA} and with ${\wrlim} = 0.1$~{\AA} to 0.7~{\AA}, in steps of 0.1~{\AA}. 

Based on the evolution exhibited in Figure~\ref{fig:dndx}, we adopted a simple first-order polynomial for $f(z)$,
\begin{equation} 
f(z) = f_0 - A z \, ,
\label{eq:fz}
\end{equation}
where $f_0 = f(z\!=\!0)$ and $A=-df(z)/dz$ is the slope, which gives the negative of the rate of change.

If we are to equate a physical meaning to the model parameters, we would interpret $A$ as an ``evolution constant'' describing the evolution rate of the ratio $n(z)\sigma(z)/\langle n \sigma \rangle$ over the redshift range {\zrangeall}.  For the zero intercept, $f_0$, we would interpret this as the ratio of the present-epoch ($z\!=\!0$) product ${n_0\sigma_0 = n(0)\sigma(0)}$, to the measured cosmic mean, $\langle n \sigma \rangle$. Our model also yields a derived quantity, which is the redshift above which the incidence of absorbers in our survey vanishes, i.e., $f(z_0) = 0$, which corresponds to $z_0 = {f_0}/{A}$. 
We would interpret $z_0$ as an ``onset redshift'', meaning the redshift at which the absorber population would first appear in the universe. Both $f_0$ and $z_0$ are based on linear extrapolation of the model, which is fit only in the redshift range {\zrangeall}.

\begin{figure*}[htbp]
\centering
\gridline{\fig{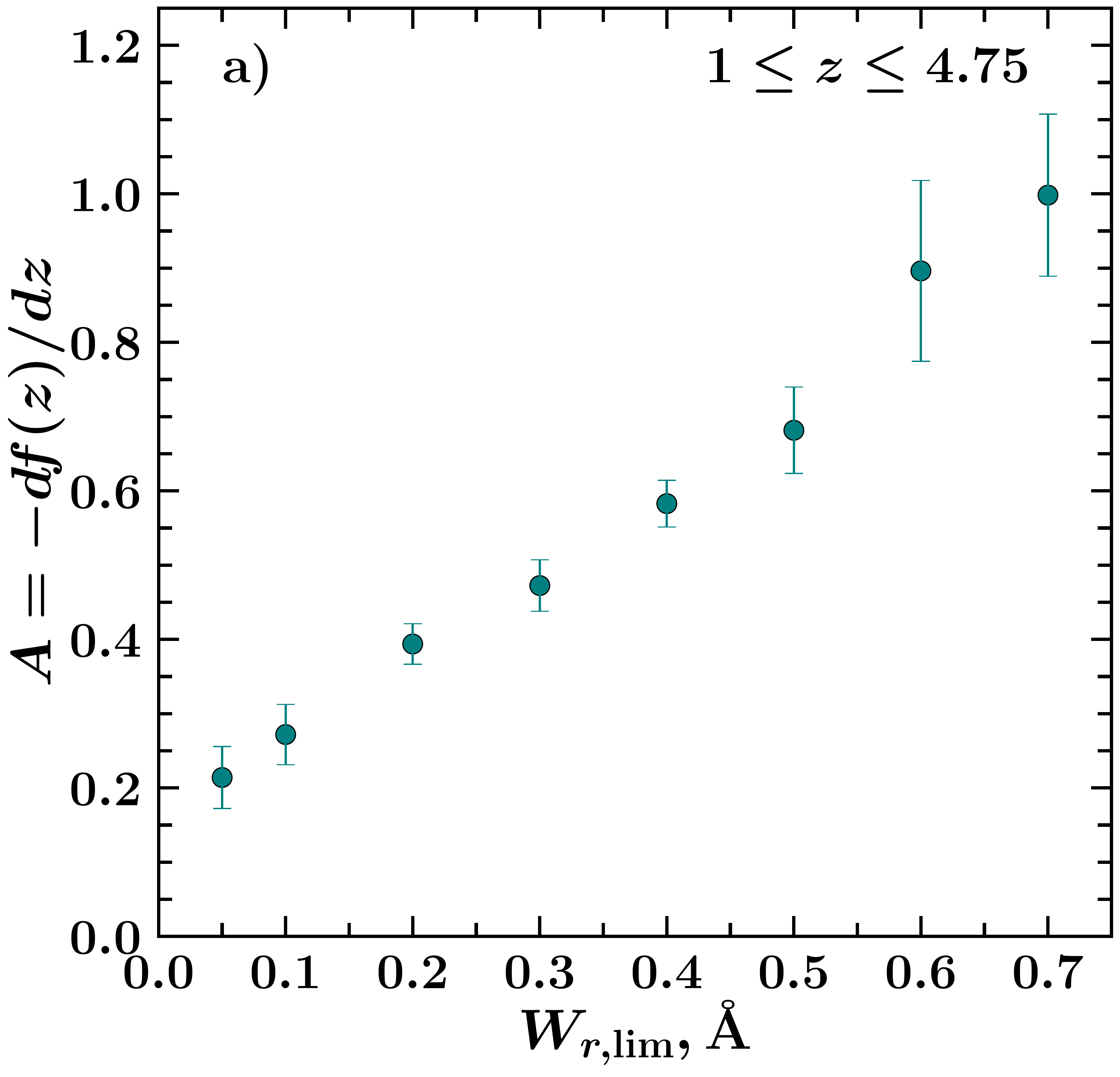}{0.31\textwidth}{}
        \fig{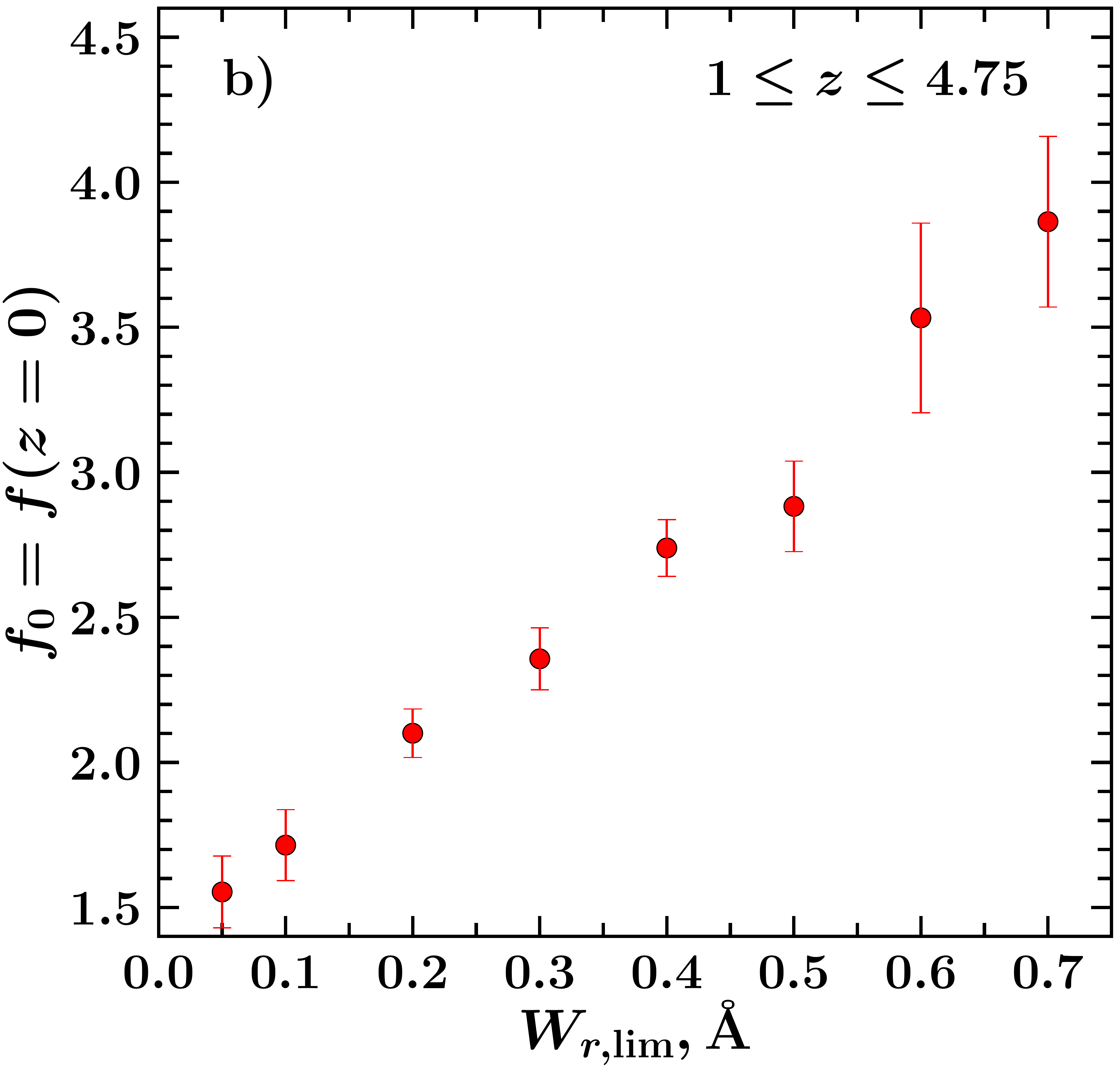}{0.31\textwidth}{}
        \fig{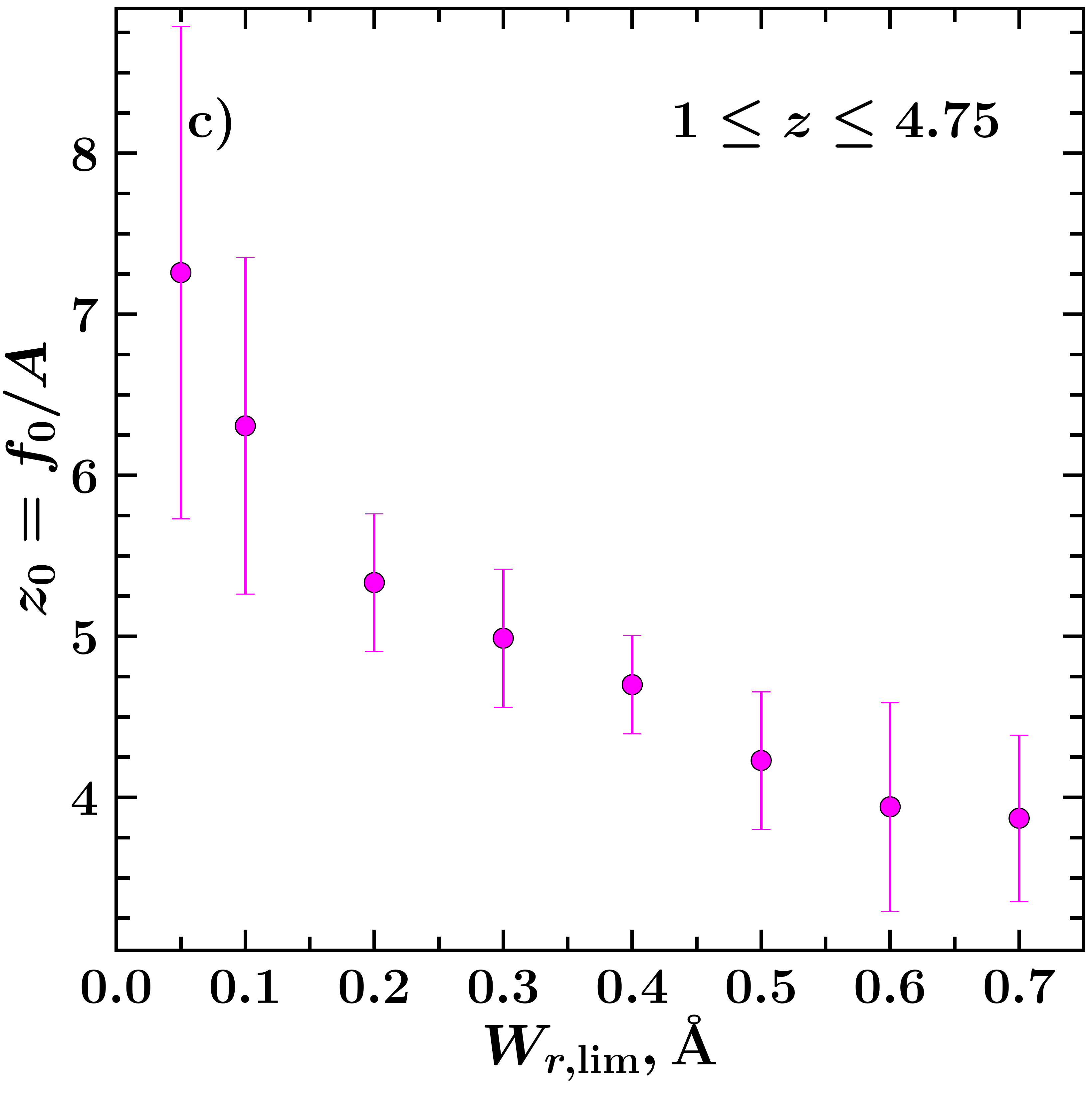}{0.3\textwidth}{}
        } \vspace{-20pt}
\caption{(a) The best-fit parameter $A = -df(z)/dz$, which we call the evolution constant. This is the rate of evolution of the product $n(z)\sigma(z)$ per unit redshift for {\CIV} absorbers having $W_r \geq {\wrlim}$. $A$ increases linearly with {\wrlim}. (b) The best-fit parameter $f_0 = N_{\hbox{\tiny X}}(z=0)/\langle N_{\hbox{\tiny X}} \rangle$, which is the ratio of the product of the present-epoch $n_0\sigma_0$ to the measured cosmic mean value $\langle n \sigma \rangle$ for a given {\wrlim}. Under the assumption that evolution remains linear from $z=1$ to $z=0$, this ratio is predicted to increase in direct proportion to {\wrlim}. (c) The best-fit parameter $z_0 = f_0/A$, which is the onset redshift for {\CIV} absorbers having $W_r \geq {\wrlim}$. The onset redshift decreases monotonically with increasing {\wrlim}, following $z_0 \propto {\wrlim}^{-0.22}$.}
\label{fig:Af0z0}
\vglue 5pt
\end{figure*}

We computed $f(z) = {\Nx}/{\meanNx}$ in each redshift bin, for each {\wrlim} threshold. The results are presented in Figures~\ref{fig:fz}(a)--(h). We performed a least-squares fit using ODR to Eq.~\ref{eq:fz} for each {\wrlim} sample and obtained the best-fit parameters $f_0$ and $A$ and their uncertainties. The fitted functions are superimposed on the data in Figures~\ref{fig:fz}(a)--(h). The fitted parameters and the onset redshift, $z_0$, are listed in Table~\ref{tab:evolve}. 
The positive evolution constant indicates that the product $n(z)\sigma(z)$ is increasing with cosmic time (decreasing with redshift).

Figures~\ref{fig:fz}(a)--(h) show that the data are well-described by a linear evolution with redshift for all populations of absorbers, with the rate of evolution being more rapid for larger {\wrlim}. 
The positive evolution constant for all {\wrlim} indicates that the product $n(z)\sigma(z)$ is increasing with cosmic time for all populations of {\CIV} absorbers.

\begin{deluxetable}{ccccc}[htbp]
\vglue 5pt 
\centering
\tablecaption{Best-Fit Evolution Model Parameters} \label{tab:evolve}
\tablehead{
\colhead{$W_{r,\mathrm{lim}}$} & 
\colhead{$\langle N_{\hbox{\tiny X}} \rangle$} & 
\colhead{$A$} & 
\colhead{$f_0$} & 
\colhead{$z_0$} \\[-5pt]
\colhead{({\AA})} &
\colhead{} &  
\colhead{} & 
\colhead{} & 
\colhead{}
} 
\startdata
$0.05$ & $1.77 \pm 0.05$ & $0.21 \pm 0.04$ & $1.55 \pm 0.12$ & $7.26 \pm 1.53$ \\ [-3pt]
$0.1$ & $1.25 \pm 0.04$ & $0.27 \pm 0.04$ & $1.71 \pm 0.12$ & $6.31 \pm 1.04$ \\ [-3pt]
$0.2$ & $0.75 \pm 0.03$ & $0.39 \pm 0.03$ & $2.10 \pm 0.08$ & $5.33 \pm 0.43$ \\ [-3pt]
$0.3$ & $0.49 \pm 0.02$ & $0.47 \pm 0.03$ & $2.36 \pm 0.11$ & $4.99 \pm 0.43$ \\ [-3pt]
$0.4$ & $0.33 \pm 0.02$ & $0.58 \pm 0.03$ & $2.74 \pm 0.10$ & $4.70 \pm 0.30$ \\ [-3pt]
$0.5$ & $0.28 \pm 0.02$ & $0.68 \pm 0.06$ & $2.88 \pm 0.16$ & $4.23 \pm 0.43$ \\ [-3pt]
$0.6$ & $0.21 \pm 0.02$ & $0.90 \pm 0.12$ & $3.53 \pm 0.33$ & $3.94 \pm 0.65$ \\ [-3pt]
$0.7$ & $0.16 \pm 0.01$ & $1.00 \pm 0.11$ & $3.86 \pm 0.29$ & $3.87 \pm 0.52$ \\
\enddata 
\end{deluxetable}

In Figure~\ref{fig:Af0z0}(a), we plot the evolution constant as a function of {\wrlim}. With the progression of cosmic time, the ratio $A = n(z)\sigma(z)/\langle n\sigma \rangle$ increases by a factor of $\sim\!\!20\%$ per unit redshift for ${\wrlim} = 0.1$~{\AA}, whereas this quantity increases by a factor of $\sim\!\!100\%$ per unit redshift for ${\wrlim} = 0.7$~{\AA}. Thus, the product $n(z)\sigma(z)$ of the larger {\EWr} absorbers rises much faster than that of the smaller {\EWr} absorbers.

In Figure~\ref{fig:Af0z0}(b), we plot $f_0 = n_0\sigma_0/\langle n \sigma \rangle$ as a function of {\wrlim}. As the minimum redshift of our survey is $z\!=\!1.0$, the value of $f_0$ has physical meaning only under the assumption that $f(z)$ can be linearly extrapolated to $z\!=\!0$. Assuming linear evolution continues to $z\!=\!0$ (over the last $\simeq\! 7.7$ Gyr of cosmic time), our model would suggest $n_0\sigma_0/\langle n \sigma \rangle$ is a factor of $\sim\!1.6$ for ${\wrlim}=0.05$~{\AA} and increases to $\sim\!3.9$ for ${\wrlim}=0.6$~{\AA}. We investigate how well the extrapolations for the ${\wrlim}=0.05$~{\AA} and ${\wrlim}=0.6$~{\AA} populations hold up against the observed $z<1$ data from the literature in Section~\ref{sec:evolve_lit}. 

In Figure~\ref{fig:Af0z0}(c), we plot the onset redshift, $z_0 = f_0/A$, as a function of {\wrlim}. 
For example, our simple model would suggest that {\CIV} absorbers with ${\wrlim} = 0.6$~{\AA} should have a cosmic onset in the approximate redshift range $3.3 \leq z \leq 4.6$.  Accounting for uncertainties in the extrapolation to obtain $z_0$, the onset redshifts for ${\wrlim} \leq 0.4$~{\AA} populations are above $z=4.75$, which are greater than the maximum redshift of our survey. Thus, for the ${\wrlim} < 0.4$~{\AA} absorber populations, $z_0$ is a valid quantity only under the condition that linear evolution of the absorber populations continues beyond the redshift range of our model fits.

\begin{deluxetable}{rlrr}[hbtp]
\vglue 5pt
\centering
\tablecaption{Model Parameters Fits \label{tab:evolveeqn}}
\tablehead{
\multicolumn{2}{l}{Parameterization }& 
\colhead{$a$} &
\colhead{$b$}
}
\startdata
${\meanNx}$\!\! & $= aW^{-b}$ & $0.16 \pm 0.02$ & $0.83 \pm 0.06$ \\ [-3pt]
$z_0$\!\! & $= aW^{-b}$ & $3.67 \pm 0.08$ & $0.24 \pm 0.02$ \\ [-3pt] 
$A$\!\! & $= aW+b$   & $1.08 \pm 0.05$ & $0.16 \pm 0.02$ \\ [-3pt]
$f_0$\!\! &  $= aW+b$ & $3.28 \pm 0.16$ & $1.40 \pm 0.05$ \\ 
\enddata 
\tablecomments{$W$ represents {\wrlim}. Fits are applicable for {\wrlim}$=0.05$ {\AA} to $0.7$ {\AA} over the redshift range {\zrangeall}.}
\vspace{-20pt}
\end{deluxetable} 

As {\wrlim} of the population is lowered, extrapolation yields onset redshifts that increase such that absorbers with ${\wrlim} = 0.3$~{\AA} would not be present above $z \sim 5.4$ and absorbers with ${\wrlim} = 0.05$~{\AA} would not be present above $z \sim 8.6$. The uncertainties in these predicted onset redshifts increase with decreasing {\wrlim}. However, given our simple linear model, allowing for cosmic variance, and considering that we have no {\it a priori\/} expectation for linear evolution to hold for $z > 4.75$, we reserve further judgement as to any predictive features of the linear model. In Section~\ref{sec:evolve_lit}, we contrast expectations of the model at higher redshifts with the available data in the literature, and in Section~\ref{sec:sims}, compare both the model and the data to cosmological simulations.

As a guide to the behavior of our fitting parameters as a function of {\wrlim}, we performed least squares fits to {\meanNx} (see Figure~\ref{fig:mean}), $A$, $f_0$, and $z_0$ (see Figure~\ref{fig:Af0z0}). The convenient functional forms of these parameters are presented in Table~\ref{tab:evolveeqn}. Both {\meanNx} and $z_0$ are well fit by a declining power law, whereas $A$ and $f_0$ increase linearly with {\wrlim}.


\section{Discussion} 
\label{disc}

This survey has provided a first opportunity to examine the properties of a sizable sample of ``weak'' ($W_r \leq 0.3$~{\AA}) {\CIV} absorbers over the redshift range {\zrangeall}, which corresponds to a cosmic time from when the universe was $\sim\! 1.5$~Gyr old ($z=4.75$) to when it was $\sim\!6$~Gyr old ($z=1.0$). Thus, we explore the evolutionary behavior of {\CIV} absorption-selected gas structures for a $\sim\!4.5$ Gyr period corresponding to $\sim\!10\%$ to $\sim\!45\%$ of the present age of the universe. 
With this work, we (1) extended previous measurements of the EWD and {\dndx} by an order of magnitude, down to ${\wrlim}\!=\!0.05$~{\AA}, and (2) characterized redshift evolution in {\dndx} as a function of equivalent width threshold over three orders of magnitude in {\EWr}. 

When analyzing the evolutionary behavior of absorption properties, such as equivalent widths and column densities, it is important to remain mindful that the measured evolution does not necessarily map directly to evolutionary behavior in the physical environments giving rise to the absorption. At different redshifts, absorbers with similar equivalent widths can potentially arise in very different astrophysical environments. For example, \citet{Dave99} showed that {\Lya} absorbers with ${\EWr} \sim 0.2$--$0.3$~{\AA} arise in gaseous environments at the cosmic mean density at $z\simeq 3$, whereas they arise in environments characterized by an overdensity of $\sim20$ at $z\simeq 0$. As a result of redshift evolution of the UVB and corresponding changes in the ionization state of absorbing gas, similar astrophysical environments likely host higher equivalent width {\CIV} absorbers at $z\sim1$ than at $z\sim0$ \citep{N05}. 

Given these considerations, the following is an exploration of the global evolution of {\CIV} absorbing properties more than it is of the astrophysical structures hosting absorbers.  We will, however, engage in some speculation with regards to astrophysical environments in our discussion. In addition to changing astrophysical environment, absorption strength is also  a complex combination of the densities, sizes, kinematics, ionization conditions, and chemical enrichment history of the gas structures.

\subsection{{\CIV} Evolution Across {\zrangeall}} 

\label{sec:evolveCIV}

In a pioneering survey, \citet{Sargent88} found that the redshift path density, {\dndz}, of strong (${\EWr} \geq 0.3$~{\AA}) absorbers increased by a factor of roughly three from ${z \sim 3.4}$ to ${z \sim 1.3}$. Extending their sample, \citet{Steidel90} reported this increase to be a factor of roughly four at ${1.3 \leq z \leq 4}$. The trend of increasing redshift path density with cosmic time for strong absorbers was later confirmed by \citet{Misawa02} (${2.3 \leq z \leq 4.5}$) and \citet{Peroux04} (${1.5 \leq z \leq 4.5}$). 
For {\wweak}, \citetalias{Cooksey13} found a factor of $\simeq\! 2.5$ increase in {\dndx} from $z=4.55$ to $z=2$ followed by a plateau at $z\leq2$ (they had only $\sim\!20\%$ completeness at ${\EWr} = 0.3$~{\AA}).  Similar to these previous works, we found {\wweak} absorbers increase monotonically by a factor of roughly three over this same cosmic period, with no sign of a plateau for $z \leq 2$ (we are $\sim\! 100\%$ complete at ${\EWr} = 0.3$~{\AA}).

As reproduced in Figure~\ref{fig:dndx}, \citetalias{Cooksey13} found that {\dndx} of {\wstrong} absorbers increases by a factor of roughly 2.5 from $z\!=\!4.55$ to $z\!=\!2$ and then are consistent with no evolution or a shallow decline from $z\!\sim\!2$ to $z\!\sim\!1.5$.  
Our measured {\dndx} for ${\wrlim} = 0.6$~{\AA} is consistent with that of \citetalias{Cooksey13} in the range $1.6 \leq z \leq 3.3$, but is roughly $\sim\! 2.5$ times lower for $3.4 \leq z \leq 4.5$. For this population of absorbers, we found that {\dndx} increases monotonically and smoothly by a factor of $\approx\!8.5$ from $z=4.75$ to $z=1$.

The new insights into {\CIV} evolution from this work are that {\CIV} absorbers are evolving linearly with redshift over the $\sim\!\!4.5$~Gyr time period from $z=4.75$ to $z=1$, such that the rate of evolution is dependent on the strength of absorption. The higher the {\wrlim} of the absorber population, the steeper the linear increase in {\dndx} with decreasing redshift. This quantifiable ``differential cosmic evolution'' with equivalent width threshold is informing us how the product $n(z)\sigma(z)$ is evolving as a function of the absorbing gas optical depth and velocity profile. We discuss physical interpretations of the observed evolution in Section~\ref{sec:phys}.

\subsection{Modeling {\CIV} Evolution} 
\label{sec:evolve_mod}

The inverse-power law (asymptotic) decline in {\meanNx} with increasing {\wrlim} suggests that the cosmic mean $\langle n \sigma \rangle$ across {\zrangeall} is smaller for populations of absorbers with progressively higher equivalent widths. In part, this behavior is reflecting the general behavior of the EWD across all redshifts, that smaller $W_r$ absorbers are more common than larger $W_r$ absorbers. For a given {\wrlim}, the linear decline in ${\Nx}(z)/{\meanNx}$ with redshift is either due to evolution in the number density, $n(z)$, the statistical absorber cross-section, $\sigma(z)$, or some combination of both. The next level of detail in understanding {\CIV} absorber evolution will consist of better understanding the evolution in $n(z)\sigma(z)$ for each absorber population. The linear model we have formulated can, in principle, be applied to constrain the nature of this evolution.

Consider the evolution of the ${\wrlim}\!\!=\!\!0.6$~{\AA} population. From Table~\ref{tab:evolve}, which gives $A\!\simeq\!0.9$, we infer that the dimensionless ratio $n(z)\sigma(z)/\langle n\sigma \rangle$ decreases by $90\%$ for each unit of redshift. Since $\langle n\sigma \rangle = (H_0/c){\meanNx}$, the evolution can be expressed in physical units. For ${\wrlim} = 0.6$~{\AA}, we measured ${\meanNx}=0.23$, which yields $\langle n\sigma \rangle \simeq  0.055$~kpc$^{-1}$. Multiplying by the evolution constant, we obtain the physical evolution rate, ${\cal A} = \langle n\sigma \rangle A = (0.055)\cdot(0.9) = 0.05$~kpc$^{-1}$ per unit redshift.

The physical evolution rate has the potential to provide powerful constraints on the absorber evolution. Adopting the definition ${\cal A} = \langle n\sigma \rangle A = -d[n(z)\sigma(z)]/dz$, we have  
\begin{equation}
{\cal A} 
         = - n(z) \frac{d\sigma(z)}{dz} \!-\! \sigma(z) \frac{dn(z)}{dz} \, .
\label{eq:real-evolve}
\end{equation}
Quantifying each term in this equation, $n(z)$, $dn(z)/dz$, $\sigma(z)$, and $d\sigma(z)/dz$, as a function of {\wrlim}, would completely specify the statistical evolution of {\CIV} absorbers. This would require extensive modeling, which
would draw on a broader set of observational constraints to inform the model. State-of-the-art theoretical simulations hold the key to addressing the physics underlying the evolution in $n(z)$ and $\sigma(z)$ and how they manifest the observed evolution of {\CIV} absorbing structures.

\subsection{{\CIV} Evolution Across $\sim\!12$ Gyr} 
\label{sec:evolve_lit}

In order to construct a broader picture of {\CIV} evolution from the first $\sim1$ Gyr after the big bang ($z\sim6$) to the present day ($z=0$) and compare to the simple linear extrapolations of our evolution model, we searched the literature for observations of {\dndx} in the redshifts $z<1$ and $z>4.75$.

There are caveats associated with comparing co-moving path densities from the literature. 
The lower the spectral resolution of a survey, the less sensitivity it has to lower {\EWr} absorbers and the more rapidly the survey incompleteness grows as {\wrlim} is decreased. High-resolution surveys, on the other hand, can be sensitive to much smaller {\EWr}. However, low-resolution surveys often comprise vastly greater numbers of quasar spectra (and therefore total redshift path coverage) than high-resolution surveys. 
This results in generally smaller uncertainties in the {\dndx} measurements of low-resolution surveys and large enough redshift path length coverage to accurately sample the larger {\EWr} absorbers, which are rarer due to the exponential drop in the EWD. Because of the typically shorter total redshift path coverage of high-resolution surveys, they can suffer from Poisson noise in the counts of the highest {\EWr} absorbers.  

Furthermore, in higher resolution spectra, the definition of what constitutes a single {\CIV} absorber can vary from study to study. For instance, \citet{Dodo13} defined absorbers within a $\dv =\pm 50$~{\kms} velocity window, whereas we used $\dv = \pm 500$~{\kms}.  
Moreover, some studies use equivalent width as a fundamental measurement to define an absorber population \citep[e.g.,][]{Sargent88, Steidel90}, whereas others use column density \citep[e.g.,][]{Dodo13, Burchett15, Codoreanu18}. Finally, how one determines their survey completeness in order to apply redshift path corrections as a function of {\wrlim} or column density threshold is central to measuring {\dndx}.

From our literature search, we found measurements of {\dndx} that extend to both $z\!<\!1$ and $z\!>\!4.75$ for only the {\wvweak} and {\wstrong} populations. For $z\!<\!1$, we found {\wvweak} measurements from \citet{Cooksey10} and \citet{Burchett15} and {\wstrong} measurements from \citet{Cooksey10}. For $z>4.75$, we obtained {\wvweak} measurements from \citet{Codoreanu18}\footnote{\citet{Codoreanu18} note that their own measured values of {\dndx} and {\dndz} roughly agree with the value they estimated from the $z \gtrsim 5$ high-resolution sample published by \citet{Dodo13}, who did not compute {\dndx} nor {\dndz}.} and {\wstrong} measurements from \citet{Simcoe+11}. These data thus provide a comparison over the full redshift range $z\sim0$ to $z\! \sim\! 6$ and enable us to assess the validity of our linear evolution model in the first $0.9-1.5$~Gyr and last $\sim\!7.7$~Gyr of cosmic time for these two {\CIV} absorber populations.

\begin{figure}[htbp]
\vglue 5pt 
\centering
\includegraphics[width=0.46\textwidth]{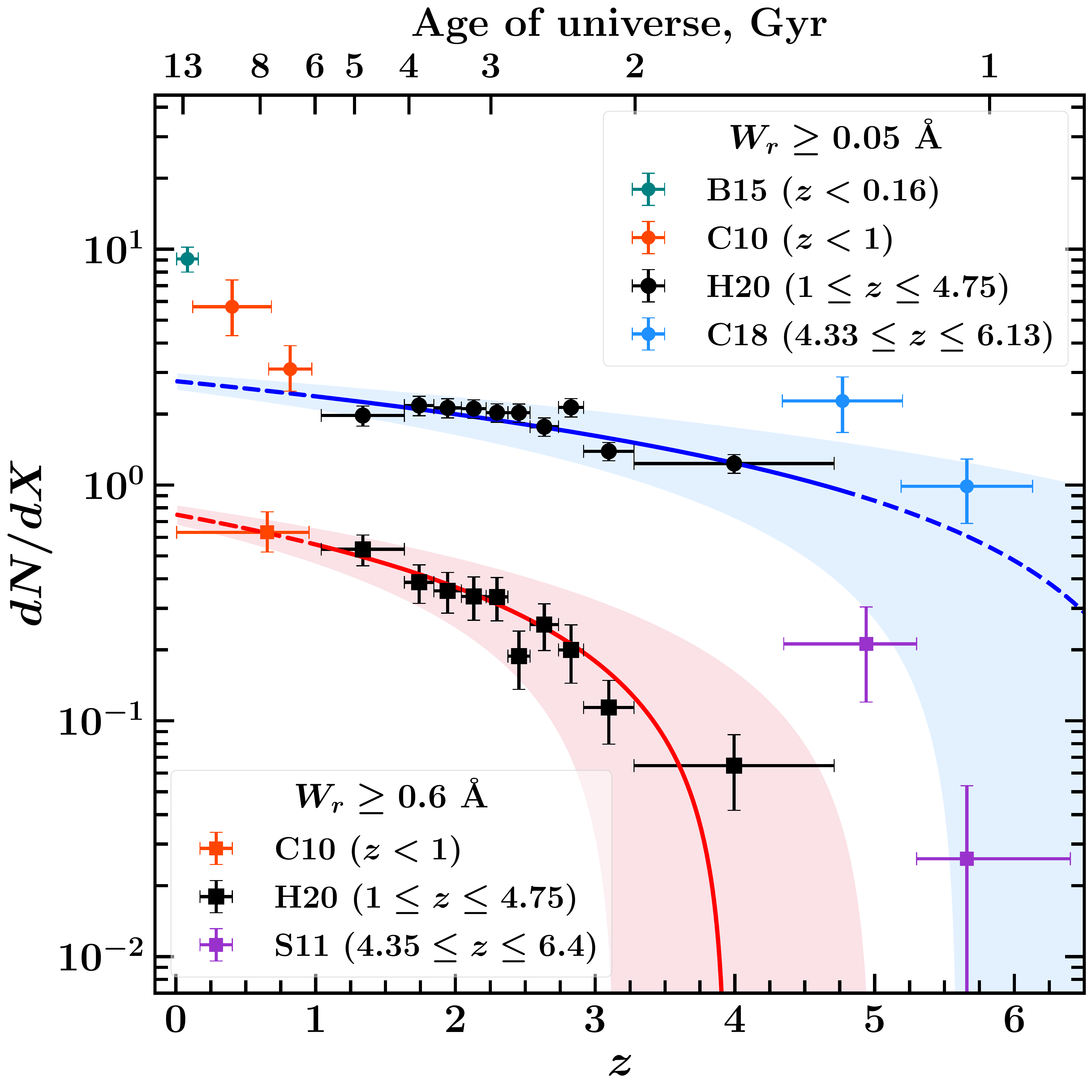}
\caption{Comparison of our measured {\dndx} (black points) over the range {\zrangeall}) with measurements at $z \leq 1$ \citep[][C10 and B15, respectively]{Cooksey10, Burchett15} and $z \geq 5$ \citep[][S11 and C18, respectively]{Simcoe+11, Codoreanu18}. Circles represent {\wvweak} and squares represent {\wstrong} absorbers, respectively. Our linear best-fit models (Eq.~\ref{eq:fz}) are overplotted as a solid blue curve for {\wvweak} absorbers and as a solid red curve for {\wstrong} absorbers. Light shadings represent the $\pm1\sigma$ confidence intervals of the fits. Model extrapolations are shown as dashed curves.}
\label{fig:dndx_comps}
\end{figure}

For the weak population, the {\wrlim} thresholds for the quoted {\dndx} values from \citet{Burchett15} and \citet{Codoreanu18} were not precisely equal to $0.05$ {\AA}. We first had to convert their column density thresholds to equivalent width thresholds (there is no uncertainty in the conversion due to thermal broadening because the absorption is on the linear part of the curve of growth). We then applied small correction factors to scale their measured {\dndx} to the threshold ${\wrlim}=0.05$ {\AA}.

For the \citet{Burchett15} data point at $z \simeq 0$, the correction factor was taken as the ratio of the areas under their fitted column density distribution (CDD), which was was normalized to {\dndx} in the same manner as our EWD (see Eq.~\ref{eq:ewdnorm}). The column density corresponding to ${\wrlim} = 0.05$ {\AA} is $\log N = 13.2$ and the correction factor is 1.21. 
However, \citet{Codoreanu18} normalized their CDD to the number of completeness-corrected {\CIV} absorbers instead of {\dndx}. Therefore, we used the ratio of the areas under our own measured EWD at {\zrangeh} (for the best-fit parameters shown in Table~\ref{tab:schparams}) for both the \citet{Codoreanu18} data points at $z>4.75$. The converted {\wrlim} from \citet{Codoreanu18} at $z \simeq 4.8$ was $0.015$~{\AA} and the correction factor is $0.61$; at $z \simeq 5.8$, the converted {\wrlim} was $0.06$~{\AA}, and the correction factor is $1.1$.

In Figure~\ref{fig:dndx_comps}, we present the co-moving path density of {\CIV} for {\wvweak} (${\wrlim}\!\!\simeq\!\!0.05$~{\AA}) and for {\wstrong} (${\wrlim}\!\!\simeq\!\!0.6$~{\AA}) over the redshift range {${0 \leq z \leq 6.4}$}. Our data ({\zrangeall}) are black points and our model fits and regions of $\pm1\sigma$ uncertainty are represented by curves and shaded regions, respectively. Absorbers with ${\wrlim}\!\simeq\!0.05$~{\AA} are represented by circles, while those with ${\wrlim}\!\simeq\!0.6$~{\AA} are squares.

\subsubsection{The {\wvweak} Population}

For the absorber population with {\wvweak} at ${z<1}$, the findings of \citet{Cooksey10} and \citet{Burchett15} indicate that {\dndx} evolves with redshift such that by $z=0$, the value of {\dndx} increases by a factor of roughly 4.5 over our linear model extrapolation. The quantity $n(z)\sigma(z)$ increased by only $\simeq1.8$ times in $4.5$~Gyr (over {\zrangeall}), while in the last $\sim\! 7.7$~Gyr ($0 \leq z \leq 1$), it increased by about a factor of $\simeq4.5$. Interestingly, this is not a dramatic change in terms of cosmic time, as $n(z)\sigma(z)$ increases at a rate of $\sim0.4$~kpc$^{-1}$~Gyr$^{-1}$ (from Eq.~\ref{eq:dndx2}) at {\zrangeall} and $\sim0.6$~kpc$^{-1}$~Gyr$^{-1}$ at $z<1$.

At  $z\!>\!4.75$, {\dndx} for {\wvweak} measured by \citet{Codoreanu18} at $5.2 \leq z \leq 6.2$ is consistent within the uncertainties of our model extrapolations. However, there is mild tension between our model and their measured {\dndx} for $4.3 \leq z \leq 5.2$. 
As previously discussed, the correction factor we applied to scale their measured $dN/dX$ was 0.61. If the weak-end slope of the true EWD at $5.2 \leq z \leq 6.2$ is steeper than at {\zrangeh} (a trend that would be consistent with the redshift evolution we have inferred for the EWD), then the corrected {\dndx} value would be in better agreement with the model extrapolation and consistent with a smooth evolution from $z\!\sim\!3$ to $z\!\sim\!6$. For example, to obtained a corrected $dN/dX \simeq\! 1.0$ to match our model extrapolation, we would require $\alpha$ to evolve from the measured $-0.98$ at $z \sim 3.5$ to $-1.8$ at $z \simeq 5$ (yielding a correction factor of 0.29).

Thus, we may infer that linear evolution could remain valid at $z\!>\!4.75$ for ${\wrlim} \simeq 0.05$~{\AA} absorbers if the EWD has a relatively higher frequency of weak absorbers as compared to the average relative frequency over $2.5 \leq z \leq 4.75$ (i.e. with a weak-end slope that is steeper by $\approx80\%$). Conversely, the data of \citet{Codoreanu18} may be indicating that {\CIV}-absorbing structures are more common than predicted from a linear extrapolation to $z\!>\!4.75$ and that linear evolution of {\dndx} with redshift does not continue to higher redshifts. Either way, additional data (or a reanalysis of the current data to ensure a uniform analysis between studies) would be required to resolve the tension between the extrapolated linear evolution and the measurement at $4.3 \leq z \leq 5.2$. 
    
\subsubsection{The {\wstrong} Population}

For the {\wstrong} population at $z<1$, the reported {\dndx} of \citet{Cooksey10} is consistent with our linear evolution model extrapolated to $z=0$. This indicates that linear evolution with redshift may be sufficient in describing the incidence of strong systems from $z=4.75$ to $z=0$, a cosmic period covering the last $\simeq\! 12.2$ Gyr of the universe. 

For $z\!>\!4.75$, the results are not as clear. The {\dndx} values from the survey of \citet{Simcoe+11}, as computed and presented by \citetalias{Cooksey13}, suggest a dramatic rise in strong {\CIV} absorbers between $4.3 \leq z \leq 5.3$ (i.e., for cosmic age $\sim\!1.0$--1.6 Gyr) such that {\dndx} is a factor of $\simeq\! 3$ higher at $z\sim\! 4.3$ than our measured {\dndx}. The measured value of {\dndx} at $5.3 \leq z \leq 6.3$ (i.e., for cosmic age $\sim\!0.8$--1.0 Gyr), is in complete conflict with the extrapolation of our linear evolution model, as the onset redshift for the ${\wrlim} \!=\! 0.6$~{\AA} absorber population is roughly $3 \leq z \leq 5.2$ (pink shaded area in Figure~\ref{fig:dndx_comps}), suggesting that we should not observe these higher redshift absorbers according to our linear model.

\subsubsection{Beyond Linear Evolution?}

In summary, there are two concrete conclusions that can be drawn from the $z<1$ data comprising Figure~\ref{fig:dndx_comps}. (1) Evolution of the {\wvweak} absorber population appears to undergo a transition at $z\simeq 1$ such that, the quantity $n(z)\sigma(z)$ increases by a factor of 4--5 from $z=1$ to $z=0$ relative to an extrapolation of the linear evolution observed for {\zrangeall}. (2) For {{\wstrong}}, linear evolution appears to hold for $0 \leq z \leq 4.75$, such that the quantity $n(z)\sigma(z)$ increases by a factor of $\sim\!10$ over the last $12.2$~Gyr.  

Interestingly, linear evolution with redshift does not translate to linear evolution with cosmic time. If the evolution in $n(z)\sigma(z)$ is examined per unit time, the rate is actually a fairly constant $\sim\! 0.4$--0.6~kpc$^{-1}$~Gyr$^{-1}$ for the {\wvweak} absorber population over the range $0 \leq z \leq 4.75$.  The ``turn up'' in redshift for $z<1$ is a consequence of a fairly steady evolution and the compression of the time axis with decreasing redshift (see Figure~\ref{fig:dndx_comps}). For the {\wstrong} population, the evolution rate is $\sim\! 1.8$~kpc$^{-1}$~Gyr$^{-1}$ for the 4.5~Gyr spanning $1.0 \leq z \leq 4.75$, and $\sim\! 0.14$~kpc$^{-1}$~Gyr$^{-1}$ for the 7.7~Gyr spanning $0 \leq z \leq 1.0$. Thus, from the standpoint of temporal evolution, the strongest systems evolve most rapidly in the early universe and slower in the recent universe, even though their evolution is linear with redshift.

When comparing our {\dndx} values with those at $z> 4.75$ for both weak and strong absorber populations, there is unexpected upturn in {\dndx} at $z\sim5$, followed by a decline at $z \sim 6$. Our measured {\dndx} (and that of \citetalias{Cooksey13} from their SDSS survey) shows a smooth and steady decrease with increasing redshift as $z=5$ is approached, and this trend might be expected to continue to higher redshifts. Thus, the elevated {\dndx} at $z\sim\! 5$ for both weak and strong absorbing structures from the infrared surveys of \citet{Simcoe+11} and \citet{Codoreanu18} places some tension on our expectations of {\CIV} evolution between the first and second billion years of the universe. 

Notably, the \citet{Simcoe+11} measurement at $4.3 \leq z \leq 5.3$ for ${\wrlim}=0.6$~{\AA} is based on only two {\CIV} absorbers and the $5.3 \leq z \leq 6.3$ measurement is based on just one absorber. Furthermore, the values reported by \citet{Codoreanu18} for ${\wrlim} \simeq0.05$~{\AA} incorporated 30 {\CIV} absorbers for the $4.3 \leq z \leq 5.2$ measurement and six for the $5.2 \leq z \leq 6.2$ measurement. However, these 36 absorbers were found in only four quasar spectra. As such, it is not implausible that there may exist some systematic uncertainties, such as cosmic variance, in these high-redshift infrared surveys.

On the other hand, the redshift path coverage of optical surveys declines at $z>3$ (see Figure~\ref{fig:gwz}), so that in order to reduce the uncertainties in {\dndx}, the redshift bin is extended. This has the effect of averaging across a larger redshift range and the loss of ``redshift resolution'' for $z\simeq3$ to $z\simeq 4.75$. This may be blurring our ability to resolve an upward trend in our data in this redshift range.

Alternatively, the infrared data may reflect a true rise in the cosmic incidence of {\CIV} absorbers in a short period of time, only to fall again in a few hundred million years. Though a physical argument for such behavior may be difficult to formulate and may appear to be contrived at face value, it cannot yet be ruled out. For example, such a scenario might require chemical enrichment in both high and low overdensity astrophysical environments in such a manner that the different ionization conditions in these different environments constrain both optically thin and optically thick absorbing structures to evolve similarly over the same narrow cosmic time period. Future observational programs to study the high-resolution spectra of a large number of high-redshift quasars would be key for resolving the tension in the measured {\CIV} absorber evolution.

\subsection{Sizes of {\CIV} Absorbing Structures} 
\label{sec:rstar}

To gain further insight into the statistical cross section of {\CIV} absorbers, we estimate the $\sigma(z)$ of the gas complexes associated with different populations of {\CIV} absorbers from ${\Nx}(z) = (c/H_0)n(z)\sigma(z)$ (Eq.~\ref{eq:dndx2}). For this exercise, we assume that these complexes are the gaseous halos of galaxies, or the CGM.
We adopt the standard \citet{Holmberg75} scaling relation between galaxy luminosity, $L$, and halo absorbing gas radius, $R(L) = R_{\star} (L/L_{\star})^{\beta}$, where $\beta \!\approx\! 0.3$--0.4 \citep[e.g.,][]{Chen01, glenn08, nikki13}, and {\Rstar} is the effective absorbing gas halo radius for an $L_{\star}$ galaxy. For a given luminosity, the gaseous halo cross section is $\sigma(z) = \pi f_c(z) R^2(L)$, where $f_c(z)$ is the covering fraction of the absorbing gas, which may change with redshift \citep[see e.g.][]{nikki13}.

The cosmic number density of gas structures, $n(z)$, is given by the observed number density of galaxies, obtained by integrating the galaxy luminosity function. 
We adopt the fitted Schechter functions of \citet{Parsa16}, who provide the functional parameters $\alpha(z)$, ${\Lstar}(z)$, and $\phi_{\star}(z)$ as fitted functions from observed UV luminosity functions over the redshift range $0\leq\! z \leq\!8$.
We then integrated the product $n(z)\sigma(z)$ over luminosity from a minimum luminosity of ${L_{min}(z) = 0.1{\Lstar}(z)}$ to infinity, and obtain
\begin{equation}
R_{\star} (z) = 
\left[ 
\frac{H_0}{\pi c f_c(z)\phi_{\star}(z)}
\frac{N_{\hbox{\tiny X}}(z)}{\Gamma\left[x(z),l(z)\right]}
\right]^{1/2} \, ,
\end{equation}
where $\Gamma(a,b)$ is the upper incomplete Gamma function \citep{Abramowitz72}, ${\Nx}(z)$ is the measured {\dndx} at redshift $z$, ${l(z) = L_{min}(z)/L_{\star}(z) = 0.1}$,  and $x(z)\!=\! 2\beta\!+\!\alpha(z)\!+\!1$. We assume $\beta=0.35$. The choice of $l(z) = 0.1$ is motivated by the findings of \citet{Burchett16} who observed that $z \simeq 0$ {\CIV} absorbers with ${\EWr} \gtrsim 0.05$~{\AA} are found around galaxies with $M_{\star} > 10^{9.5}$~{\Msun}, which corresponds to $\sim0.1{\Lstar}$.  Though the luminosity corresponding to a given stellar mass is known to evolve with redshift at $z>4$ \citep{Behroozi19}, and also considering that the minimum $M_{\star}$ for {\wvweak} {\CIV} absorbers could evolve, at the present time there are no observational data to further constrain $l(z)$ in the redshift range of our study.

In Table~\ref{tab:rstar}, we present selected rounded values of {\Rstar} for three redshifts in the range $1.0\leq z\leq 4.75$ for three populations, ${\wrlim}\!=\!0.05$, 0.3, and 0.6~{\AA}. We assume a unity covering fraction, where ${\Rstar} \propto f_c(z)^{-1/2}$.  For {\wvweak} absorbers, {\Rstar} ranges from $\sim\!205$~kpc at $z\simeq4$ to $\sim\! 240$~kpc at $z\simeq2.5$, which holds roughly constant down to $z\simeq1.3$. For {\wweak} absorbers, we obtained $\sim\!90$~kpc at $z\simeq4$ to $\sim\!145$~kpc at $z\simeq1.3$. And for {\wstrong} absorbers, {\Rstar} ranges from $\sim\! 45$~kpc at $z\simeq4$ to $\sim\!115$~kpc at $z\simeq 1.3$. The typical uncertainties on {\Rstar} from propagating errors on $\Nx(z)$ are on the order of $\sim\! 10$--$20\%$.

\begin{deluxetable}{cccc}[htbp]
\vglue 5pt 
\centering
\tablewidth{0pt}
\tablecaption{Characteristic Absorbing Halo Sizes for an {\Lstar} Galaxy}
\label{tab:rstar}
\tablehead{
\colhead{{\wrlim}} & 
\colhead{{\Rstar}$(z\!\simeq\!1.3)$} &
\colhead{{\Rstar}$(z\!\simeq\!2.5)$} &
\colhead{{\Rstar}$(z\!\simeq\!4.0)$} \\[-5pt]
\colhead{({\AA})} &
\colhead{(kpc)} &
\colhead{(kpc)} &
\colhead{(kpc)}
}
\startdata
0.05 & 230 & 240 & 205 \\ [-3pt]
0.3 & 145 & 125 & 90 \\ [-3pt]
0.6 & 115 & 75 & 45 \\
\enddata 
\end{deluxetable}

Under the assumption that $n(z)$ of {\CIV} absorbers is given by the galaxy luminosity function, the exercise indicates that {\Rstar} for {\wvweak} absorber structures grows at a rate of $\sim\! 35$ kpc in $\approx\!1.5$ Gyr (from $z=4$ to $z=2.5$) and then remains constant (within the uncertainties) until $z\sim1$. For the {\wweak} absorbers, {\Rstar} increases steadily and monotonically by $\approx\!12$~kpc~Gyr$^{-1}$ for the 4.5 Gyr period covered by our survey, while the rate of increase of the {\wstrong} absorbers is $\approx\!16$~kpc~Gyr$^{-1}$. These estimates can be considered lower limits, as {\Rstar} increases as the inverse square root of the covering fraction.

We did not take into account the possibility of different absorber populations being primarily associated with different types of galaxies; e.g. stronger absorbers associated with brighter galaxies. For example, \citetalias{Cooksey13} estimated {\Rstar} for {\wstrong} absorbers assuming these absorbers reside only in the halos of galaxies brighter than $L_{min} = 0.5L_{\star}$. They obtained {\Rstar} $\approx 50$ kpc at {\zrangec} (noting $\sim\!\!20-60\%$ uncertainties in their calculated galaxy number densities). If we adopt $L_{min} = 0.5L_{\star}$, our estimated {\Rstar} increases by $\sim50\%$ at $z\simeq 1.3$, yielding $\sim\!180$~kpc, and by $\sim60\%$ at $z\simeq 4$, yielding $\sim\!80$~kpc.
Any differences between our estimate and that of \citetalias{Cooksey13} are likely due to the adoption of different luminosity functions and $L_{min}$.

There is a well-established anti-correlation between the strength of {\CIV} absorption and impact parameter from the host galaxy \citep{Chen01, Chen12, LC14, Bordoloi14}, which is interpreted as the absorption strength decreasing with galactocentric distance. At $z\sim 2.5$, \citet{Adelberger05} and \citet{Steidel10} found {\CIV} absorption out to $\sim\! 100$~kpc around Lyman Break Galaxies (LBG), with the strength of absorption decreasing with increasing impact parameter.
\citet{Adelberger05} notes that a $W_r=0.6$~{\AA} {\CIV} absorber would have a typical impact parameter of 80--85~kpc. Our estimates of {\Rstar} for the absorber populations defined by {\wweak} and {\wstrong} are in good agreement with these findings. 
However, the comparison is not direct, as LBGs represent the bright end of the luminosity function and are thus not fully representative of the field galaxy population at the redshifts we study.

The virial radius of a Milky Way-sized galaxy halo is $\approx200$ kpc at $z=0$ \citep{Navarro10}. Given our estimates for {\Rstar} of 210 to 240~kpc from $z\simeq4$ to $z\simeq 1.3$, and since the virial radius of Milky Way-like galaxies is substantially smaller at higher redshift, we may infer that some fraction of the weakest absorbers reside outside the virial radius of galaxies and perhaps even in the IGM. This may also hold true at $z\simeq 0$, where we estimate $R_{\star} \approx 450$~kpc for {\wvweak} based on the data of \citet{Burchett15}. An understanding of how metals escape hundreds of kiloparsecs from the galaxies wherein they are produced will likely be developed through detailed theoretical modeling with physically accurate feedback prescriptions.

\subsection{Physical Interpretations} 
\label{sec:phys}

As discussed above, the statistical evolution of {\CIV} absorbers can be understood in terms of the evolution of the product $n(z)\sigma(z)$.  Physically, these quantities evolve due to changes in the column densities of the absorbing clouds and the kinematic and dynamical motions of absorbing gas structures. The monotonic increase in the evolution constant $A$ with increasing {\wrlim} suggests that the evolution of absorbing clouds is more rapid for more optically thick and/or kinematically complex absorbers. That is, the rate of increase of the product $n(z)\sigma(z)$ with time is higher for these absorbers than for optically thin and/or kinematically simple absorbers (which dominate the overall {\CIV} absorber population). 

There are two factors that govern the evolution of {\CIV} absorbers. The first is the metal enrichment history of the IGM and CGM and the second is the nature of the ionization of carbon, which is dependent on both the intensity and spectral energy distribution of the ionizing spectrum local to the absorbing structures and the optical depth of the absorbing structures themselves. We first consider the chemical enrichment history and then the nature of the ionization physics. 

\citet{Schaye03} used pixel statistics obtained from UVES and HIRES data to derive a model of the metallicity distribution ([C/H]) in the IGM and concluded that there was little enrichment across $2 \leq z  \leq 4$, implying that most of the enrichment must have taken place prior to $z\sim4$. Indeed, the {\CIV} mass density, $\Omega_{\hbox{\tiny CIV}}$, is observed to be effectively unchanging across $1.5 \leq z \leq 4.5$ \citep{Pettini03,Songaila05,Scanna06,BS15}. \citet{Cooksey10} combined their $\Omega_{\hbox{\tiny CIV}}$ measurements at $z<1$ with those at higher redshift and found only a shallow rise in $\Omega_{\hbox{\tiny CIV}}$ from $z\sim5$ to the present epoch. 

Unfortunately, $\Omega_{\hbox{\tiny CIV}}$ is not an ideal tracer of the total carbon mass density, $\Omega_{\hbox{\tiny C}}$, because $\Omega_{\hbox{\tiny CIV}} \!\ll\! \Omega_{\hbox{\tiny C}}$ \citep{Schaye03}. Furthermore, $\Omega_{\hbox{\tiny CIV}}$ is dominated by the few highest column density {\CIV} absorbers and is thus sensitive to the range of column densities observed, which can be dependent on the total redshift path covered in a survey.  
Similarly, {\dndx} is a good statistic for the target ion of a given atomic species, but not for the atomic species itself. In order to obtain a robust count of all the carbon in the universe, one needs to measure the incidence of neutral carbon as well as all of its ionized phases (of which {\CIV} is but one). Nonetheless, there is an important difference between $\Omega_{\rm ion}$ and {\dndx} for said ion. $\Omega_{\hbox{\tiny CIV}}$ is most sensitive to and dominated by the highest column density {\CIV} absorbers, which are the most rare by number. Thus, $\Omega_{\hbox{\tiny CIV}}$ can be subject to cosmic variance. {\dndx} is most sensitive to and dominated by the most common or numerous absorbers, which are the weakest ones (as shown by the shape of the EWD).  The incidence of these abundant weak absorbers is far less subject to cosmic variance.  As such, the two metrics are complementary in the information they provide about absorber populations.

Based on $\Omega_{\hbox{\tiny CIV}}$ and applying ionization corrections, \citet{Simcoe11} found that $\Omega_{\hbox{\tiny C}}$ increased a factor of no more than two from $z\sim4.3$ to $z\sim2.4$, which indicates that there has been a small enrichment of carbon in this time period. This result is consistent with the estimated amount of carbon ejected into the CGM and IGM via supernovae feedback.

Given that the cosmic star formation rate density peaked around $z\sim2$ \citep[e.g.,][]{MD14}, it would seem reasonable that metal enrichment of the IGM and CGM occurred at redshifts much lower than $z\!\simeq\!4$. Indeed, the period known as ``cosmic noon'', defined as the epoch at which galaxies assembled roughly half of their stellar mass, occurred at $1.5 \leq z \leq 3$ \citep[e.g.,][]{Murphy11, Behroozi13, Feldman16}. This epoch also saw the peak in galactic-scale outflows from stellar processes \citep[e.g.,][]{Rupke18}. Large observational programs are currently underway to characterize the CGM, including {\CIV} absorbers, at cosmic noon \citep[e.g.][]{nikki20}.
The measured $dN/dX$ of {\CIV} absorbers exhibits only a slow, steady increase during cosmic noon, which at face value, does not suggest a substantial rise in the carbon abundance in the CGM and IGM in this period. The slow rise in $dN/dX$ is, however, consistent with a steady increase in the mean metallicity of the universe from $z\sim5$ to $z\sim0$ \citep[e.g.,][]{Rafelski12, Lehner16, Mcquinn16, DeCia18}.

For {\zrangeall}, it is an open question as to what degree the evolution of the metagalactic ionizing background, both globally and local to {\CIV}-absorbing structures, drives the observed evolution of $dN/dX$. Of particular interest in this regard is to pose the question of how the UVB may govern the linear evolution of $n(z)\sigma(z)$ with redshift such that the rate of evolution increases for progressively larger {\wrlim} populations (recall that we find $A \propto {\wrlim}$, see Table~\ref{tab:evolveeqn}). According to linear evolution, for a given {\wrlim}, the evolution of $n(z)$ and $\sigma(z)$ are constrained by Eq.~\ref{eq:real-evolve}.

The average global UVB has been known to become harder (increased relative proportion of higher energy photons) with time \citep[e.g.,][]{FG09,HM12,Puchwein19}, allowing the {\CIV} transition to dominate over lower-ionization transitions. Theoretical models have often assumed the UVB is spatially homogeneous and externally imposed \citep[e.g.,][]{Oppenheimer10,vdv11b,Keating16,Rahmati16}, but such models have often been unsuccessful in reproducing observed properties of {\CIV} absorbers at $z \geq 4$ \citep{ODF09,Rahmati16,KF16,KF18}. High redshift observational studies have also identified a need for variations in the local ionizing radiation field around absorbers \citep[e.g.,][]{Dodo13, BS15, Morrison19}.

While the reionization of hydrogen produces inhomogeneities in the UVB only at $z>5$, helium reionization creates significant spatial variations that persist to $z\simeq 3$ \citep[e.g.,][]{Becker11}. This is due to the shorter mean free path of {\HeII}-ionizing photons and the high ionization potential of {\HeII} (4 Ryd) which requires highly energetic sources with ``hard'' spectra, such as quasars \citep{Faucher08}. By the end of {\HeII} reionization near $z\simeq3$, isolated regions as large as $\sim50$~Mpc,  characterized as ``{\HeIII} bubbles'', can have a harder UVB spectral energy distribution relative to the average UVB spectral distribution \citep{Mcquinn09, Fur09}. 

Since the C$^{+3}$ and He$^{+}$ ions have very similar ground-state ionization energies, the fraction of carbon in the C$^{+3}$ ionization stage will vary in relation to the {\HeII} ionization edge opacity, which governs the hardness of the local UVB for energies above 4 Ryd.  \citet{Worseck16} show that these {\HeIII} bubbles persist for over $\sim\!600$ Myr past the end of {\HeII} reionization, lasting until at least $z\sim2.7$. These {\HeIII} bubbles therefore persist into the early period of cosmic noon, when galaxies begin to contribute {\HI} and {\HeII} ionizing photons to the UVB \citep[e.g.,][]{HM12}. As cosmic noon is dialing down by $z=1.5$, we see that the behavior of the UVB and its influence on the ionization balance of carbon, and thus the incidence of {\CIV} absorbers, is dramatically evolving in both its homogeneity and spectral energy distribution over the redshift range {\zrangeall}. 

Given the complexity of the evolution of the UVB, it is remarkable that $dN/dX$ of {\CIV} absorbers shows a monotonic linear increase with decreasing redshift over the 4.5~Gyr cosmic time covered by our survey. Interestingly, \citet{Simcoe11} found that {\CIV} and higher-ionization species are more abundant relative to {\CIII} and lower ionization species from $z=4.3$ to $z=2.4$; a result that generally supports a trend in which higher ionization states are favored at later times \citep[see also][]{Becker19}. This is consistent with a rise in the incidence of {\CIV} absorbers with cosmic time.

Our results for the onset redshift $z_0$ suggest that the population dominated by weak absorbers, i.e., ${\wrlim}=0.05$~{\AA}, would be present as soon as $z\sim\!8$ (when the universe was $\sim\!600$~Myr old), whereas absorbers with ${\wrlim}=0.6$~{\AA} would appear no earlier than $z\sim\!4$ (when the universe was about 1.8~Gyr old). This would imply that optically thin {\CIV} clouds came into existence in the midst of the epoch of {\HI} reionization \citep[e.g.,][]{Gnedin97, Gnedin00, BL01}, whereas optically thick and/or kinematically complex clouds arose following the completion of this epoch \citep[e.g.,][]{Bolton07, Robertson13}. Establishing the redshifts at which {\CIV} absorbers first arise would provide insights into the origins of cosmic carbon enrichment, the early mechanisms by which metals are distributed into the IGM and CGM, and details into the transmission and spectral shape of the ionizing radiation field during the epoch of {\HI} reionization \citep[e.g.,][]{Becker15, KF15}.

Whether the origins and physical processes affecting the weakest absorbers is distinct from that of the strongest absorbers is an open question. Though some studies find evidence to relate {\CIV} to galactic winds \citep[e.g.,][]{Fox07,Steidel10}, strong {\MgII} is likely a better tracer of such outflows as their evolution mimics the star formation and stellar-driven outflow activity of the universe \citep[e.g.,][]{Menard11,MS12,Zhu13,Chen15}. \citet{Songaila06} points out that very weak {\CIV} absorbers would require unreasonably large outflow velocities if they were associated with galactic outflows. Alternatively, they argue that these weak absorbers could be ionized by active galactic nuclei (AGN) spectra, rather than by local galaxies or the global UVB.

However, if local AGN were indeed the primary culprits behind the incidence of weak {\CIV} absorbers, then we would expect $dN/dX$ for {\wvweak} to trace the AGN activity of the universe. As shown in Figure~\ref{fig:dndx_comps}, {\dndx} for weak absorbers does not peak around $z\sim2$ nor does it decline at later times that would mirror the observed AGN luminosity function or cosmic black hole accretion rate \citep[e.g.,][]{Shankar09, Kulkarni19, Shen20}. On the contrary, {\dndx} for weak absorbers rises rapidly with decreasing redshift only below $z =1$, well after the epoch of peak AGN activity. This is also further evidence against outflows being the source of weak {\CIV} since star formation activity peaks around $z\sim2$. At the very least, we would expect a relatively more rapid increase in {\CIV} {\dndx} from higher redshifts to $z\sim2$ if outflows were the source of {\CIV}.

\begin{figure*}[thbp]
\centering
\gridline{\fig{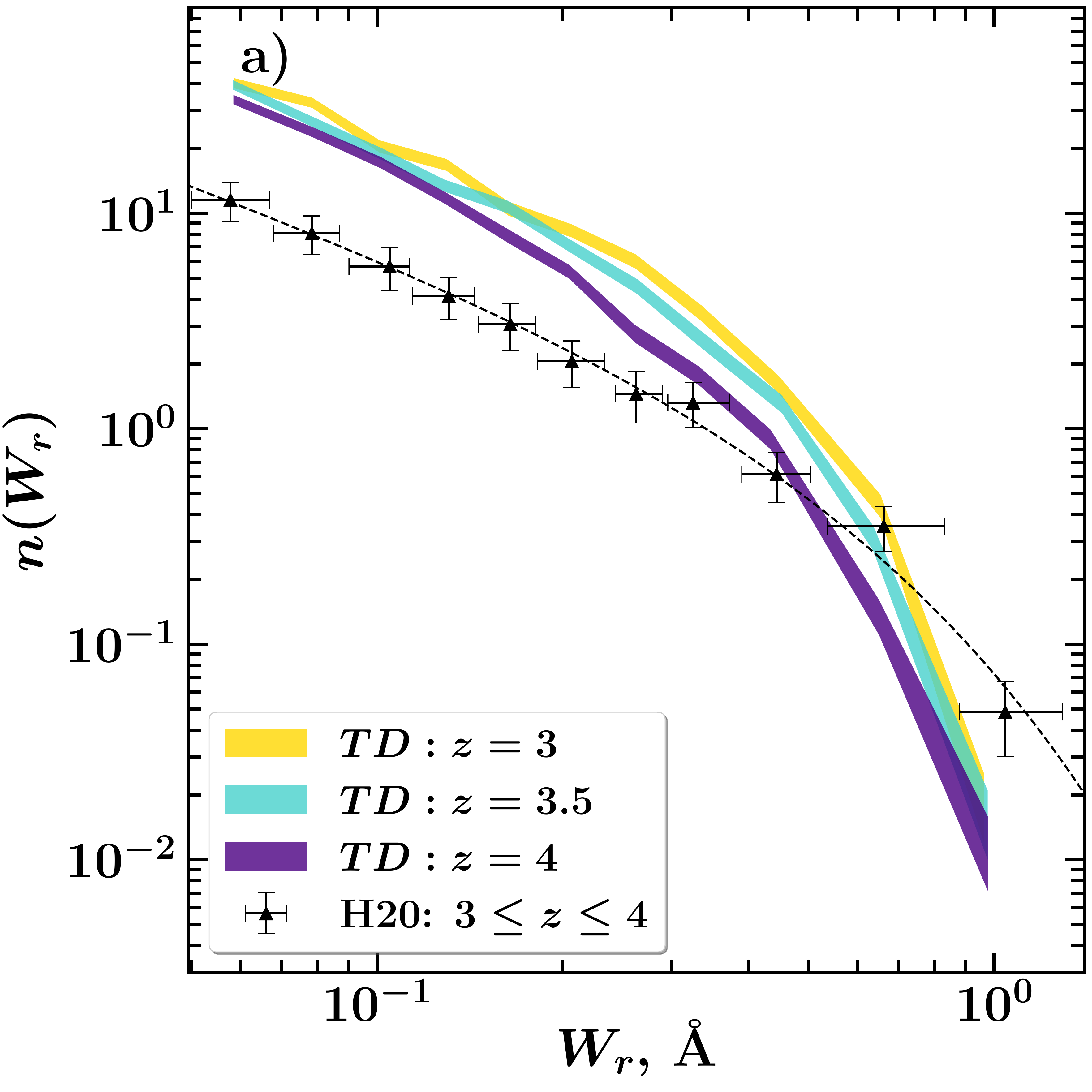}{0.32\textwidth}{}
        \fig{dndx_kf_comp_new.pdf}{0.32\textwidth}{}
        \fig{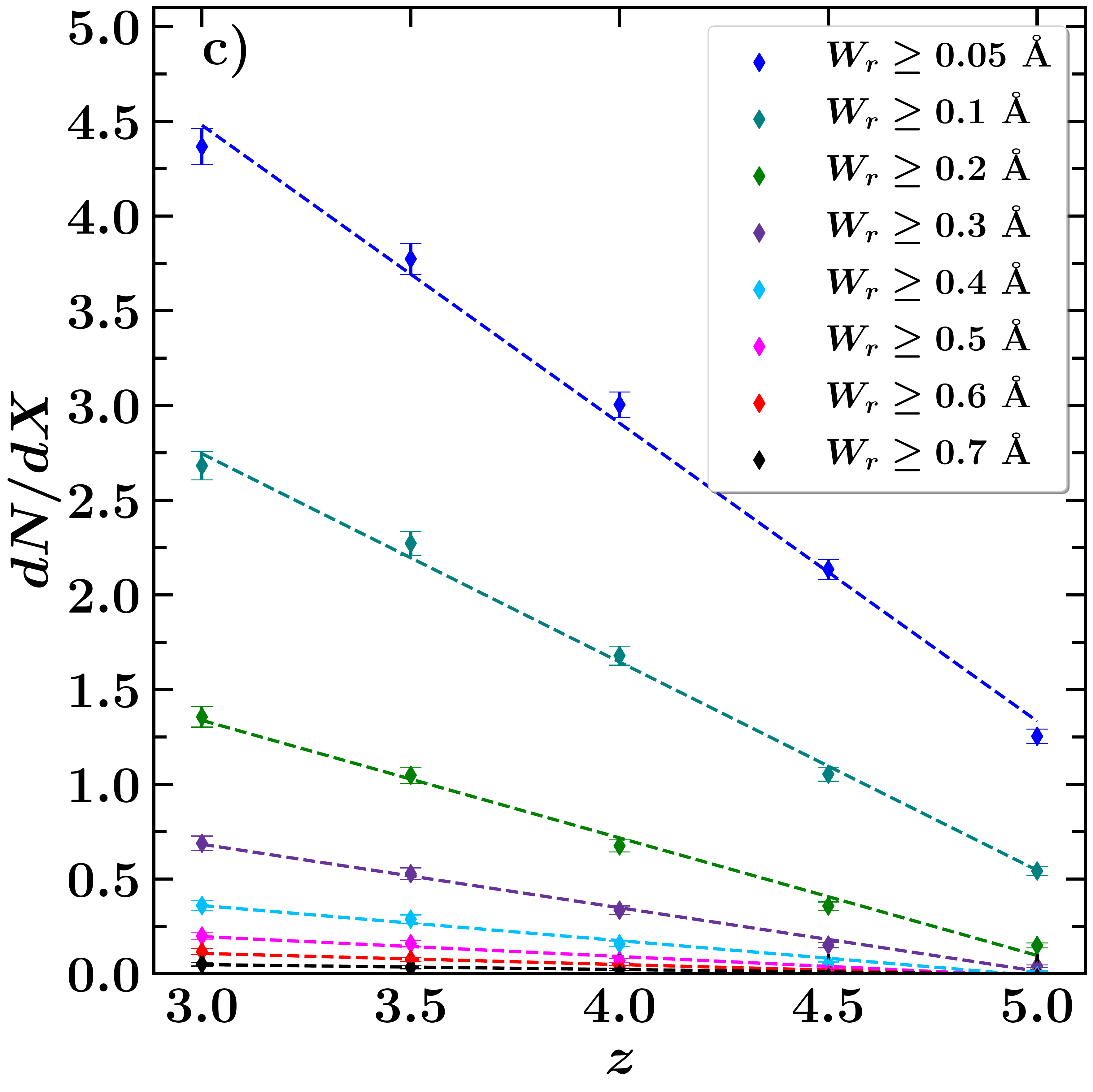}{0.32\textwidth}{}
        } \vspace{-20pt}
\caption{(a) Our observed EWD data points, with best-fit Schechter function parameterization represented by a dashed black curve, at $3 \leq z \leq 4$, compared to the {\it TD} simulations at $z=3$, $z=3.5$, and $z=4$ (represented by colored shading). (b) Our observed {\dndx} (data points and error bars), compared to the {\it TD} simulations at $z\geq3$ (opaque bands).
Our linear best-fit evolution models (Eq.~\ref{eq:fz}) are overplotted in blue, purple, and red dashed lines for {\wvweak}, {\wweak}, and {\wstrong} absorbers, respectively, with light shading depicting the $\pm1\sigma$ confidence intervals for those fits. (c) {\dndx} as a function of redshift, for different {\wrlim} in the TD simulations. The dashed lines show the linear best-fit models for each absorber species in the simulations.}
\vglue 5pt
\label{fig:kfcomp}
\end{figure*}

\subsection{Theoretical Explorations} 
\label{sec:sims}

To augment our theoretical understanding of the observed characteristics and evolution of {\CIV} at high redshift, we compared our observations to the predictions of the cosmological radiation hydrodynamic simulations {\sc Technicolor Dawn} \citep[TD,][]{KF18}. The TD simulations model galaxy growth and the baryon cycle, while self-consistently modeling an evolving multi-frequency, spatially-inhomogeneous UVB. Though the TD simulations accurately reproduce observations of the galaxy stellar mass function, the co-moving path density of {\CII} and {\SiIV} absorbers, and the mean transmission in the {\Lya} forest, at $z \geq 5$ they underproduce the {\dndx} of strong {\CIV} absorbers \citep{KF18,Doughty19,KF20}. Similar results have been reported for the {\sc Eagle} simulations \citep{Rahmati16}.

Though \citet{KF20} showed that this under-abundance of {\CIV} absorbers could be evidence for density-bounded ionizing escape, this interpretation is not unique in that it is degenerate with the possibility that the intrinsic stellar emission is too ``soft'' \citep[deficient in {\CIV} ionizing photons; see][]{Z13}. Either way, the cosmic epoch of $z\geq5$ is complicated to simulate as it covers the immediate aftermath of {\HI} reionization and the earliest stages of {\HeII} reionization. 
Previous comparisons of {\CIV} observations to the TD simulations have been restricted to $z\gtrsim5$ and focused on column densities \citep{KF18, KF20}. Here, with our improved observational constraints on the EWD and {\dndx} of {\CIV} for $z < 5$, we investigate whether the TD simulations reproduce observed {\CIV} statistics at the slightly lower redshift range $z\!=\!3$ to $z\!=\!5$, under the assumption of ionization-bounded escape. 

We generated a sample of {\CIV} absorbers from the TD simulations at $z=3.0$, 3.5, 4.0, 4.5, and 5.0. The TD spectra and {\CIV} absorption lines were generated as described in \citet{KF18} using instrumental parameters (resolution, pixel sizes, and pixels per resolution element) consistent with the HIRES and UVES spectrographs. The signal-to-noise ratio of the spectra was fixed at 35 per pixel, which ensures that the TD spectra are 100\% complete to a $5\sigma$ detection threshold of $W_r = 0.05$~{\AA}. To ensure that both the absorber incidence and EWDs are consistent with our survey, we define a single absorber in the TD spectra to include all absorption components within a $\pm 500$~{\kms} window (see Section~\ref{sec:detect}). The TD simulated {\CIV} absorber catalogs contain a total of $\sim\!10,000$ absorbers with {{\wvweak}}.

In Figure~\ref{fig:kfcomp}(a), we compare our observed EWD at $3 \leq z \leq 4$ to the EWDs from TD at $z=3.0, 3.5,$ and $4.0$. We limited the comparison for the EWD to this redshift range because (1) we have robust statistics in this range of our survey, and (2) we aim to examine a regime cleanly segregated from $z \geq 5$. The TD EWD was computed from the simulated {\CIV} absorber catalog using methods identical to those we applied to our observational data (see Section~\ref{sec:ewd}). 

For a statistical comparison between our observed EWD and the simulated EWDs, we perform Monte Carlo realizations similar to those described in section~\ref{sec:ewd}. We draw a random sample of {\EWr} values from the simulated TD catalogs at $3 \leq z \leq 4$ with the same number of absorbers as in our observed sample. In this redshift range, we have 193 absorbers with {\wvweak} and 56 absorbers with {\wweak}. We then perform a K-S test comparing the randomly drawn sample with our observed Schechter probability distribution function\footnote{The reason we compare to the observed probability distribution function and not to the observed data is that the raw data are not completeness-corrected, whereas the Schechter function is fit to the completeness corrected data.} and conduct 1 million realizations of this experiment. These realizations were performed for both {\wvweak} and {\wweak} absorbers from TD.

\begin{figure}[htb]
\centering
\vglue +10pt
\includegraphics[width=0.42\textwidth]{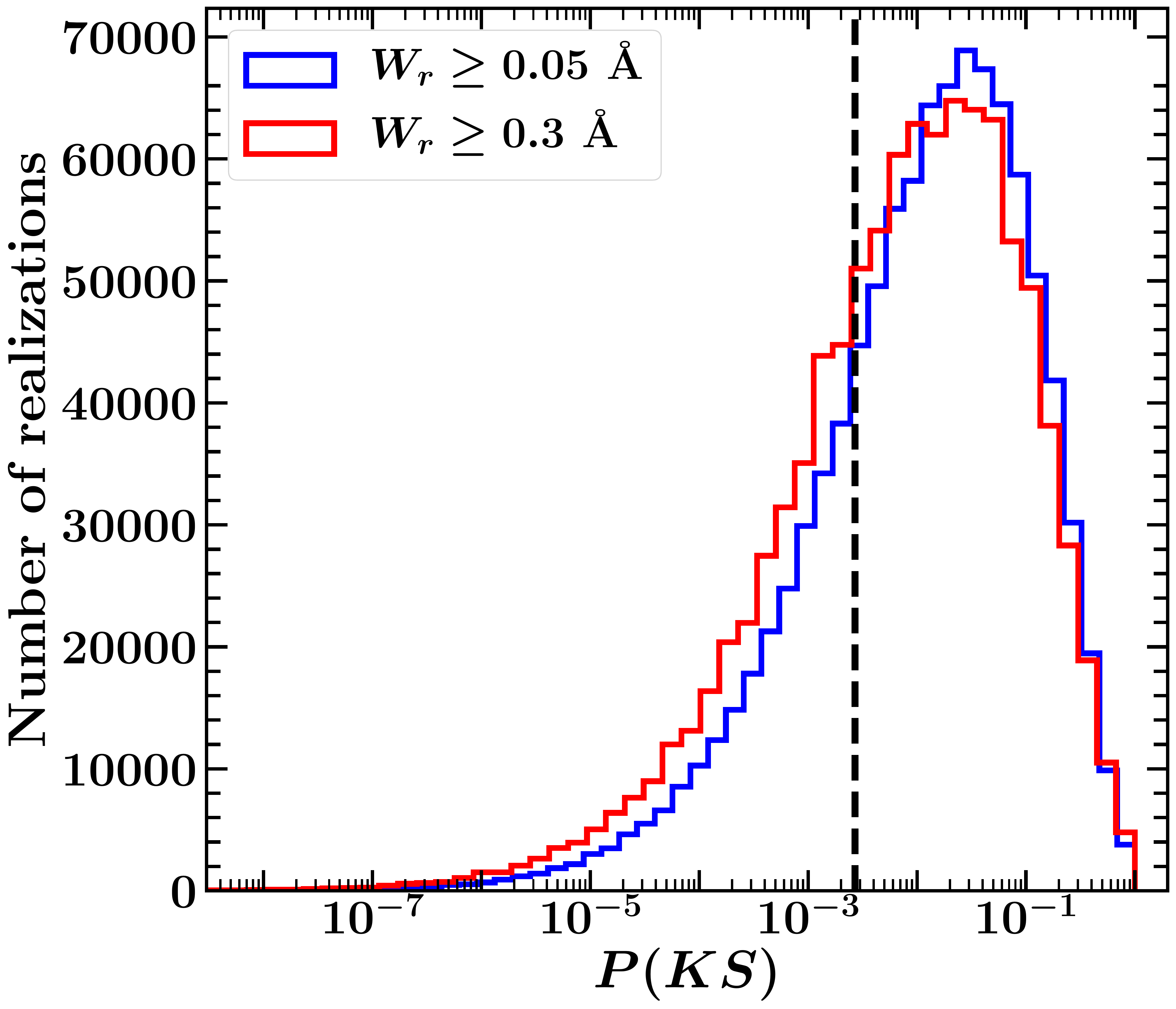}
\caption{Distribution of $P(KS)$ values from 1 million realizations of K-S tests comparing simulated {\EWr} values from TD to our observed Schechter function at $3 \leq z \leq 4$, for {\wvweak} absorbers (blue) and {\wweak} absorbers (red). The vertical black dashed line shows $P(KS)=0.0027$. See text for details.}
\label{fig:ks_td}
\vglue +5pt
\end{figure}

The distributions of $P(KS)$ values from these realizations are presented in Figure~\ref{fig:ks_td}. We adopt ${P(KS)\leq 0.0027}$, or a 99.73\% confidence level corresponding to a $3\sigma$ significance as the criterion for ruling out the null hypothesis.  For {\wvweak} absorbers, we obtained $P(KS)\leq 0.0027$ for $26\%$ of the realizations; for {\wweak} absorbers, we obtained $P(KS)\leq 0.0027$ for $32\%$ of the realizations. These results imply the null hypothesis that the simulated {\EWr} values are drawn from the Schechter function describing the observed data cannot be ruled out at the $\sim74\%$ confidence level for {\wvweak} and the $\sim 68\%$ confidence level for {\wweak}. Neither of these confidence levels are high enough to reject the null hypothesis.  

Our tests show that the majority of the realizations are statistically indistinguishable from the realization of the observed data. The degree of agreement between the TD predictions and the observed {\CIV} absorber statistics at $z\!=\!3$ to $z\!=\!4$ is encouraging with regards 
to the physical modelling.

TD is most successful at reproducing the frequency of {\CIV} absorbers in the equivalent width range ${0.4 \leq W_r \leq 0.8}$~{\AA}. At the strong end, simulations underproduce absorbers with $W_r \!\sim\! 1$ {\AA} by roughly a factor of four. Moreover, they entirely fail to produce stronger absorbers with $W_r \!>\! 1.1$ {\AA}. 
Our unprecedented dynamic range in {\EWr} now reveals that the problem is not a simple offset in normalization. TD \emph{over}produces weak absorbers with $W_r\leq 0.4$ {\AA} by a factor that increases from $\sim2$ for $W_r\leq 0.3$ {\AA} to $\sim3$ for $W_r\leq 0.15$ {\AA}. The higher frequency of ${W_r\leq 0.3}$ {\AA} absorbers in TD relative to observations suggests that the solution to the {\CIV} mismatch may not be as simple as uniformly boosting the overall carbon yield or hardening the UVB.

In Figure~\ref{fig:kfcomp}(b), we compare the $3 \leq z \leq 5$ observed co-moving path densities, {\dndx}, with the TD {\dndx} predictions. The data points show the observational measurements for {\wvweak}, {\wweak}, and {\wstrong}. The dashed and shaded regions show our best-fit linear evolution models and uncertainties (see Section~\ref{sec:quantevolve}). The opaque bands show the predicted {\dndx} from the TD {\CIV} absorber catalog, where the thickness reflects Poisson uncertainties. These values and their uncertainties were computed using identical methods to those we applied to the observational data (see Section~\ref{sec:dndx}). 
At $3 \leq z \leq 5$, the behavior of {\dndx} as predicted by TD is more or less consistent with the uncertainties in the data for {\wweak}  and {\wstrong} absorbers, though there is an over-abundance of {\wweak} absorbers at $z=3$. On the other hand, TD overpredicts {\dndx} of {\wvweak} absorbers across the full range of $z\!=\!3$--5 with the discrepancy increasing up to a factor of roughly three by $z=3$. 
Interestingly, the redshift evolution of the TD {\CIV} absorbers is such that {\dndx} for a given {\wrlim} declines toward higher redshifts in a qualitatively similar manner as the observations. To quantify this behavior, we examined whether a model of linear decrease in {\dndx} with increasing redshift describes the evolution in the simulated catalogs, as it did our observations in Section~\ref{sec:quantevolve}. 

In Figure~\ref{fig:kfcomp}(c), we show that the {\dndx} of different species of absorbers defined by progressively increasing {\wrlim} all decrease linearly with increasing redshift, with the rate of this evolution (represented by the slope of each best-fit line) increasing towards lower {\wrlim}. Though the linear fits to the simulated data were performed over a smaller redshift range than the observed data, they agree with the qualitative outcomes that (1) the co-moving path density of {\CIV} absorbers increases with cosmic time, and (2) the evolution is progressively more rapid as the optical depth and/or kinematic complexity of the {\CIV} gas clouds increase. 

Using the {\sc Illustris} simulations \citep{Genel14,Vogelsberger14,Bird14,Nelson15}, \citet{Bird16} conducted similar comparisons between simulated and observed {\CIV} statistics. They examined the observed EWD and {\dndx} of {\wweak} absorbers at $2 \leq z \leq 4$ from \citetalias{Cooksey13}. While their simulations reproduce the observed abundance of {\wweak} absorbers from \citetalias{Cooksey13} fairly well, they underproduce the abundance for {\wstrong}, with up to an order of magnitude mismatch for ${\EWr} \sim 1$~{\AA}. This result is commensurate with the under-prediction of strong {\CIV} absorbers at higher redshifts, in TD \citep[]{KF18,KF20} and in other simulations \citep[e.g.,][]{Rahmati16,Keating16}.

\subsubsection{Discussion of TD Results}

Given the results of the TD simulations, we are prompted to ask two generalized questions: (1) Even though the incidence of the strongest {\CIV} absorbers is underpredicted by TD at $z<5$, why do TD predictions better match the observed incidence of strong absorbers at $z<5$ than at $z>5$?  (2) Why does TD dramatically overproduce weak absorbers at $z<5$?

One reason TD may underproduce the strongest absorbers is that rare, massive galaxies and their associated absorbers are missing from the TD simulation volume ($3375 ~h^{-3} ~\mathrm{Mpc}^3$), which may contribute to very few strong {\CIV} absorbers with ${\EWr} >1$~{\AA} in the simulations.  Additionally, in the real universe there are hard UVB fluctuations owing to quasars, which have a space density of $\sim10^{-6} ~\mathrm{Mpc}^{-3}$. The simulations only subtend a fraction of that volume, so TD has neither the quasars nor the hard UVB fluctuations; it is difficult to speculate about the influence of those large-scale UVB fluctuations on the incidence rate of the strongest {\CIV} absorbers.  On the other hand, one possibility for why the simulated abundances of stronger systems better match the observations at $z<5$ is that a reduction in spatial fluctuations in the simulated UVB as the epoch of {\HI} reionization reaches completion and as {\HeII} reionization advances, may result in higher incidences of stronger {\CIV} absorbers in the simulations. Further work would be required to examine this supposition.

It is more difficult to understand the predicted over-abundance of the weaker {\CIV} absorbers. The simulations resolve dark matter halos down to $10^8 ~{\Msun}$ with at least 100 particles, which \citet{KF20} showed is enough to be resolution-convergent for column densities of $10^{12} ~\mathrm{cm}^{-2}$, corresponding to ${\EWr} = 0.004$~{\AA}. Thus, TD is resolution-convergent to an order of magnitude lower than the minimum {\EWr} studied in this paper.

With regards to the ionization conditions affecting the weakest absorbers, we note that TD yields a UVB mean intensity at $z<5$ that is weaker than what is observed \citep{Hassan20}. Though a higher mean intensity would improve the slight discrepancy of TD with the observed mean transmission in the {\Lya} forest \citep{KF18}, it would further increase the incidence of weak {\CIV}. 

Alternatively, the overprediction of weak {\CIV} by TD may indicate problems with the feedback model in that the simulated outflows could eject too many metals or expel them too far from galaxies. If outflows eject metals too far from galaxies, then we might expect too little power in the velocity clustering function of weak {\CIV} absorbers in TD \citep[see][]{KF18,KF20}. If outflows remove too much (or not enough) gas from low-mass galaxies, then we would expect the UV luminosity function and/or stellar mass function to depart from observations at the faint end.  Again, further work would be required to examine this supposition.

Finally, because the simulations are quite successful at reproducing the observed linear redshift-evolution of {\CIV} absorbers from $z=5$ to $z=3$, the TD predictions place further tension on the apparent observed ``bump'' in the {\dndx} at $z\sim5$ \citep[][see Figure~\ref{fig:dndx_comps}]{Simcoe+11,Codoreanu18}. A theoretical physical mechanism for such cosmically brief enhancement is difficult to understand.

\section{Conclusion} \label{conc}

We conducted a survey of archival high-resolution Keck/HIRES and VLT/UVES spectra ($\sim\!6.6$~{\kms}) of 369 quasars spanning the range of emission redshifts $1.1 \leq {\zem} \leq 5.3$. The high survey sensitivity allowed us to characterize, for the first time, the distribution and evolution of the weakest {\CIV} absorption systems with ${\EWr} < 0.3$~{\AA}, as well as stronger absorbers with ${\EWr} \geq 0.3$~{\AA}. Using automated {\CIV} doublet detection, combined with visual inspections to verify each candidate {\CIV} absorber, we measured the equivalent widths of the absorbers and quantified the detection thresholds, redshift path lengths, and completeness limits of our survey. We find that we are $\sim50\%$ complete at ${\EWr}=0.05$~{\AA}, and limit our scientific analysis to absorbers with equivalent widths above this threshold; our survey is $\sim100\%$ complete to absorbers with $W_r \geq 0.3$~{\AA}.

Over a total co-moving redshift path of $\Delta X \simeq 803$, we detected 1268 {\CIV} absorbers with {\wvweak}, within the redshift range {\zrangeall}. This corresponds to a cosmic age from $\sim\! 1.5$ Gyr ($z=4.75$) to  $\sim\! 6$ Gyr ($z=1.0$), a roughly $\sim\!4.5$ Gyr period from when the universe was $\sim\! 10$\% to $\sim\!45$\% of its present age. 
Using data from the literature, we extended our study to cover the redshift range $0 \leq z \leq 6.4$. We then compared our observational results to theoretical predictions from hydrodynamic cosmological simulations.

\begin{enumerate}
\itemsep0em
\item The EWD of {\CIV} is well-fit by a Schechter function (see Figure~\ref{fig:ewd}), with a power law slope of $\alpha \simeq -0.9$ and a characteristic equivalent width of $W_{\star} \simeq 0.5$~{\AA}. In the range {\zrangeall}, the EWD evolves from $z<2.5$ to $z\geq2.5$ such $\alpha$ increases by $\sim\! 7$\% and $W_{\star}$ decreases by $\sim\! 20\%$. This suggests that the cosmic incidence of weaker {\CIV} absorbers relative to stronger {\CIV} absorbers increases toward higher redshift.

\item The co-moving redshift path density, {\dndx}, increases with cosmic time (decreases with redshift), with the rate of the increase being larger as the population is limited to progressively higher {\EWr} absorbers (see Figure~\ref{fig:dndx}). From ${z=4.75}$ to ${z=1}$, {\dndx} rises by a factor of $\sim\!\!1.8$ for {\wvweak} absorbers, $\sim\!3.3$ for {\wweak} absorbers, and $\sim\! 8.5$ for {\wstrong} absorbers.

\item The evolution of {\dndx} is well-described by a linear model. We successfully parameterize the linear evolution as a function of the minimum {\EWr} of a population, {\wrlim}, using three parameters (see Section~\ref{sec:quantevolve}), {\meanNx}, the cosmic mean of {\dndx}, $A$, the evolution constant, and $f_0$, the extrapolated value of {\dndx} at $z=0$. An additional parameter is $z_0$, the onset redshift. The model fits yield a quantitative picture of linear {\CIV} absorption in which populations with progressively higher {\wrlim} evolve faster and have smaller cosmic mean incidence. Simple extrapolation of the model yields the expectation that populations with progressively higher {\wrlim} might make their first appearance in the universe at later times than populations defined by lower {\wrlim}.

\item For absorbers with {\wvweak} and absorbers with {\wstrong}, we extended our {\dndx} measurements for {\zrangeall} with those from the literature for $z<1$ and for $z>4.75$, thus examining {\CIV} evolution across $0 \leq z \leq 6.4$ (see Figure~\ref{fig:dndx_comps}). For $z\leq1$, the evolution remains consistent with our linear model for {\wstrong} absorbers.  However, for {\wvweak} absorbers, the evolution rapidly increases relative to the linear extrapolation. For $z>4.75$, both populations of absorbers have higher $dN/dX$ than what would be expected based on extrapolation of linear evolution. It is not clear if this is due to small number statistics in the high redshift measurements, or if there is an increase in {\CIV} absorbing structure over a brief $\simeq 0.5$~Gyr period around $z\simeq 5$. We argue that it is difficult to understand such rapid evolution at that epoch.

\item Assuming the gas structures hosting {\CIV} absorbers are the halos of galaxies, we estimate the {\CIV} absorbing halo radius, {\Rstar}, for an $L_{\star}$ galaxy. For {\wvweak} absorbers, {\Rstar} grows from ${\sim\!205}$~kpc at $z\simeq4$ to $\sim\!240$~kpc at $z\simeq2.5$, becoming relatively constant until $z\simeq1$. For {\wstrong} absorbers, {\Rstar} grows from ${\sim\! 45}$~kpc at $z\simeq4$ to ${\sim\!115}$~kpc at $z\simeq 1.3$. The large halo sizes for the weakest absorbers would suggest that they could be found in the outer extremes of the CGM and that some fraction may even reside in the IGM.

\item A $3 \!\leq\! z \!\leq\! 5$ mock survey of {\CIV} absorbers using the {\sc Technicolor Dawn} simulations of \citet{KF18} indicates that the simulations match the observed EWD and {\dndx} of {\CIV} absorbers more accurately in this redshift range than they do at $z>5$ \citep[see][]{KF20}. Furthermore, consistent with our linear evolution model, the simulations yield {\CIV} absorber populations for which {\dndx} decreases linearly with increasing redshift and for which the rate of evolution increases as {\wrlim} is increased. The major discrepancy between our observations and the simulations at $z<5$ is the over-abundance of weak absorbers with ${\EWr} \leq 0.3$~{\AA}, implying the need for modifications to the simulated UVB and/or the feedback model.

\end{enumerate}

We discussed the observed {\CIV} evolution in terms of a changing metal content and a changing metagalactic UV background of the universe. Whatever the details of the physical mechanisms manifesting the observed evolution may be, we can still infer that the rise in cosmic incidence of all populations of {\CIV} absorbers implies that the structures they trace become increasingly more common with cosmic time due to higher cosmic number density and/or larger physical size (physical cross section). 

Our estimates of the sizes of the gas structures that comprise {\CIV} absorbers would suggest that, though the very strongest {\CIV} absorbers reside well within the virial radii of galaxies, {\CIV} absorbing gas clouds with $W_r \leq 0.3$~{\AA} likely persist out to the extreme limits of the CGM and may bridge the interface with the IGM. The weakest absorbing clouds could possibly even reside in the IGM, perhaps residing in the gaseous filamentary structures that interconnect galaxies. We wish to explore this in future work, as well as explore the evolution in the kinematics of {\CIV}.


\acknowledgements
We thank Kathy Cooksey, Valentina D'Odorico, Jane Charlton, and Joe Burchett for helpful discussions regarding this work. We also thank the anonymous referee for several helpful comments to improve the manuscript. C.W.C. thanks the National Science Foundation for the grants AST-0708210 and AST-1517816, which partially supported this work. N.M.N., G.G.K., and M.T.M. acknowledge the support of the Australian Research Council through {\it Discovery Project} grant DP170103470. M.T.M. also thanks the Australian Research Council for {\it Discovery Project} grant DP130100568 for their partial support of this work. Parts of this research were supported by the Australian Research Council Centre of Excellence for All Sky Astrophysics in 3 Dimensions (ASTRO 3D), through project number CE170100012.

We dedicate this paper to the memory of Dr.\ Wallace Leslie William Sargent, who was a pioneer of the field of quasar absorption lines and so positively influenced the lives and careers of multiple generations of astronomers. This research has made use of the services of the ESO Science Archive Facility. Some of the data presented herein were obtained at the W. M. Keck Observatory, which is operated as a scientific partnership among the California Institute of Technology, the University of California, and the National Aeronautics and Space Administration and made possible by support of the W. M. Keck Foundation. The authors wish to recognize and acknowledge the very significant cultural role and reverence that the summit of Maunakea has always had within the indigenous Hawaiian community. We are most fortunate to have the opportunity to conduct observations from this mountain.


\facilities VLT: Kueyen (UVES), Keck: I (HIRES).

\software{ {\sc Astropy} \citep{astropy1}, {\sc Matplotlib} \citep{mpl}, {\sc Numpy} \citep{numpy}, {\sc Scipy} \citep{scipy}, {\sc Ipython} \citep{ipython}, {\sc Sysanal} \citep{cwcthesis}, {\sc Search} \citep{cwc99}, {\sc IRAF} \citep{iraf}, {\sc UVES\_POPLER} \citep{Murphy16}, {\sc MAKEE} \citep{barlow05}.  }



\bibliographystyle{aasjournal_nikki}
\bibliography{Refs}

\begin{thebibliography}{}
\expandafter\ifx\csname natexlab\endcsname\relax\def\natexlab#1{#1}\fi
\providecommand{\url}[1]{\href{#1}{#1}}
\providecommand{\dodoi}[1]{doi:~\href{http://doi.org/#1}{\nolinkurl{#1}}}
\providecommand{\doeprint}[1]{\href{http://ascl.net/#1}{\nolinkurl{http://ascl.net/#1}}}
\providecommand{\doarXiv}[1]{\href{https://arxiv.org/abs/#1}{\nolinkurl{https://arxiv.org/abs/#1}}}

\bibitem[{{Abramowitz} \& {Stegun}(1972)}]{Abramowitz72}
{Abramowitz}, M., \& {Stegun}, I.~A. 1972, {Handbook of Mathematical Functions}

\bibitem[{{Adelberger} {et~al.}(2005){Adelberger}, {Shapley}, {Steidel},
  {Pettini}, {Erb}, \& {Reddy}}]{Adelberger05}
{Adelberger}, K.~L., {Shapley}, A.~E., {Steidel}, C.~C., {et~al.} 2005,
  \href{http://dx.doi.org/10.1086/431753}{\color{magenta}\apj},
  \href{https://ui.adsabs.harvard.edu/abs/2005ApJ...629..636A}{629},
  \href{https://ui.adsabs.harvard.edu/abs/2005ApJ...629..636A}{636}

\bibitem[{{Angl{\'e}s-Alc{\'a}zar} {et~al.}(2017){Angl{\'e}s-Alc{\'a}zar},
  {Faucher-Gigu{\`e}re}, {Kere{\v{s}}}, {Hopkins}, {Quataert}, \&
  {Murray}}]{AA17}
{Angl{\'e}s-Alc{\'a}zar}, D., {Faucher-Gigu{\`e}re}, C.-A., {Kere{\v{s}}}, D.,
  {et~al.} 2017,
  \href{http://dx.doi.org/10.1093/mnras/stx1517}{\color{magenta}\mnras},
  \href{https://ui.adsabs.harvard.edu/abs/2017MNRAS.470.4698A}{470},
  \href{https://ui.adsabs.harvard.edu/abs/2017MNRAS.470.4698A}{4698}

\bibitem[{{Astropy Collaboration} {et~al.}(2013){Astropy Collaboration},
  {Robitaille}, {Tollerud}, {Greenfield}, {Droettboom}, {Bray}, {Aldcroft},
  {Davis}, {Ginsburg}, {Price-Whelan}, {Kerzendorf}, {Conley}, {Crighton},
  {Barbary}, {Muna}, {Ferguson}, {Grollier}, {Parikh}, {Nair}, {Unther},
  {Deil}, {Woillez}, {Conseil}, {Kramer}, {Turner}, {Singer}, {Fox}, {Weaver},
  {Zabalza}, {Edwards}, {Azalee Bostroem}, {Burke}, {Casey}, {Crawford},
  {Dencheva}, {Ely}, {Jenness}, {Labrie}, {Lim}, {Pierfederici}, {Pontzen},
  {Ptak}, {Refsdal}, {Servillat}, \& {Streicher}}]{astropy1}
{Astropy Collaboration}, {Robitaille}, T.~P., {Tollerud}, E.~J., {et~al.} 2013,
  \href{http://dx.doi.org/10.1051/0004-6361/201322068}{\color{magenta}\aap},
  \href{http://adsabs.harvard.edu/abs/2013A%26A...558A..33A}{558},
  \href{http://adsabs.harvard.edu/abs/2013A%26A...558A..33A}{A33}

\bibitem[{{Bahcall} \& {Peebles}(1969)}]{BP69}
{Bahcall}, J.~N., \& {Peebles}, P.~J.~E. 1969,
  \href{http://dx.doi.org/10.1086/180337}{\color{magenta}\apjl},
  \href{https://ui.adsabs.harvard.edu/abs/1969ApJ...156L...7B}{156},
  \href{https://ui.adsabs.harvard.edu/abs/1969ApJ...156L...7B}{L7}

\bibitem[{{Barkana} \& {Loeb}(2001)}]{BL01}
{Barkana}, R., \& {Loeb}, A. 2001,
  \href{http://dx.doi.org/10.1016/S0370-1573(01)00019-9}{\color{magenta}\physrep},
  \href{https://ui.adsabs.harvard.edu/abs/2001PhR...349..125B}{349},
  \href{https://ui.adsabs.harvard.edu/abs/2001PhR...349..125B}{125}

\bibitem[{{Barlow}(2005)}]{barlow05}
{Barlow}, T.~A. 2005, MAKEE Data Reduction Package.
\newblock
  \url{https://www.astro.caltech.edu/~tb/ipac_staff/tab/makee/index.html}

\bibitem[{{Becker} \& {Bolton}(2013)}]{Becker13}
{Becker}, G.~D., \& {Bolton}, J.~S. 2013,
  \href{http://dx.doi.org/10.1093/mnras/stt1610}{\color{magenta}\mnras},
  \href{https://ui.adsabs.harvard.edu/abs/2013MNRAS.436.1023B}{436},
  \href{https://ui.adsabs.harvard.edu/abs/2013MNRAS.436.1023B}{1023}

\bibitem[{{Becker} {et~al.}(2011){Becker}, {Bolton}, {Haehnelt}, \&
  {Sargent}}]{Becker11}
{Becker}, G.~D., {Bolton}, J.~S., {Haehnelt}, M.~G., \& {Sargent}, W. L.~W.
  2011,
  \href{http://dx.doi.org/10.1111/j.1365-2966.2010.17507.x}{\color{magenta}\mnras},
  \href{https://ui.adsabs.harvard.edu/abs/2011MNRAS.410.1096B}{410},
  \href{https://ui.adsabs.harvard.edu/abs/2011MNRAS.410.1096B}{1096}

\bibitem[{{Becker} {et~al.}(2015){Becker}, {Bolton}, \& {Lidz}}]{Becker15}
{Becker}, G.~D., {Bolton}, J.~S., \& {Lidz}, A. 2015,
  \href{http://dx.doi.org/10.1017/pasa.2015.45}{\color{magenta}\pasa},
  \href{https://ui.adsabs.harvard.edu/abs/2015PASA...32...45B}{32},
  \href{https://ui.adsabs.harvard.edu/abs/2015PASA...32...45B}{e045}

\bibitem[{{Becker} {et~al.}(2009){Becker}, {Rauch}, \& {Sargent}}]{Becker09}
{Becker}, G.~D., {Rauch}, M., \& {Sargent}, W. L.~W. 2009,
  \href{http://dx.doi.org/10.1088/0004-637X/698/2/1010}{\color{magenta}\apj},
  \href{https://ui.adsabs.harvard.edu/abs/2009ApJ...698.1010B}{698},
  \href{https://ui.adsabs.harvard.edu/abs/2009ApJ...698.1010B}{1010}

\bibitem[{{Becker} {et~al.}(2019){Becker}, {Pettini}, {Rafelski}, {D'Odorico},
  {Boera}, {Christensen}, {Cupani}, {Ellison}, {Farina}, {Fumagalli},
  {L{\'o}pez}, {Neeleman}, {Ryan-Weber}, \& {Worseck}}]{Becker19}
{Becker}, G.~D., {Pettini}, M., {Rafelski}, M., {et~al.} 2019,
  \href{http://dx.doi.org/10.3847/1538-4357/ab3eb5}{\color{magenta}\apj},
  \href{https://ui.adsabs.harvard.edu/abs/2019ApJ...883..163B}{883},
  \href{https://ui.adsabs.harvard.edu/abs/2019ApJ...883..163B}{163}

\bibitem[{{Behroozi} {et~al.}(2019){Behroozi}, {Wechsler}, {Hearin}, \&
  {Conroy}}]{Behroozi19}
{Behroozi}, P., {Wechsler}, R.~H., {Hearin}, A.~P., \& {Conroy}, C. 2019,
  \href{http://dx.doi.org/10.1093/mnras/stz1182}{\color{magenta}\mnras},
  \href{https://ui.adsabs.harvard.edu/abs/2019MNRAS.488.3143B}{488},
  \href{https://ui.adsabs.harvard.edu/abs/2019MNRAS.488.3143B}{3143}

\bibitem[{{Behroozi} {et~al.}(2013){Behroozi}, {Wechsler}, \&
  {Conroy}}]{Behroozi13}
{Behroozi}, P.~S., {Wechsler}, R.~H., \& {Conroy}, C. 2013,
  \href{http://dx.doi.org/10.1088/0004-637X/770/1/57}{\color{magenta}\apj},
  \href{https://ui.adsabs.harvard.edu/abs/2013ApJ...770...57B}{770},
  \href{https://ui.adsabs.harvard.edu/abs/2013ApJ...770...57B}{57}

\bibitem[{{Bergeron} \& {Herbert-Fort}(2005)}]{Bergeron05}
{Bergeron}, J., \& {Herbert-Fort}, S. 2005, in IAU Colloq. 199: Probing
  Galaxies through Quasar Absorption Lines, ed. P.~{Williams}, C.-G. {Shu}, \&
  B.~{Menard}, 265--280

\bibitem[{{Bird} {et~al.}(2016){Bird}, {Rubin}, {Suresh}, \&
  {Hernquist}}]{Bird16}
{Bird}, S., {Rubin}, K. H.~R., {Suresh}, J., \& {Hernquist}, L. 2016,
  \href{http://dx.doi.org/10.1093/mnras/stw1582}{\color{magenta}\mnras},
  \href{https://ui.adsabs.harvard.edu/abs/2016MNRAS.462..307B}{462},
  \href{https://ui.adsabs.harvard.edu/abs/2016MNRAS.462..307B}{307}

\bibitem[{{Bird} {et~al.}(2014){Bird}, {Vogelsberger}, {Haehnelt}, {Sijacki},
  {Genel}, {Torrey}, {Springel}, \& {Hernquist}}]{Bird14}
{Bird}, S., {Vogelsberger}, M., {Haehnelt}, M., {et~al.} 2014,
  \href{http://dx.doi.org/10.1093/mnras/stu1923}{\color{magenta}\mnras},
  \href{https://ui.adsabs.harvard.edu/abs/2014MNRAS.445.2313B}{445},
  \href{https://ui.adsabs.harvard.edu/abs/2014MNRAS.445.2313B}{2313}

\bibitem[{{Boksenberg} \& {Sargent}(2015)}]{BS15}
{Boksenberg}, A., \& {Sargent}, W. L.~W. 2015,
  \href{http://dx.doi.org/10.1088/0067-0049/218/1/7}{\color{magenta}\apjs},
  218, 7

\bibitem[{{Bolton} \& {Haehnelt}(2007)}]{Bolton07}
{Bolton}, J.~S., \& {Haehnelt}, M.~G. 2007,
  \href{http://dx.doi.org/10.1111/j.1365-2966.2007.12372.x}{\color{magenta}\mnras},
  \href{https://ui.adsabs.harvard.edu/abs/2007MNRAS.382..325B}{382},
  \href{https://ui.adsabs.harvard.edu/abs/2007MNRAS.382..325B}{325}

\bibitem[{{Bordoloi} {et~al.}(2014){Bordoloi}, {Tumlinson}, {Werk},
  {Oppenheimer}, {Peeples}, {Prochaska}, {Tripp}, {Katz}, {Dav{\'e}}, {Fox},
  {Thom}, {Ford}, {Weinberg}, {Burchett}, \& {Kollmeier}}]{Bordoloi14}
{Bordoloi}, R., {Tumlinson}, J., {Werk}, J.~K., {et~al.} 2014,
  \href{http://dx.doi.org/10.1088/0004-637X/796/2/136}{\color{magenta}\apj},
  796, 136

\bibitem[{{Brook} {et~al.}(2014){Brook}, {Stinson}, {Gibson}, {Shen},
  {Macci{\`o}}, {Obreja}, {Wadsley}, \& {Quinn}}]{Brook14}
{Brook}, C.~B., {Stinson}, G., {Gibson}, B.~K., {et~al.} 2014,
  \href{http://dx.doi.org/10.1093/mnras/stu1406}{\color{magenta}\mnras},
  \href{https://ui.adsabs.harvard.edu/abs/2014MNRAS.443.3809B}{443},
  \href{https://ui.adsabs.harvard.edu/abs/2014MNRAS.443.3809B}{3809}

\bibitem[{{Burchett} {et~al.}(2015){Burchett}, {Tripp}, {Prochaska}, {Werk},
  {Tumlinson}, {O'Meara}, {Bordoloi}, {Katz}, \& {Willmer}}]{Burchett15}
{Burchett}, J.~N., {Tripp}, T.~M., {Prochaska}, J.~X., {et~al.} 2015,
  \href{http://dx.doi.org/10.1088/0004-637X/815/2/91}{\color{magenta}\apj},
  \href{https://ui.adsabs.harvard.edu/abs/2015ApJ...815...91B}{815},
  \href{https://ui.adsabs.harvard.edu/abs/2015ApJ...815...91B}{91}

\bibitem[{{Burchett} {et~al.}(2016){Burchett}, {Tripp}, {Bordoloi}, {Werk},
  {Prochaska}, {Tumlinson}, {Willmer}, {O'Meara}, \& {Katz}}]{Burchett16}
{Burchett}, J.~N., {Tripp}, T.~M., {Bordoloi}, R., {et~al.} 2016,
  \href{http://dx.doi.org/10.3847/0004-637X/832/2/124}{\color{magenta}\apj},
  \href{https://ui.adsabs.harvard.edu/abs/2016ApJ...832..124B}{832},
  \href{https://ui.adsabs.harvard.edu/abs/2016ApJ...832..124B}{124}

\bibitem[{{Cen} \& {Chisari}(2011)}]{CC11}
{Cen}, R., \& {Chisari}, N.~E. 2011,
  \href{http://dx.doi.org/10.1088/0004-637X/731/1/11}{\color{magenta}\apj},
  \href{https://ui.adsabs.harvard.edu/abs/2011ApJ...731...11C}{731},
  \href{https://ui.adsabs.harvard.edu/abs/2011ApJ...731...11C}{11}

\bibitem[{{Chen}(2012)}]{Chen12}
{Chen}, H.-W. 2012,
  \href{http://dx.doi.org/10.1111/j.1365-2966.2012.22053.x}{\color{magenta}\mnras},
  \href{https://ui.adsabs.harvard.edu/abs/2012MNRAS.427.1238C}{427},
  \href{https://ui.adsabs.harvard.edu/abs/2012MNRAS.427.1238C}{1238}

\bibitem[{{Chen} {et~al.}(2001){Chen}, {Lanzetta}, \& {Webb}}]{Chen01}
{Chen}, H.-W., {Lanzetta}, K.~M., \& {Webb}, J.~K. 2001,
  \href{http://dx.doi.org/10.1086/321537}{\color{magenta}\apj}, 556, 158-163

\bibitem[{{Chen} {et~al.}(2015){Chen}, {Gu}, \& {Chen}}]{Chen15}
{Chen}, Z.-F., {Gu}, Q.-S., \& {Chen}, Y.-M. 2015,
  \href{http://dx.doi.org/10.1088/0067-0049/221/2/32}{\color{magenta}\apjs},
  \href{https://ui.adsabs.harvard.edu/abs/2015ApJS..221...32C}{221},
  \href{https://ui.adsabs.harvard.edu/abs/2015ApJS..221...32C}{32}

\bibitem[{{Christensen} {et~al.}(2016){Christensen}, {Dav{\'e}}, {Governato},
  {Pontzen}, {Brooks}, {Munshi}, {Quinn}, \& {Wadsley}}]{Christensen16}
{Christensen}, C.~R., {Dav{\'e}}, R., {Governato}, F., {et~al.} 2016,
  \href{http://dx.doi.org/10.3847/0004-637X/824/1/57}{\color{magenta}\apj},
  \href{https://ui.adsabs.harvard.edu/abs/2016ApJ...824...57C}{824},
  \href{https://ui.adsabs.harvard.edu/abs/2016ApJ...824...57C}{57}

\bibitem[{{Churchill}(1997)}]{cwcthesis}
{Churchill}, C.~W. 1997, PhD thesis, University of California, Santa Cruz

\bibitem[{{Churchill} {et~al.}(2020){Churchill}, {Evans}, {Stemock}, {Nielsen},
  {Kacprzak}, \& {Murphy}}]{cwc20}
{Churchill}, C.~W., {Evans}, J.~L., {Stemock}, B., {et~al.} 2020, arXiv
  e-prints,
  \href{https://ui.adsabs.harvard.edu/abs/2020arXiv200808487C}{arXiv:2008.08487}

\bibitem[{{Churchill} {et~al.}(1999){Churchill}, {Rigby}, {Charlton}, \&
  {Vogt}}]{cwc99}
{Churchill}, C.~W., {Rigby}, J.~R., {Charlton}, J.~C., \& {Vogt}, S.~S. 1999,
  \href{http://dx.doi.org/10.1086/313168}{\color{magenta}\apjs},
  \href{https://ui.adsabs.harvard.edu/abs/1999ApJS..120...51C}{120},
  \href{https://ui.adsabs.harvard.edu/abs/1999ApJS..120...51C}{51}

\bibitem[{{Churchill} {et~al.}(2015){Churchill}, {Vander Vliet},
  {Trujillo-Gomez}, {Kacprzak}, \& {Klypin}}]{cwc15}
{Churchill}, C.~W., {Vander Vliet}, J.~R., {Trujillo-Gomez}, S., {Kacprzak},
  G.~G., \& {Klypin}, A. 2015,
  \href{http://dx.doi.org/10.1088/0004-637X/802/1/10}{\color{magenta}\apj},
  \href{https://ui.adsabs.harvard.edu/abs/2015ApJ...802...10C}{802},
  \href{https://ui.adsabs.harvard.edu/abs/2015ApJ...802...10C}{10}

\bibitem[{{Churchill} \& {Vogt}(2001)}]{cwc01}
{Churchill}, C.~W., \& {Vogt}, S.~S. 2001,
  \href{http://dx.doi.org/10.1086/321174}{\color{magenta}\aj},
  \href{http://adsabs.harvard.edu/abs/2001AJ....122..679C}{122},
  \href{http://adsabs.harvard.edu/abs/2001AJ....122..679C}{679}

\bibitem[{{Codoreanu} {et~al.}(2018){Codoreanu}, {Ryan-Weber}, {Garc{\'\i}a},
  {Crighton}, {Becker}, {Pettini}, {Madau}, \& {Venemans}}]{Codoreanu18}
{Codoreanu}, A., {Ryan-Weber}, E.~V., {Garc{\'\i}a}, L.~{\'A}., {et~al.} 2018,
  \href{http://dx.doi.org/10.1093/mnras/sty2576}{\color{magenta}\mnras},
  \href{https://ui.adsabs.harvard.edu/abs/2018MNRAS.481.4940C}{481},
  \href{https://ui.adsabs.harvard.edu/abs/2018MNRAS.481.4940C}{4940}

\bibitem[{{Cooksey} {et~al.}(2013){Cooksey}, {Kao}, {Simcoe}, {O'Meara}, \&
  {Prochaska}}]{Cooksey13}
{Cooksey}, K.~L., {Kao}, M.~M., {Simcoe}, R.~A., {O'Meara}, J.~M., \&
  {Prochaska}, J.~X. 2013,
  \href{http://dx.doi.org/10.1088/0004-637X/763/1/37}{\color{magenta}\apj},
  \href{http://adsabs.harvard.edu/abs/2013ApJ...763...37C}{763},
  \href{http://adsabs.harvard.edu/abs/2013ApJ...763...37C}{37}

\bibitem[{{Cooksey} {et~al.}(2010){Cooksey}, {Thom}, {Prochaska}, \&
  {Chen}}]{Cooksey10}
{Cooksey}, K.~L., {Thom}, C., {Prochaska}, J.~X., \& {Chen}, H.-W. 2010,
  \href{http://dx.doi.org/10.1088/0004-637X/708/1/868}{\color{magenta}\apj},
  708, 868-908

\bibitem[{{Cooper} {et~al.}(2019){Cooper}, {Simcoe}, {Cooksey}, {Bordoloi},
  {Miller}, {Furesz}, {Turner}, \& {Ba{\~n}ados}}]{Cooper19}
{Cooper}, T.~J., {Simcoe}, R.~A., {Cooksey}, K.~L., {et~al.} 2019,
  \href{http://dx.doi.org/10.3847/1538-4357/ab3402}{\color{magenta}\apj},
  \href{https://ui.adsabs.harvard.edu/abs/2019ApJ...882...77C}{882},
  \href{https://ui.adsabs.harvard.edu/abs/2019ApJ...882...77C}{77}

\bibitem[{{Danforth} \& {Shull}(2008)}]{DS08}
{Danforth}, C.~W., \& {Shull}, J.~M. 2008,
  \href{http://dx.doi.org/10.1086/587127}{\color{magenta}\apj},
  \href{https://ui.adsabs.harvard.edu/abs/2008ApJ...679..194D}{679},
  \href{https://ui.adsabs.harvard.edu/abs/2008ApJ...679..194D}{194}

\bibitem[{{Dav{\'e}} {et~al.}(1999){Dav{\'e}}, {Hernquist}, {Katz}, \&
  {Weinberg}}]{Dave99}
{Dav{\'e}}, R., {Hernquist}, L., {Katz}, N., \& {Weinberg}, D.~H. 1999,
  \href{http://dx.doi.org/10.1086/306722}{\color{magenta}\apj},
  \href{https://ui.adsabs.harvard.edu/abs/1999ApJ...511..521D}{511},
  \href{https://ui.adsabs.harvard.edu/abs/1999ApJ...511..521D}{521}

\bibitem[{{De Cia} {et~al.}(2018){De Cia}, {Ledoux}, {Petitjean}, \&
  {Savaglio}}]{DeCia18}
{De Cia}, A., {Ledoux}, C., {Petitjean}, P., \& {Savaglio}, S. 2018,
  \href{http://dx.doi.org/10.1051/0004-6361/201731970}{\color{magenta}\aap},
  \href{https://ui.adsabs.harvard.edu/abs/2018A&A...611A..76D}{611},
  \href{https://ui.adsabs.harvard.edu/abs/2018A&A...611A..76D}{A76}

\bibitem[{{Dekel} \& {Mandelker}(2014)}]{DM14}
{Dekel}, A., \& {Mandelker}, N. 2014,
  \href{http://dx.doi.org/10.1093/mnras/stu1427}{\color{magenta}\mnras},
  \href{https://ui.adsabs.harvard.edu/abs/2014MNRAS.444.2071D}{444},
  \href{https://ui.adsabs.harvard.edu/abs/2014MNRAS.444.2071D}{2071}

\bibitem[{{Dekker} {et~al.}(2000)}]{uves00}
{Dekker}, H., {et~al.} 2000, in \procspie, ed. M.~{Iye} \& A.~F. {Moorwood},
  Vol. 4008, 534--545

\bibitem[{{D'Odorico} {et~al.}(2010){D'Odorico}, {Calura}, {Cristiani}, \&
  {Viel}}]{Dodo10}
{D'Odorico}, V., {Calura}, F., {Cristiani}, S., \& {Viel}, M. 2010,
  \href{http://dx.doi.org/10.1111/j.1365-2966.2009.15856.x}{\color{magenta}\mnras},
  \href{https://ui.adsabs.harvard.edu/abs/2010MNRAS.401.2715D}{401},
  \href{https://ui.adsabs.harvard.edu/abs/2010MNRAS.401.2715D}{2715}

\bibitem[{{D'Odorico} {et~al.}(2016){D'Odorico}, {Calura}, {Cristiani}, \&
  {Viel}}]{Dodo16}
---. 2016,
  \href{http://dx.doi.org/10.1093/mnras/stw633}{\color{magenta}\mnras},
  \href{https://ui.adsabs.harvard.edu/abs/2016MNRAS.459..232D}{459},
  \href{https://ui.adsabs.harvard.edu/abs/2016MNRAS.459..232D}{232}

\bibitem[{{D'Odorico} {et~al.}(2013){D'Odorico}, {Cupani}, {Cristiani},
  {Maiolino}, {Molaro}, {Nonino}, {Centuri{\'o}n}, {Cimatti}, {di Serego
  Alighieri}, {Fiore}, {Fontana}, {Gallerani}, {Giallongo}, {Mannucci},
  {Marconi}, {Pentericci}, {Viel}, \& {Vladilo}}]{Dodo13}
{D'Odorico}, V., {Cupani}, G., {Cristiani}, S., {et~al.} 2013,
  \href{http://dx.doi.org/10.1093/mnras/stt1365}{\color{magenta}\mnras},
  \href{https://ui.adsabs.harvard.edu/abs/2013MNRAS.435.1198D}{435},
  \href{https://ui.adsabs.harvard.edu/abs/2013MNRAS.435.1198D}{1198}

\bibitem[{{Doughty} \& {Finlator}(2019)}]{Doughty19}
{Doughty}, C., \& {Finlator}, K. 2019,
  \href{http://dx.doi.org/10.1093/mnras/stz2331}{\color{magenta}\mnras},
  \href{https://ui.adsabs.harvard.edu/abs/2019MNRAS.489.2755D}{489},
  \href{https://ui.adsabs.harvard.edu/abs/2019MNRAS.489.2755D}{2755}

\bibitem[{{Doughty} {et~al.}(2018){Doughty}, {Finlator}, {Oppenheimer},
  {Dav{\'e}}, \& {Zackrisson}}]{Doughty18}
{Doughty}, C., {Finlator}, K., {Oppenheimer}, B.~D., {Dav{\'e}}, R., \&
  {Zackrisson}, E. 2018,
  \href{http://dx.doi.org/10.1093/mnras/sty156}{\color{magenta}\mnras},
  \href{https://ui.adsabs.harvard.edu/abs/2018MNRAS.475.4717D}{475},
  \href{https://ui.adsabs.harvard.edu/abs/2018MNRAS.475.4717D}{4717}

\bibitem[{{Faucher-Gigu{\`e}re} {et~al.}(2008){Faucher-Gigu{\`e}re}, {Lidz},
  {Hernquist}, \& {Zaldarriaga}}]{Faucher08}
{Faucher-Gigu{\`e}re}, C.-A., {Lidz}, A., {Hernquist}, L., \& {Zaldarriaga}, M.
  2008, \href{http://dx.doi.org/10.1086/592289}{\color{magenta}\apj},
  \href{https://ui.adsabs.harvard.edu/abs/2008ApJ...688...85F}{688},
  \href{https://ui.adsabs.harvard.edu/abs/2008ApJ...688...85F}{85}

\bibitem[{{Faucher-Gigu{\`e}re} {et~al.}(2009){Faucher-Gigu{\`e}re}, {Lidz},
  {Zaldarriaga}, \& {Hernquist}}]{FG09}
{Faucher-Gigu{\`e}re}, C.-A., {Lidz}, A., {Zaldarriaga}, M., \& {Hernquist}, L.
  2009,
  \href{http://dx.doi.org/10.1088/0004-637X/703/2/1416}{\color{magenta}\apj},
  \href{https://ui.adsabs.harvard.edu/abs/2009ApJ...703.1416F}{703},
  \href{https://ui.adsabs.harvard.edu/abs/2009ApJ...703.1416F}{1416}

\bibitem[{{Feldmann} {et~al.}(2016){Feldmann}, {Hopkins}, {Quataert},
  {Faucher-Gigu{\`e}re}, \& {Kere{\v{s}}}}]{Feldman16}
{Feldmann}, R., {Hopkins}, P.~F., {Quataert}, E., {Faucher-Gigu{\`e}re}, C.-A.,
  \& {Kere{\v{s}}}, D. 2016,
  \href{http://dx.doi.org/10.1093/mnrasl/slw014}{\color{magenta}\mnras},
  \href{https://ui.adsabs.harvard.edu/abs/2016MNRAS.458L..14F}{458},
  \href{https://ui.adsabs.harvard.edu/abs/2016MNRAS.458L..14F}{L14}

\bibitem[{{Finlator}(2017)}]{KF17}
{Finlator}, K. 2017, in Gas Accretion onto Galaxies, ed. A.~{Fox} \&
  R.~{Dav{\'e}}, Vol. 430

\bibitem[{{Finlator} {et~al.}(2020){Finlator}, {Doughty}, {Cai}, \&
  {D{\'\i}az}}]{KF20}
{Finlator}, K., {Doughty}, C., {Cai}, Z., \& {D{\'\i}az}, G. 2020,
  \href{http://dx.doi.org/10.1093/mnras/staa377}{\color{magenta}\mnras},
  \href{https://ui.adsabs.harvard.edu/abs/2020MNRAS.493.3223F}{493},
  \href{https://ui.adsabs.harvard.edu/abs/2020MNRAS.493.3223F}{3223}

\bibitem[{{Finlator} {et~al.}(2018){Finlator}, {Keating}, {Oppenheimer},
  {Dav{\'e}}, \& {Zackrisson}}]{KF18}
{Finlator}, K., {Keating}, L., {Oppenheimer}, B.~D., {Dav{\'e}}, R., \&
  {Zackrisson}, E. 2018,
  \href{http://dx.doi.org/10.1093/mnras/sty1949}{\color{magenta}\mnras},
  \href{https://ui.adsabs.harvard.edu/abs/2018MNRAS.480.2628F}{480},
  \href{https://ui.adsabs.harvard.edu/abs/2018MNRAS.480.2628F}{2628}

\bibitem[{Finlator {et~al.}(2016)Finlator, Oppenheimer, Davé, Zackrisson,
  Thompson, \& Huang}]{KF16}
Finlator, K., Oppenheimer, B.~D., Davé, R., {et~al.} 2016,
  \href{http://dx.doi.org/10.1093/mnras/stw805}{\color{magenta}\mnras}, 459,
  2299-2310

\bibitem[{{Finlator} {et~al.}(2015){Finlator}, {Thompson}, {Huang}, {Dav{\'e}},
  {Zackrisson}, \& {Oppenheimer}}]{KF15}
{Finlator}, K., {Thompson}, R., {Huang}, S., {et~al.} 2015,
  \href{http://dx.doi.org/10.1093/mnras/stu2668}{\color{magenta}\mnras},
  \href{https://ui.adsabs.harvard.edu/abs/2015MNRAS.447.2526F}{447},
  \href{https://ui.adsabs.harvard.edu/abs/2015MNRAS.447.2526F}{2526}

\bibitem[{{Ford} {et~al.}(2014){Ford}, {Dav{\'e}}, {Oppenheimer}, {Katz},
  {Kollmeier}, {Thompson}, \& {Weinberg}}]{Ford14}
{Ford}, A.~B., {Dav{\'e}}, R., {Oppenheimer}, B.~D., {et~al.} 2014,
  \href{http://dx.doi.org/10.1093/mnras/stu1418}{\color{magenta}\mnras},
  \href{https://ui.adsabs.harvard.edu/abs/2014MNRAS.444.1260F}{444},
  \href{https://ui.adsabs.harvard.edu/abs/2014MNRAS.444.1260F}{1260}

\bibitem[{{Fox} {et~al.}(2007){Fox}, {Ledoux}, {Petitjean}, \&
  {Srianand}}]{Fox07}
{Fox}, A.~J., {Ledoux}, C., {Petitjean}, P., \& {Srianand}, R. 2007,
  \href{http://dx.doi.org/10.1051/0004-6361:20077640}{\color{magenta}\aap},
  \href{https://ui.adsabs.harvard.edu/abs/2007A&A...473..791F}{473},
  \href{https://ui.adsabs.harvard.edu/abs/2007A&A...473..791F}{791}

\bibitem[{{Furlanetto}(2009)}]{Fur09}
{Furlanetto}, S.~R. 2009,
  \href{http://dx.doi.org/10.1088/0004-637X/703/1/702}{\color{magenta}\apj},
  \href{https://ui.adsabs.harvard.edu/abs/2009ApJ...703..702F}{703},
  \href{https://ui.adsabs.harvard.edu/abs/2009ApJ...703..702F}{702}

\bibitem[{{Genel} {et~al.}(2014){Genel}, {Vogelsberger}, {Springel}, {Sijacki},
  {Nelson}, {Snyder}, {Rodriguez-Gomez}, {Torrey}, \& {Hernquist}}]{Genel14}
{Genel}, S., {Vogelsberger}, M., {Springel}, V., {et~al.} 2014,
  \href{http://dx.doi.org/10.1093/mnras/stu1654}{\color{magenta}\mnras},
  \href{https://ui.adsabs.harvard.edu/abs/2014MNRAS.445..175G}{445},
  \href{https://ui.adsabs.harvard.edu/abs/2014MNRAS.445..175G}{175}

\bibitem[{{Gnedin}(2000)}]{Gnedin00}
{Gnedin}, N.~Y. 2000,
  \href{http://dx.doi.org/10.1086/308876}{\color{magenta}\apj},
  \href{https://ui.adsabs.harvard.edu/abs/2000ApJ...535..530G}{535},
  \href{https://ui.adsabs.harvard.edu/abs/2000ApJ...535..530G}{530}

\bibitem[{{Gnedin} \& {Ostriker}(1997)}]{Gnedin97}
{Gnedin}, N.~Y., \& {Ostriker}, J.~P. 1997,
  \href{http://dx.doi.org/10.1086/304548}{\color{magenta}\apj},
  \href{https://ui.adsabs.harvard.edu/abs/1997ApJ...486..581G}{486},
  \href{https://ui.adsabs.harvard.edu/abs/1997ApJ...486..581G}{581}

\bibitem[{{Haardt} \& {Madau}(2012)}]{HM12}
{Haardt}, F., \& {Madau}, P. 2012,
  \href{http://dx.doi.org/10.1088/0004-637X/746/2/125}{\color{magenta}\apj},
  \href{https://ui.adsabs.harvard.edu/abs/2012ApJ...746..125H}{746},
  \href{https://ui.adsabs.harvard.edu/abs/2012ApJ...746..125H}{125}

\bibitem[{{Hassan} {et~al.}(2020){Hassan}, {Finlator}, {Dav{\'e}}, {Churchill},
  \& {Prochaska}}]{Hassan20}
{Hassan}, S., {Finlator}, K., {Dav{\'e}}, R., {Churchill}, C.~W., \&
  {Prochaska}, J.~X. 2020,
  \href{http://dx.doi.org/10.1093/mnras/staa056}{\color{magenta}\mnras},
  \href{https://ui.adsabs.harvard.edu/abs/2020MNRAS.492.2835H}{492},
  \href{https://ui.adsabs.harvard.edu/abs/2020MNRAS.492.2835H}{2835}

\bibitem[{{Holmberg}(1975)}]{Holmberg75}
{Holmberg}, E. 1975, in Galaxies and the Universe, ed. A.~{Sandage},
  M.~{Sandage}, \& J.~{Kristian}, 123

\bibitem[{Hunter(2007)}]{mpl}
Hunter, J.~D. 2007,
  \href{http://dx.doi.org/10.1109/MCSE.2007.55}{\color{magenta}Computing in
  Science \& Engineering}, 9, 90--95

\bibitem[{{Kacprzak} \& {Churchill}(2011)}]{glenncwc11}
{Kacprzak}, G.~G., \& {Churchill}, C.~W. 2011,
  \href{http://dx.doi.org/10.1088/2041-8205/743/2/L34}{\color{magenta}\apjl},
  \href{https://ui.adsabs.harvard.edu/abs/2011ApJ...743L..34K}{743},
  \href{https://ui.adsabs.harvard.edu/abs/2011ApJ...743L..34K}{L34}

\bibitem[{{Kacprzak} {et~al.}(2012){Kacprzak}, {Churchill}, \&
  {Nielsen}}]{glenn12}
{Kacprzak}, G.~G., {Churchill}, C.~W., \& {Nielsen}, N.~M. 2012,
  \href{http://dx.doi.org/10.1088/2041-8205/760/1/L7}{\color{magenta}\apjl},
  \href{https://ui.adsabs.harvard.edu/abs/2012ApJ...760L...7K}{760},
  \href{https://ui.adsabs.harvard.edu/abs/2012ApJ...760L...7K}{L7}

\bibitem[{{Kacprzak} {et~al.}(2008){Kacprzak}, {Churchill}, {Steidel}, \&
  {Murphy}}]{glenn08}
{Kacprzak}, G.~G., {Churchill}, C.~W., {Steidel}, C.~C., \& {Murphy}, M.~T.
  2008,
  \href{http://dx.doi.org/10.1088/0004-6256/135/3/922}{\color{magenta}\aj},
  \href{https://ui.adsabs.harvard.edu/abs/2008AJ....135..922K}{135},
  \href{https://ui.adsabs.harvard.edu/abs/2008AJ....135..922K}{922}

\bibitem[{{Kacprzak} {et~al.}(2019){Kacprzak}, {Vander Vliet}, {Nielsen},
  {Muzahid}, {Pointon}, {Churchill}, {Ceverino}, {Arraki}, {Klypin},
  {Charlton}, \& {Lewis}}]{glenn19}
{Kacprzak}, G.~G., {Vander Vliet}, J.~R., {Nielsen}, N.~M., {et~al.} 2019,
  \href{http://dx.doi.org/10.3847/1538-4357/aaf1a6}{\color{magenta}\apj},
  \href{https://ui.adsabs.harvard.edu/abs/2019ApJ...870..137K}{870},
  \href{https://ui.adsabs.harvard.edu/abs/2019ApJ...870..137K}{137}

\bibitem[{Keating {et~al.}(2016)Keating, Puchwein, Haehnelt, Bird, \&
  Bolton}]{Keating16}
Keating, L.~C., Puchwein, E., Haehnelt, M.~G., Bird, S., \& Bolton, J.~S. 2016,
  \href{http://dx.doi.org/10.1093/mnras/stw1306}{\color{magenta}\mnras}, 461,
  606

\bibitem[{{King} {et~al.}(2012){King}, {Webb}, {Murphy}, {Flambaum},
  {Carswell}, {Bainbridge}, {Wilczynska}, \& {Koch}}]{King12}
{King}, J.~A., {Webb}, J.~K., {Murphy}, M.~T., {et~al.} 2012,
  \href{http://dx.doi.org/10.1111/j.1365-2966.2012.20852.x}{\color{magenta}\mnras},
  \href{https://ui.adsabs.harvard.edu/abs/2012MNRAS.422.3370K}{422},
  \href{https://ui.adsabs.harvard.edu/abs/2012MNRAS.422.3370K}{3370}

\bibitem[{{Komatsu} {et~al.}(2009){Komatsu}, {Dunkley}, {Nolta}, {Bennett},
  {Gold}, {Hinshaw}, {Jarosik}, {Larson}, {Limon}, {Page}, {Spergel},
  {Halpern}, {Hill}, {Kogut}, {Meyer}, {Tucker}, {Weiland}, {Wollack}, \&
  {Wright}}]{WMAP09}
{Komatsu}, E., {Dunkley}, J., {Nolta}, M.~R., {et~al.} 2009,
  \href{http://dx.doi.org/10.1088/0067-0049/180/2/330}{\color{magenta}\apjs},
  \href{https://ui.adsabs.harvard.edu/abs/2009ApJS..180..330K}{180},
  \href{https://ui.adsabs.harvard.edu/abs/2009ApJS..180..330K}{330}

\bibitem[{{Kulkarni} {et~al.}(2019){Kulkarni}, {Worseck}, \&
  {Hennawi}}]{Kulkarni19}
{Kulkarni}, G., {Worseck}, G., \& {Hennawi}, J.~F. 2019,
  \href{http://dx.doi.org/10.1093/mnras/stz1493}{\color{magenta}\mnras},
  \href{https://ui.adsabs.harvard.edu/abs/2019MNRAS.488.1035K}{488},
  \href{https://ui.adsabs.harvard.edu/abs/2019MNRAS.488.1035K}{1035}

\bibitem[{{Lanzetta} {et~al.}(1987){Lanzetta}, {Turnshek}, \&
  {Wolfe}}]{Lanzetta87}
{Lanzetta}, K.~M., {Turnshek}, D.~A., \& {Wolfe}, A.~M. 1987,
  \href{http://dx.doi.org/10.1086/165769}{\color{magenta}\apj},
  \href{http://adsabs.harvard.edu/abs/1987ApJ...322..739L}{322},
  \href{http://adsabs.harvard.edu/abs/1987ApJ...322..739L}{739}

\bibitem[{{Lehner} {et~al.}(2019){Lehner}, {Burchett}, {Howk}, {O'Meara},
  {Peeples}, {Rafelski}, {Ribaudo}, \& {Tuttle}}]{lehner19}
{Lehner}, N., {Burchett}, J.~N., {Howk}, J.~C., {et~al.} 2019, \baas,
  \href{https://ui.adsabs.harvard.edu/abs/2019BAAS...51c.473L}{51},
  \href{https://ui.adsabs.harvard.edu/abs/2019BAAS...51c.473L}{473}

\bibitem[{{Lehner} {et~al.}(2016){Lehner}, {O'Meara}, {Howk}, {Prochaska}, \&
  {Fumagalli}}]{Lehner16}
{Lehner}, N., {O'Meara}, J.~M., {Howk}, J.~C., {Prochaska}, J.~X., \&
  {Fumagalli}, M. 2016,
  \href{http://dx.doi.org/10.3847/1538-4357/833/2/283}{\color{magenta}\apj},
  \href{https://ui.adsabs.harvard.edu/abs/2016ApJ...833..283L}{833},
  \href{https://ui.adsabs.harvard.edu/abs/2016ApJ...833..283L}{283}

\bibitem[{{Liang} \& {Chen}(2014)}]{LC14}
{Liang}, C.~J., \& {Chen}, H.-W. 2014,
  \href{http://dx.doi.org/10.1093/mnras/stu1901}{\color{magenta}\mnras},
  \href{https://ui.adsabs.harvard.edu/abs/2014MNRAS.445.2061L}{445},
  \href{https://ui.adsabs.harvard.edu/abs/2014MNRAS.445.2061L}{2061}

\bibitem[{{Lilly} {et~al.}(2013){Lilly}, {Carollo}, {Pipino}, {Renzini}, \&
  {Peng}}]{Lilly13}
{Lilly}, S.~J., {Carollo}, C.~M., {Pipino}, A., {Renzini}, A., \& {Peng}, Y.
  2013,
  \href{http://dx.doi.org/10.1088/0004-637X/772/2/119}{\color{magenta}\apj},
  \href{https://ui.adsabs.harvard.edu/abs/2013ApJ...772..119L}{772},
  \href{https://ui.adsabs.harvard.edu/abs/2013ApJ...772..119L}{119}

\bibitem[{{Madau} \& {Dickinson}(2014)}]{MD14}
{Madau}, P., \& {Dickinson}, M. 2014,
  \href{http://dx.doi.org/10.1146/annurev-astro-081811-125615}{\color{magenta}\araa},
  \href{https://ui.adsabs.harvard.edu/abs/2014ARA&A..52..415M}{52},
  \href{https://ui.adsabs.harvard.edu/abs/2014ARA&A..52..415M}{415}

\bibitem[{{Matejek} \& {Simcoe}(2012)}]{MS12}
{Matejek}, M.~S., \& {Simcoe}, R.~A. 2012,
  \href{http://dx.doi.org/10.1088/0004-637X/761/2/112}{\color{magenta}\apj},
  \href{https://ui.adsabs.harvard.edu/abs/2012ApJ...761..112M}{761},
  \href{https://ui.adsabs.harvard.edu/abs/2012ApJ...761..112M}{112}

\bibitem[{{Mathes}(2017)}]{nigelthesis}
{Mathes}, N.~L. 2017, PhD thesis, New Mexico State University

\bibitem[{{McQuinn}(2016)}]{Mcquinn16}
{McQuinn}, M. 2016,
  \href{http://dx.doi.org/10.1146/annurev-astro-082214-122355}{\color{magenta}\araa},
  \href{https://ui.adsabs.harvard.edu/abs/2016ARA&A..54..313M}{54},
  \href{https://ui.adsabs.harvard.edu/abs/2016ARA&A..54..313M}{313}

\bibitem[{{McQuinn} {et~al.}(2009){McQuinn}, {Lidz}, {Zaldarriaga},
  {Hernquist}, {Hopkins}, {Dutta}, \& {Faucher-Gigu{\`e}re}}]{Mcquinn09}
{McQuinn}, M., {Lidz}, A., {Zaldarriaga}, M., {et~al.} 2009,
  \href{http://dx.doi.org/10.1088/0004-637X/694/2/842}{\color{magenta}\apj},
  694, 842-866

\bibitem[{{M{\'e}nard} {et~al.}(2011){M{\'e}nard}, {Wild}, {Nestor}, {Quider},
  {Zibetti}, {Rao}, \& {Turnshek}}]{Menard11}
{M{\'e}nard}, B., {Wild}, V., {Nestor}, D., {et~al.} 2011,
  \href{http://dx.doi.org/10.1111/j.1365-2966.2011.18227.x}{\color{magenta}\mnras},
  \href{https://ui.adsabs.harvard.edu/abs/2011MNRAS.417..801M}{417},
  \href{https://ui.adsabs.harvard.edu/abs/2011MNRAS.417..801M}{801}

\bibitem[{{Misawa} {et~al.}(2002){Misawa}, {Tytler}, {Iye}, {Storrie-Lombardi},
  {Suzuki}, \& {Wolfe}}]{Misawa02}
{Misawa}, T., {Tytler}, D., {Iye}, M., {et~al.} 2002,
  \href{http://dx.doi.org/10.1086/339313}{\color{magenta}\aj},
  \href{https://ui.adsabs.harvard.edu/abs/2002AJ....123.1847M}{123},
  \href{https://ui.adsabs.harvard.edu/abs/2002AJ....123.1847M}{1847}

\bibitem[{{Morrison} {et~al.}(2019){Morrison}, {Pieri}, {Syphers}, \&
  {Kim}}]{Morrison19}
{Morrison}, S., {Pieri}, M.~M., {Syphers}, D., \& {Kim}, T.-S. 2019,
  \href{http://dx.doi.org/10.1093/mnras/stz2187}{\color{magenta}\mnras},
  \href{https://ui.adsabs.harvard.edu/abs/2019MNRAS.489..868M}{489},
  \href{https://ui.adsabs.harvard.edu/abs/2019MNRAS.489..868M}{868}

\bibitem[{{Muratov} {et~al.}(2015){Muratov}, {Kere{\v{s}}},
  {Faucher-Gigu{\`e}re}, {Hopkins}, {Quataert}, \& {Murray}}]{Muratov15}
{Muratov}, A.~L., {Kere{\v{s}}}, D., {Faucher-Gigu{\`e}re}, C.-A., {et~al.}
  2015, \href{http://dx.doi.org/10.1093/mnras/stv2126}{\color{magenta}\mnras},
  \href{https://ui.adsabs.harvard.edu/abs/2015MNRAS.454.2691M}{454},
  \href{https://ui.adsabs.harvard.edu/abs/2015MNRAS.454.2691M}{2691}

\bibitem[{{Muratov} {et~al.}(2017){Muratov}, {Kere{\v{s}}},
  {Faucher-Gigu{\`e}re}, {Hopkins}, {Ma}, {Angl{\'e}s-Alc{\'a}zar}, {Chan},
  {Torrey}, {Hafen}, {Quataert}, \& {Murray}}]{Muratov17}
---. 2017,
  \href{http://dx.doi.org/10.1093/mnras/stx667}{\color{magenta}\mnras},
  \href{https://ui.adsabs.harvard.edu/abs/2017MNRAS.468.4170M}{468},
  \href{https://ui.adsabs.harvard.edu/abs/2017MNRAS.468.4170M}{4170}

\bibitem[{{Murphy} {et~al.}(2011){Murphy}, {Chary}, {Dickinson}, {Pope},
  {Frayer}, \& {Lin}}]{Murphy11}
{Murphy}, E.~J., {Chary}, R.~R., {Dickinson}, M., {et~al.} 2011,
  \href{http://dx.doi.org/10.1088/0004-637X/732/2/126}{\color{magenta}\apj},
  \href{https://ui.adsabs.harvard.edu/abs/2011ApJ...732..126M}{732},
  \href{https://ui.adsabs.harvard.edu/abs/2011ApJ...732..126M}{126}

\bibitem[{{Murphy}(2016)}]{MurphyPOPLER}
{Murphy}, M.~T. 2016, {{UVES\_popler}: {POst} {PipeLine} {Echelle} {Reduction}
  software}.
\newblock \url{https://doi.org/10.5281/zenodo.56158}

\bibitem[{{Murphy} {et~al.}(2019){Murphy}, {Kacprzak}, {Savorgnan}, \&
  {Carswell}}]{Murphy19}
{Murphy}, M.~T., {Kacprzak}, G.~G., {Savorgnan}, G. A.~D., \& {Carswell}, R.~F.
  2019, \href{http://dx.doi.org/10.1093/mnras/sty2834}{\color{magenta}\mnras},
  \href{https://ui.adsabs.harvard.edu/abs/2019MNRAS.482.3458M}{482},
  \href{https://ui.adsabs.harvard.edu/abs/2019MNRAS.482.3458M}{3458}

\bibitem[{{Murphy} {et~al.}(2016){Murphy}, {Malec}, \& {Prochaska}}]{Murphy16}
{Murphy}, M.~T., {Malec}, A.~L., \& {Prochaska}, J.~X. 2016,
  \href{http://dx.doi.org/10.1093/mnras/stw1482}{\color{magenta}\mnras},
  \href{https://ui.adsabs.harvard.edu/abs/2016MNRAS.461.2461M}{461},
  \href{https://ui.adsabs.harvard.edu/abs/2016MNRAS.461.2461M}{2461}

\bibitem[{{Narayanan} {et~al.}(2005){Narayanan}, {Charlton}, {Masiero}, \&
  {Lynch}}]{N05}
{Narayanan}, A., {Charlton}, J.~C., {Masiero}, J.~R., \& {Lynch}, R. 2005,
  \href{http://dx.doi.org/10.1086/432750}{\color{magenta}\apj},
  \href{https://ui.adsabs.harvard.edu/abs/2005ApJ...632...92N}{632},
  \href{https://ui.adsabs.harvard.edu/abs/2005ApJ...632...92N}{92}

\bibitem[{{Navarro} {et~al.}(2010){Navarro}, {Ludlow}, {Springel}, {Wang},
  {Vogelsberger}, {White}, {Jenkins}, {Frenk}, \& {Helmi}}]{Navarro10}
{Navarro}, J.~F., {Ludlow}, A., {Springel}, V., {et~al.} 2010,
  \href{http://dx.doi.org/10.1111/j.1365-2966.2009.15878.x}{\color{magenta}\mnras},
  \href{https://ui.adsabs.harvard.edu/abs/2010MNRAS.402...21N}{402},
  \href{https://ui.adsabs.harvard.edu/abs/2010MNRAS.402...21N}{21}

\bibitem[{{Nelson} {et~al.}(2015){Nelson}, {Pillepich}, {Genel},
  {Vogelsberger}, {Springel}, {Torrey}, {Rodriguez-Gomez}, {Sijacki}, {Snyder},
  {Griffen}, {Marinacci}, {Blecha}, {Sales}, {Xu}, \& {Hernquist}}]{Nelson15}
{Nelson}, D., {Pillepich}, A., {Genel}, S., {et~al.} 2015,
  \href{http://dx.doi.org/10.1016/j.ascom.2015.09.003}{\color{magenta}Astronomy
  and Computing},
  \href{https://ui.adsabs.harvard.edu/abs/2015A&C....13...12N}{13},
  \href{https://ui.adsabs.harvard.edu/abs/2015A&C....13...12N}{12}

\bibitem[{{Ng} {et~al.}(2019){Ng}, {Nielsen}, {Kacprzak}, {Pointon}, {Muzahid},
  {Churchill}, \& {Charlton}}]{mason19}
{Ng}, M., {Nielsen}, N.~M., {Kacprzak}, G.~G., {et~al.} 2019,
  \href{http://dx.doi.org/10.3847/1538-4357/ab48eb}{\color{magenta}\apj},
  \href{https://ui.adsabs.harvard.edu/abs/2019ApJ...886...66N}{886},
  \href{https://ui.adsabs.harvard.edu/abs/2019ApJ...886...66N}{66}

\bibitem[{{Nielsen} {et~al.}(2013){Nielsen}, {Churchill}, {Kacprzak}, \&
  {Murphy}}]{nikki13}
{Nielsen}, N.~M., {Churchill}, C.~W., {Kacprzak}, G.~G., \& {Murphy}, M.~T.
  2013,
  \href{http://dx.doi.org/10.1088/0004-637X/776/2/114}{\color{magenta}\apj},
  \href{https://ui.adsabs.harvard.edu/abs/2013ApJ...776..114N}{776},
  \href{https://ui.adsabs.harvard.edu/abs/2013ApJ...776..114N}{114}

\bibitem[{{Nielsen} {et~al.}(2015){Nielsen}, {Churchill}, {Kacprzak}, {Murphy},
  \& {Evans}}]{nikki15}
{Nielsen}, N.~M., {Churchill}, C.~W., {Kacprzak}, G.~G., {Murphy}, M.~T., \&
  {Evans}, J.~L. 2015,
  \href{http://dx.doi.org/10.1088/0004-637X/812/1/83}{\color{magenta}\apj},
  \href{https://ui.adsabs.harvard.edu/abs/2015ApJ...812...83N}{812},
  \href{https://ui.adsabs.harvard.edu/abs/2015ApJ...812...83N}{83}

\bibitem[{{Nielsen} {et~al.}(2016){Nielsen}, {Churchill}, {Kacprzak}, {Murphy},
  \& {Evans}}]{nikki16}
---. 2016,
  \href{http://dx.doi.org/10.3847/0004-637X/818/2/171}{\color{magenta}\apj},
  \href{https://ui.adsabs.harvard.edu/abs/2016ApJ...818..171N}{818},
  \href{https://ui.adsabs.harvard.edu/abs/2016ApJ...818..171N}{171}

\bibitem[{{Nielsen} {et~al.}(2017){Nielsen}, {Kacprzak}, {Muzahid},
  {Churchill}, {Murphy}, \& {Charlton}}]{nikki17}
{Nielsen}, N.~M., {Kacprzak}, G.~G., {Muzahid}, S., {et~al.} 2017,
  \href{http://dx.doi.org/10.3847/1538-4357/834/2/148}{\color{magenta}\apj},
  \href{https://ui.adsabs.harvard.edu/abs/2017ApJ...834..148N}{834},
  \href{https://ui.adsabs.harvard.edu/abs/2017ApJ...834..148N}{148}

\bibitem[{{Nielsen} {et~al.}(2018){Nielsen}, {Kacprzak}, {Pointon},
  {Churchill}, \& {Murphy}}]{nikki18}
{Nielsen}, N.~M., {Kacprzak}, G.~G., {Pointon}, S.~K., {Churchill}, C.~W., \&
  {Murphy}, M.~T. 2018,
  \href{http://dx.doi.org/10.3847/1538-4357/aaedbd}{\color{magenta}\apj},
  \href{https://ui.adsabs.harvard.edu/abs/2018ApJ...869..153N}{869},
  \href{https://ui.adsabs.harvard.edu/abs/2018ApJ...869..153N}{153}

\bibitem[{{Nielsen} {et~al.}(2020){Nielsen}, {Kacprzak}, {Pointon}, {Murphy},
  {Churchill}, \& {Dav{\'e}}}]{nikki20}
{Nielsen}, N.~M., {Kacprzak}, G.~G., {Pointon}, S.~K., {et~al.} 2020, arXiv
  e-prints,
  \href{https://ui.adsabs.harvard.edu/abs/2020arXiv200208516N}{arXiv:2002.08516}

\bibitem[{Oliphant(2006)}]{numpy}
Oliphant, T.~E. 2006, A guide to NumPy, Vol.~1 (Trelgol Publishing USA)

\bibitem[{{O'Meara} {et~al.}(2015){O'Meara}, {Lehner}, {Howk}, {Prochaska},
  {Fox}, {Swain}, {Gelino}, {Berriman}, \& {Tran}}]{OMeara15}
{O'Meara}, J.~M., {Lehner}, N., {Howk}, J.~C., {et~al.} 2015,
  \href{http://dx.doi.org/10.1088/0004-6256/150/4/111}{\color{magenta}\aj},
  \href{https://ui.adsabs.harvard.edu/abs/2015AJ....150..111O}{150},
  \href{https://ui.adsabs.harvard.edu/abs/2015AJ....150..111O}{111}

\bibitem[{{Oppenheimer} \& {Dav{\'e}}(2008)}]{OD08}
{Oppenheimer}, B.~D., \& {Dav{\'e}}, R. 2008,
  \href{http://dx.doi.org/10.1111/j.1365-2966.2008.13280.x}{\color{magenta}\mnras},
  \href{https://ui.adsabs.harvard.edu/abs/2008MNRAS.387..577O}{387},
  \href{https://ui.adsabs.harvard.edu/abs/2008MNRAS.387..577O}{577}

\bibitem[{{Oppenheimer} {et~al.}(2010){Oppenheimer}, {Dav{\'e}}, {Kere{\v{s}}},
  {Fardal}, {Katz}, {Kollmeier}, \& {Weinberg}}]{Oppenheimer10}
{Oppenheimer}, B.~D., {Dav{\'e}}, R., {Kere{\v{s}}}, D., {et~al.} 2010,
  \href{http://dx.doi.org/10.1111/j.1365-2966.2010.16872.x}{\color{magenta}\mnras},
  \href{https://ui.adsabs.harvard.edu/abs/2010MNRAS.406.2325O}{406},
  \href{https://ui.adsabs.harvard.edu/abs/2010MNRAS.406.2325O}{2325}

\bibitem[{Oppenheimer {et~al.}(2009)Oppenheimer, Davé, \& Finlator}]{ODF09}
Oppenheimer, B.~D., Davé, R., \& Finlator, K. 2009,
  \href{http://dx.doi.org/10.1111/j.1365-2966.2009.14771.x}{\color{magenta}\mnras},
  396, 729-758

\bibitem[{{Parsa} {et~al.}(2016){Parsa}, {Dunlop}, {McLure}, \&
  {Mortlock}}]{Parsa16}
{Parsa}, S., {Dunlop}, J.~S., {McLure}, R.~J., \& {Mortlock}, A. 2016,
  \href{http://dx.doi.org/10.1093/mnras/stv2857}{\color{magenta}\mnras},
  \href{https://ui.adsabs.harvard.edu/abs/2016MNRAS.456.3194P}{456},
  \href{https://ui.adsabs.harvard.edu/abs/2016MNRAS.456.3194P}{3194}

\bibitem[{{Perez} \& {Granger}(2007)}]{ipython}
{Perez}, F., \& {Granger}, B.~E. 2007, Computing in Science Engineering, 9,
  21-29

\bibitem[{{P{\'e}roux} {et~al.}(2004){P{\'e}roux}, {Petitjean}, {Aracil},
  {Irwin}, \& {McMahon}}]{Peroux04}
{P{\'e}roux}, C., {Petitjean}, P., {Aracil}, B., {Irwin}, M., \& {McMahon},
  R.~G. 2004,
  \href{http://dx.doi.org/10.1051/0004-6361:20034213}{\color{magenta}\aap},
  \href{https://ui.adsabs.harvard.edu/abs/2004A&A...417..443P}{417},
  \href{https://ui.adsabs.harvard.edu/abs/2004A&A...417..443P}{443}

\bibitem[{{Petitjean} \& {Bergeron}(1990)}]{PB90}
{Petitjean}, P., \& {Bergeron}, J. 1990, \aap,
  \href{https://ui.adsabs.harvard.edu/abs/1990A&A...231..309P}{231},
  \href{https://ui.adsabs.harvard.edu/abs/1990A&A...231..309P}{309}

\bibitem[{{Petitjean} \& {Bergeron}(1994)}]{PB94}
---. 1994, \aap,
  \href{https://ui.adsabs.harvard.edu/abs/1994A&A...283..759P}{283},
  \href{https://ui.adsabs.harvard.edu/abs/1994A&A...283..759P}{759}

\bibitem[{{Pettini} {et~al.}(2003){Pettini}, {Madau}, {Bolte}, {Prochaska},
  {Ellison}, \& {Fan}}]{Pettini03}
{Pettini}, M., {Madau}, P., {Bolte}, M., {et~al.} 2003,
  \href{http://dx.doi.org/10.1086/377043}{\color{magenta}\apj},
  \href{https://ui.adsabs.harvard.edu/abs/2003ApJ...594..695P}{594},
  \href{https://ui.adsabs.harvard.edu/abs/2003ApJ...594..695P}{695}

\bibitem[{{Pointon} {et~al.}(2017){Pointon}, {Nielsen}, {Kacprzak}, {Muzahid},
  {Churchill}, \& {Charlton}}]{pointon17}
{Pointon}, S.~K., {Nielsen}, N.~M., {Kacprzak}, G.~G., {et~al.} 2017,
  \href{http://dx.doi.org/10.3847/1538-4357/aa7743}{\color{magenta}\apj},
  \href{https://ui.adsabs.harvard.edu/abs/2017ApJ...844...23P}{844},
  \href{https://ui.adsabs.harvard.edu/abs/2017ApJ...844...23P}{23}

\bibitem[{{Puchwein} {et~al.}(2019){Puchwein}, {Haardt}, {Haehnelt}, \&
  {Madau}}]{Puchwein19}
{Puchwein}, E., {Haardt}, F., {Haehnelt}, M.~G., \& {Madau}, P. 2019,
  \href{http://dx.doi.org/10.1093/mnras/stz222}{\color{magenta}\mnras},
  \href{https://ui.adsabs.harvard.edu/abs/2019MNRAS.485...47P}{485},
  \href{https://ui.adsabs.harvard.edu/abs/2019MNRAS.485...47P}{47}

\bibitem[{{Rafelski} {et~al.}(2012){Rafelski}, {Wolfe}, {Prochaska},
  {Neeleman}, \& {Mendez}}]{Rafelski12}
{Rafelski}, M., {Wolfe}, A.~M., {Prochaska}, J.~X., {Neeleman}, M., \&
  {Mendez}, A.~J. 2012,
  \href{http://dx.doi.org/10.1088/0004-637X/755/2/89}{\color{magenta}\apj},
  \href{https://ui.adsabs.harvard.edu/abs/2012ApJ...755...89R}{755},
  \href{https://ui.adsabs.harvard.edu/abs/2012ApJ...755...89R}{89}

\bibitem[{{Rahmati} {et~al.}(2016){Rahmati}, {Schaye}, {Crain}, {Oppenheimer},
  {Schaller}, \& {Theuns}}]{Rahmati16}
{Rahmati}, A., {Schaye}, J., {Crain}, R.~A., {et~al.} 2016,
  \href{http://dx.doi.org/10.1093/mnras/stw453}{\color{magenta}\mnras},
  \href{https://ui.adsabs.harvard.edu/abs/2016MNRAS.459..310R}{459},
  \href{https://ui.adsabs.harvard.edu/abs/2016MNRAS.459..310R}{310}

\bibitem[{{Rauch} {et~al.}(1996){Rauch}, {Sargent}, {Womble}, \&
  {Barlow}}]{Rauch96}
{Rauch}, M., {Sargent}, W.~L.~W., {Womble}, D.~S., \& {Barlow}, T.~A. 1996,
  \href{http://dx.doi.org/10.1086/310187}{\color{magenta}\apjl},
  \href{https://ui.adsabs.harvard.edu/abs/1996ApJ...467L...5R}{467},
  \href{https://ui.adsabs.harvard.edu/abs/1996ApJ...467L...5R}{L5}

\bibitem[{{Robertson} {et~al.}(2013){Robertson}, {Furlanetto}, {Schneider},
  {Charlot}, {Ellis}, {Stark}, {McLure}, {Dunlop}, {Koekemoer}, {Schenker},
  {Ouchi}, {Ono}, {Curtis-Lake}, {Rogers}, {Bowler}, \&
  {Cirasuolo}}]{Robertson13}
{Robertson}, B.~E., {Furlanetto}, S.~R., {Schneider}, E., {et~al.} 2013,
  \href{http://dx.doi.org/10.1088/0004-637X/768/1/71}{\color{magenta}\apj},
  \href{https://ui.adsabs.harvard.edu/abs/2013ApJ...768...71R}{768},
  \href{https://ui.adsabs.harvard.edu/abs/2013ApJ...768...71R}{71}

\bibitem[{{Rupke}(2018)}]{Rupke18}
{Rupke}, D. 2018,
  \href{http://dx.doi.org/10.3390/galaxies6040138}{\color{magenta}Galaxies},
  \href{https://ui.adsabs.harvard.edu/abs/2018Galax...6..138R}{6},
  \href{https://ui.adsabs.harvard.edu/abs/2018Galax...6..138R}{138}

\bibitem[{{Sargent} {et~al.}(1988){Sargent}, {Boksenberg}, \&
  {Steidel}}]{Sargent88}
{Sargent}, W. L.~W., {Boksenberg}, A., \& {Steidel}, C.~C. 1988,
  \href{http://dx.doi.org/10.1086/191300}{\color{magenta}\apjs},
  \href{https://ui.adsabs.harvard.edu/abs/1988ApJS...68..539S}{68},
  \href{https://ui.adsabs.harvard.edu/abs/1988ApJS...68..539S}{539}

\bibitem[{{Scannapieco} {et~al.}(2006){Scannapieco}, {Pichon}, {Aracil},
  {Petitjean}, {Thacker}, {Pogosyan}, {Bergeron}, \& {Couchman}}]{Scanna06}
{Scannapieco}, E., {Pichon}, C., {Aracil}, B., {et~al.} 2006,
  \href{http://dx.doi.org/10.1111/j.1365-2966.2005.09753.x}{\color{magenta}\mnras},
  \href{https://ui.adsabs.harvard.edu/abs/2006MNRAS.365..615S}{365},
  \href{https://ui.adsabs.harvard.edu/abs/2006MNRAS.365..615S}{615}

\bibitem[{{Schaye} {et~al.}(2003){Schaye}, {Aguirre}, {Kim}, {Theuns}, {Rauch},
  \& {Sargent}}]{Schaye03}
{Schaye}, J., {Aguirre}, A., {Kim}, T.-S., {et~al.} 2003,
  \href{http://dx.doi.org/10.1086/378044}{\color{magenta}\apj},
  \href{https://ui.adsabs.harvard.edu/abs/2003ApJ...596..768S}{596},
  \href{https://ui.adsabs.harvard.edu/abs/2003ApJ...596..768S}{768}

\bibitem[{{Schaye} {et~al.}(2007){Schaye}, {Carswell}, \& {Kim}}]{Schaye07}
{Schaye}, J., {Carswell}, R.~F., \& {Kim}, T.-S. 2007,
  \href{http://dx.doi.org/10.1111/j.1365-2966.2007.12005.x}{\color{magenta}\mnras},
  \href{https://ui.adsabs.harvard.edu/abs/2007MNRAS.379.1169S}{379},
  \href{https://ui.adsabs.harvard.edu/abs/2007MNRAS.379.1169S}{1169}

\bibitem[{{Schechter}(1976)}]{Sch76}
{Schechter}, P. 1976,
  \href{http://dx.doi.org/10.1086/154079}{\color{magenta}\apj},
  \href{https://ui.adsabs.harvard.edu/abs/1976ApJ...203..297S}{203},
  \href{https://ui.adsabs.harvard.edu/abs/1976ApJ...203..297S}{297}

\bibitem[{{Schneider} {et~al.}(1993){Schneider}, {Hartig}, {Jannuzi},
  {Kirhakos}, {Saxe}, {Weymann}, {Bahcall}, {Bergeron}, {Boksenberg},
  {Sargent}, {Savage}, {Turnshek}, \& {Wolfe}}]{Schneider93}
{Schneider}, D.~P., {Hartig}, G.~F., {Jannuzi}, B.~T., {et~al.} 1993,
  \href{http://dx.doi.org/10.1086/191798}{\color{magenta}\apjs},
  \href{https://ui.adsabs.harvard.edu/abs/1993ApJS...87...45S}{87},
  \href{https://ui.adsabs.harvard.edu/abs/1993ApJS...87...45S}{45}

\bibitem[{{Shankar} {et~al.}(2009){Shankar}, {Weinberg}, \&
  {Miralda-Escud{\'e}}}]{Shankar09}
{Shankar}, F., {Weinberg}, D.~H., \& {Miralda-Escud{\'e}}, J. 2009,
  \href{http://dx.doi.org/10.1088/0004-637X/690/1/20}{\color{magenta}\apj},
  \href{https://ui.adsabs.harvard.edu/abs/2009ApJ...690...20S}{690},
  \href{https://ui.adsabs.harvard.edu/abs/2009ApJ...690...20S}{20}

\bibitem[{{Shen} {et~al.}(2020){Shen}, {Hopkins}, {Faucher-Gigu{\`e}re},
  {Alexander}, {Richards}, {Ross}, \& {Hickox}}]{Shen20}
{Shen}, X., {Hopkins}, P.~F., {Faucher-Gigu{\`e}re}, C.-A., {et~al.} 2020,
  \href{http://dx.doi.org/10.1093/mnras/staa1381}{\color{magenta}\mnras},
  \href{https://ui.adsabs.harvard.edu/abs/2020MNRAS.495.3252S}{495},
  \href{https://ui.adsabs.harvard.edu/abs/2020MNRAS.495.3252S}{3252}

\bibitem[{{Shull} {et~al.}(2014){Shull}, {Danforth}, \& {Tilton}}]{Shull14}
{Shull}, J.~M., {Danforth}, C.~W., \& {Tilton}, E.~M. 2014,
  \href{http://dx.doi.org/10.1088/0004-637X/796/1/49}{\color{magenta}\apj},
  \href{https://ui.adsabs.harvard.edu/abs/2014ApJ...796...49S}{796},
  \href{https://ui.adsabs.harvard.edu/abs/2014ApJ...796...49S}{49}

\bibitem[{{Simcoe}(2011)}]{Simcoe11}
{Simcoe}, R.~A. 2011,
  \href{http://dx.doi.org/10.1088/0004-637X/738/2/159}{\color{magenta}\apj},
  \href{https://ui.adsabs.harvard.edu/abs/2011ApJ...738..159S}{738},
  \href{https://ui.adsabs.harvard.edu/abs/2011ApJ...738..159S}{159}

\bibitem[{{Simcoe} {et~al.}(2011){Simcoe}, {Cooksey}, {Matejek}, {Burgasser},
  {Bochanski}, {Lovegrove}, {Bernstein}, {Pipher}, {Forrest}, {McMurtry},
  {Fan}, \& {O'Meara}}]{Simcoe+11}
{Simcoe}, R.~A., {Cooksey}, K.~L., {Matejek}, M., {et~al.} 2011,
  \href{http://dx.doi.org/10.1088/0004-637X/743/1/21}{\color{magenta}\apj},
  \href{https://ui.adsabs.harvard.edu/abs/2011ApJ...743...21S}{743},
  \href{https://ui.adsabs.harvard.edu/abs/2011ApJ...743...21S}{21}

\bibitem[{{Somerville} \& {Dav{\'e}}(2015)}]{SD15}
{Somerville}, R.~S., \& {Dav{\'e}}, R. 2015,
  \href{http://dx.doi.org/10.1146/annurev-astro-082812-140951}{\color{magenta}\araa},
  \href{https://ui.adsabs.harvard.edu/abs/2015ARA&A..53...51S}{53},
  \href{https://ui.adsabs.harvard.edu/abs/2015ARA&A..53...51S}{51}

\bibitem[{{Songaila}(2001)}]{Songaila01}
{Songaila}, A. 2001,
  \href{http://dx.doi.org/10.1086/324761}{\color{magenta}\apjl},
  \href{https://ui.adsabs.harvard.edu/abs/2001ApJ...561L.153S}{561},
  \href{https://ui.adsabs.harvard.edu/abs/2001ApJ...561L.153S}{L153}

\bibitem[{{Songaila}(2005)}]{Songaila05}
---. 2005, \href{http://dx.doi.org/10.1086/491704}{\color{magenta}\aj},
  \href{https://ui.adsabs.harvard.edu/abs/2005AJ....130.1996S}{130},
  \href{https://ui.adsabs.harvard.edu/abs/2005AJ....130.1996S}{1996}

\bibitem[{{Songaila}(2006)}]{Songaila06}
---. 2006, \href{http://dx.doi.org/10.1086/498692}{\color{magenta}\aj},
  \href{https://ui.adsabs.harvard.edu/abs/2006AJ....131...24S}{131},
  \href{https://ui.adsabs.harvard.edu/abs/2006AJ....131...24S}{24}

\bibitem[{{Steidel}(1990)}]{Steidel90}
{Steidel}, C.~C. 1990,
  \href{http://dx.doi.org/10.1086/191407}{\color{magenta}\apjs},
  \href{https://ui.adsabs.harvard.edu/abs/1990ApJS...72....1S}{72},
  \href{https://ui.adsabs.harvard.edu/abs/1990ApJS...72....1S}{1}

\bibitem[{{Steidel}(1993)}]{Steidel93}
{Steidel}, C.~C. 1993, in Galaxy Evolution. The Milky Way Perspective, ed.
  S.~R. {Majewski}, Vol.~49, 227

\bibitem[{{Steidel} {et~al.}(2010){Steidel}, {Erb}, {Shapley}, {Pettini},
  {Reddy}, {Bogosavljevi{\'c}}, {Rudie}, \& {Rakic}}]{Steidel10}
{Steidel}, C.~C., {Erb}, D.~K., {Shapley}, A.~E., {et~al.} 2010,
  \href{http://dx.doi.org/10.1088/0004-637X/717/1/289}{\color{magenta}\apj},
  \href{https://ui.adsabs.harvard.edu/abs/2010ApJ...717..289S}{717},
  \href{https://ui.adsabs.harvard.edu/abs/2010ApJ...717..289S}{289}

\bibitem[{{Tody}(1986)}]{iraf}
{Tody}, D. 1986, in Society of Photo-Optical Instrumentation Engineers (SPIE)
  Conference Series, Vol. 627, \procspie, ed. D.~L. {Crawford}, 733

\bibitem[{{Tumlinson} {et~al.}(2017){Tumlinson}, {Peeples}, \&
  {Werk}}]{Tumlinson17}
{Tumlinson}, J., {Peeples}, M.~S., \& {Werk}, J.~K. 2017,
  \href{http://dx.doi.org/10.1146/annurev-astro-091916-055240}{\color{magenta}\araa},
  \href{https://ui.adsabs.harvard.edu/abs/2017ARA&A..55..389T}{55},
  \href{https://ui.adsabs.harvard.edu/abs/2017ARA&A..55..389T}{389}

\bibitem[{{Tumlinson} {et~al.}(2011){Tumlinson}, {Thom}, {Werk}, {Prochaska},
  {Tripp}, {Weinberg}, {Peeples}, {O'Meara}, {Oppenheimer}, {Meiring}, {Katz},
  {Dav{\'e}}, {Ford}, \& {Sembach}}]{cos-halos11}
{Tumlinson}, J., {Thom}, C., {Werk}, J.~K., {et~al.} 2011,
  \href{http://dx.doi.org/10.1126/science.1209840}{\color{magenta}Science},
  \href{https://ui.adsabs.harvard.edu/abs/2011Sci...334..948T}{334},
  \href{https://ui.adsabs.harvard.edu/abs/2011Sci...334..948T}{948}

\bibitem[{{van de Voort} {et~al.}(2011){van de Voort}, {Schaye}, {Booth}, \&
  {Dalla Vecchia}}]{vdv11b}
{van de Voort}, F., {Schaye}, J., {Booth}, C.~M., \& {Dalla Vecchia}, C. 2011,
  \href{http://dx.doi.org/10.1111/j.1365-2966.2011.18896.x}{\color{magenta}\mnras},
  \href{https://ui.adsabs.harvard.edu/abs/2011MNRAS.415.2782V}{415},
  \href{https://ui.adsabs.harvard.edu/abs/2011MNRAS.415.2782V}{2782}

\bibitem[{{V{\'e}ron-Cetty} \& {V{\'e}ron}(2001)}]{vcv01}
{V{\'e}ron-Cetty}, M.~P., \& {V{\'e}ron}, P. 2001,
  \href{http://dx.doi.org/10.1051/0004-6361:20010718}{\color{magenta}\aap},
  \href{https://ui.adsabs.harvard.edu/abs/2001A&A...374...92V}{374},
  \href{https://ui.adsabs.harvard.edu/abs/2001A&A...374...92V}{92}

\bibitem[{{Virtanen} {et~al.}(2020){Virtanen}, {Gommers}, {Oliphant},
  {Haberland}, {Reddy}, {Cournapeau}, {Burovski}, {Peterson}, {Weckesser},
  {Bright}, {van der Walt}, {Brett}, {Wilson}, {Jarrod Millman}, {Mayorov},
  {Nelson}, {Jones}, {Kern}, {Larson}, {Carey}, {Polat}, {Feng}, {Moore}, {Vand
  erPlas}, {Laxalde}, {Perktold}, {Cimrman}, {Henriksen}, {Quintero}, {Harris},
  {Archibald}, {Ribeiro}, {Pedregosa}, {van Mulbregt}, \&
  {Contributors}}]{scipy}
{Virtanen}, P., {Gommers}, R., {Oliphant}, T.~E., {et~al.} 2020,
  \href{http://dx.doi.org/https://doi.org/10.1038/s41592-019-0686-2}{\color{magenta}Nature
  Methods}, \href{https://rdcu.be/b08Wh}{17}, \href{https://rdcu.be/b08Wh}{261}

\bibitem[{{Vogelsberger} {et~al.}(2014){Vogelsberger}, {Genel}, {Springel},
  {Torrey}, {Sijacki}, {Xu}, {Snyder}, {Bird}, {Nelson}, \&
  {Hernquist}}]{Vogelsberger14}
{Vogelsberger}, M., {Genel}, S., {Springel}, V., {et~al.} 2014,
  \href{http://dx.doi.org/10.1038/nature13316}{\color{magenta}\nat},
  \href{https://ui.adsabs.harvard.edu/abs/2014Natur.509..177V}{509},
  \href{https://ui.adsabs.harvard.edu/abs/2014Natur.509..177V}{177}

\bibitem[{{Vogt} {et~al.}(1994)}]{hires94}
{Vogt}, S.~S., {et~al.} 1994, in \procspie, ed. D.~L. {Crawford} \& E.~R.
  {Craine}, Vol. 2198, 362

\bibitem[{{Worseck} {et~al.}(2016){Worseck}, {Prochaska}, {Hennawi}, \&
  {McQuinn}}]{Worseck16}
{Worseck}, G., {Prochaska}, J.~X., {Hennawi}, J.~F., \& {McQuinn}, M. 2016,
  \href{http://dx.doi.org/10.3847/0004-637X/825/2/144}{\color{magenta}\apj},
  \href{https://ui.adsabs.harvard.edu/abs/2016ApJ...825..144W}{825},
  \href{https://ui.adsabs.harvard.edu/abs/2016ApJ...825..144W}{144}

\bibitem[{{York} {et~al.}(2000)}]{York00}
{York}, D.~G., {et~al.} 2000,
  \href{http://dx.doi.org/10.1086/301513}{\color{magenta}\aj},
  \href{https://ui.adsabs.harvard.edu/abs/2000AJ....120.1579Y}{120},
  \href{https://ui.adsabs.harvard.edu/abs/2000AJ....120.1579Y}{1579}

\bibitem[{{Zackrisson} {et~al.}(2013){Zackrisson}, {Inoue}, \& {Jensen}}]{Z13}
{Zackrisson}, E., {Inoue}, A.~K., \& {Jensen}, H. 2013,
  \href{http://dx.doi.org/10.1088/0004-637X/777/1/39}{\color{magenta}\apj},
  \href{https://ui.adsabs.harvard.edu/abs/2013ApJ...777...39Z}{777},
  \href{https://ui.adsabs.harvard.edu/abs/2013ApJ...777...39Z}{39}

\bibitem[{{Zhu} \& {M{\'e}nard}(2013)}]{Zhu13}
{Zhu}, G., \& {M{\'e}nard}, B. 2013,
  \href{http://dx.doi.org/10.1088/0004-637X/770/2/130}{\color{magenta}\apj},
  \href{http://adsabs.harvard.edu/abs/2013ApJ...770..130Z}{770},
  \href{http://adsabs.harvard.edu/abs/2013ApJ...770..130Z}{130}

\end{thebibliography}

\end{document}